\documentclass[aps,prx,twocolumn,amsmath,amssymb,nofootinbib,superscriptaddress,floatfix]{revtex4-1}

\usepackage{amsmath,amsfonts,amssymb}
\usepackage{amssymb}
\usepackage{amsthm}
\usepackage[dvipdf]{color}
\usepackage{graphicx}
\usepackage{dcolumn}
\usepackage{bm}

\begin{document}

\title{Topological Invariant and Anomalous Edge States of Strongly Nonlinear Systems}

\author{Di Zhou}
\email{dizhou@bit.edu.cn}
\affiliation{Key Lab of Advanced Optoelectronic Quantum Architecture and Measurement (MOE) and School of Physics, Beijing Institute of Technology, Beijing 100081, China}

\author{D. Zeb Rocklin}
\affiliation{School of Physics, Georgia Institute of Technology, Atlanta, GA 30332, USA}

\author{Michael Leamy}
\affiliation{School of Mechanical Engineering, Georgia Institute of Technology, Atlanta, GA 30332, USA}

\author{Yugui Yao}
\email{ygyao@bit.edu.cn}
\affiliation{Key Lab of Advanced Optoelectronic Quantum Architecture and Measurement (MOE) and School of Physics, Beijing Institute of Technology, Beijing 100081, China}

\begin{abstract}
Despite the extensive studies of topological states, their characterization in strongly nonlinear classical systems has been lacking. In this work, we identify the proper definition of Berry phase for nonlinear bulk modes and characterize topological phases in one-dimensional (1D) generalized nonlinear Schr\"{o}dinger equations in the strongly nonlinear regime. We develop an analytic strategy to demonstrate the quantization of nonlinear Berry phase due to reflection symmetry. 
Mode amplitude itself plays a key role in nonlinear modes and controls topological phase transitions.
 We then show bulk-boundary correspondence by identifying the associated nonlinear topological edge modes. Interestingly, anomalous topological modes decay away from lattice boundaries to  plateaus governed by fixed points of nonlinearities. We propose passive photonic and active electrical systems that can be experimentally implemented. Our work opens the door to the rich physics between topological phases of matter and nonlinear dynamics. 
\end{abstract}

\maketitle

\section{Introduction}
The advent of topological band theory has led to the burgeoning field of ``topological phases of matter'' which manifest exotic properties, such as surface conduction of electronic states, and wave propagation insensitive to backscattering and disorder~\cite{PhysRevLett.61.2015, PhysRevLett.98.106803, RevModPhys.82.3045, ryu2010topological}. In classical structures~\cite{kane2014topological, susstrunk2015observation, hadad2018self, li2018weyl, PhysRevB.95.125104, PhysRevResearch.2.023173, PhysRevB.99.125116, PhysRevLett.103.248101}, enormous efforts have been devoted to topological states that emulate their quantum analogs and enable many pioneering applications~\cite{PhysRevLett.121.094301, smirnova2020nonlinear, li2014granular, nash2015topological, PhysRevLett.120.068003, luo2021observation, PhysRevX.9.021054, 2021arXiv210412778L, 2020arXiv201201639Z}.


To date, most of the studies of classical structures have been limited to linear topological band theory, with a few exceptions in the weakly nonlinear regime~\cite{PhysRevB.93.155112, hadad2018self, PhysRevB.100.014302, PhysRevE.97.032209} where perturbation theory is available. In 1D problems, the topological invariant called Berry phase~\cite{PhysRevLett.62.2747} is quantized by symmetries expressed as matrix operators. Due to bulk-boundary correspondence, topologically protected evanescent modes emerge on system boundaries. Although varied topological physics has been explored in linear systems, nonlinear dynamics are more ubiquitous in nature, such as biochemical processes~\cite{van1992stochastic}, fluid dynamics~\cite{RevModPhys.68.215}, and metamaterials~\cite{chen2014nonlinear, PhysRevB.101.104106}, etc. They give rise to rich properties like bifurcation~\cite{RevModPhys.63.991}, instability, solitons~\cite{PhysRevLett.42.1698}, and chaos~\cite{RevModPhys.57.617, RevModPhys.85.869}. The question naturally arises: can topological invariants and phases be extended to nonlinear systems?

In this paper, we present a systematic study of topological attributes in 1D generalized nonlinear Schr\"{o}dinger equations 
beyond Kerr-nonlinearities~\cite{PhysRevB.93.155112}. The nonlinear parts of interactions are comparable to the linear ones and perturbation theory breaks down, which we designate the ``strongly nonlinear regime". We limit our considerations within the amplitude range~\cite{narisetti2010perturbation, zaera2018propagation} that chaos does not occur. Consequently, nonlinear bulk modes~\cite{vakakis2001normal, fronk2017higher} are remarkably distinct from sinusoidal waves (e.g., fig.\ref{fig1}(b) and fig.\ref{SIfig8}(c)). We develop the proper definition of Berry phase in nonlinear bulk modes. By adopting a symmetry-based analytic treatment, we demonstrate the quantization of Berry phase in reflection-symmetric systems, regardless of the availability of linear analysis. The emergence of nonlinear topological edge modes is associated with a quantized Berry phase that protects them from defects. Finally, instead of exponentially localizing on lattice boundaries, topological edge modes exhibit anomalous behaviors that decay to a plateau governed by the stable fixed points of nonlinearities.

\section{Quantized Berry phase of nonlinear bulk modes}
Generalized nonlinear Schr\"{o}dinger equations are widely studied in classical systems like nonlinear optics~\cite{smirnova2020nonlinear, PhysRevLett.95.013902} and electrical circuits~\cite{hadad2018self}. Their equations of motion are summarized as the general form in Eqs.(\ref{1}) below. We study nonlinear bulk modes, from which we define Berry phase and demonstrate its quantization in reflection-symmetric models. 

\begin{figure}[htbp]
\includegraphics[scale=0.52]{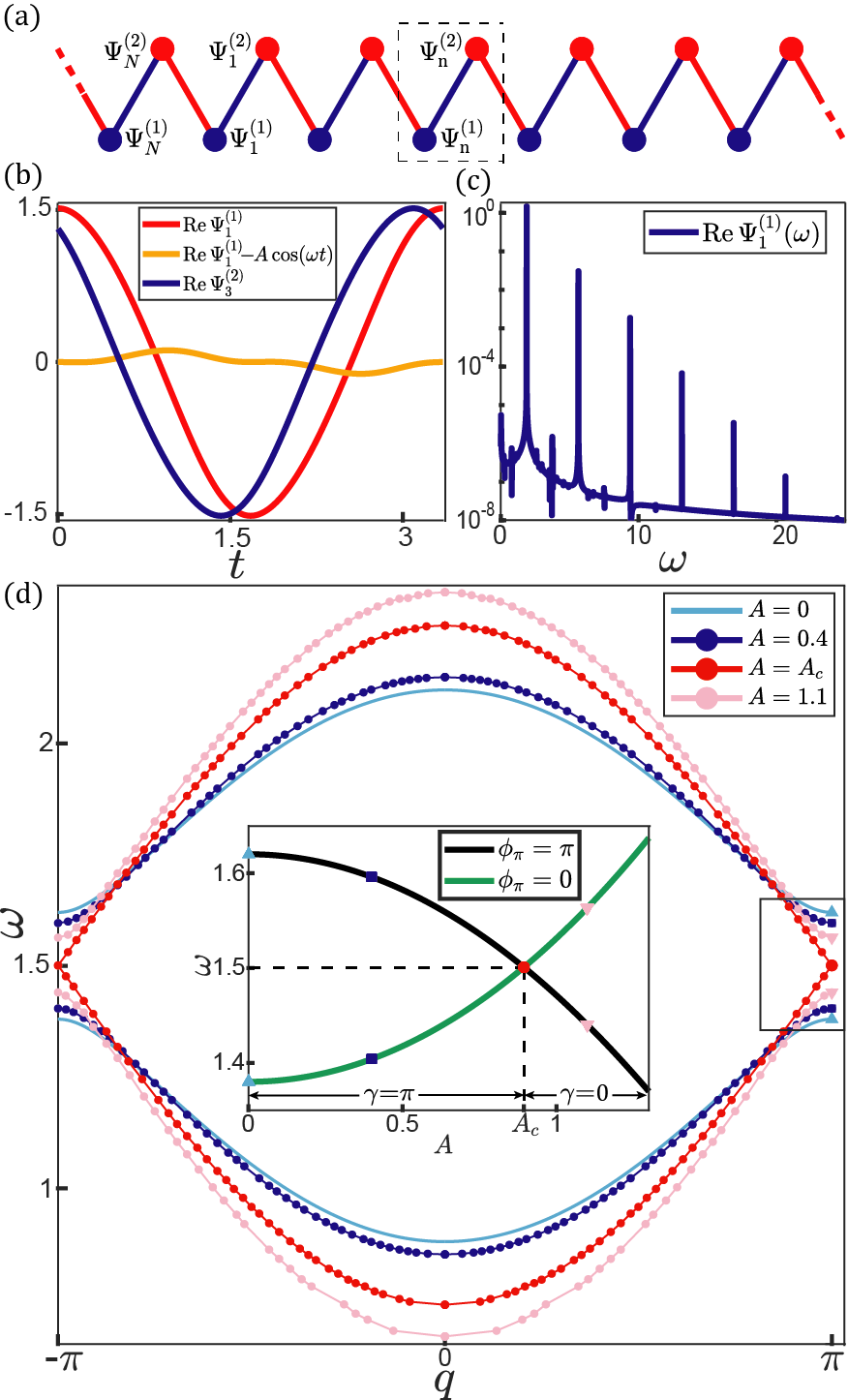}
\caption{The minimal model of nonlinear SSH chain. (a), schematic illustration of the lattice subjected to PBC. Unit cell is enclosed by black dashed box. Red and blue bonds represent intracell and intercell couplings. (b), a nonlinear bulk mode computed by \emph{shooting method}~\cite{renson2016numerical, ha2001nonlinear, peeters2008nonlinear} with amplitude $A=1.5$ and wavenumber $q=4\pi/5$. Red and blue curves are the wave functions of $n=1$ and $3$ sites, respectively. Orange curve shows the noticeable difference between nonlinear mode and sinusoidal function. (c), frequency profile of nonlinear bulk mode in (b). (d), nonlinear band structures $\omega = \omega(q,A)$ plotted for bulk mode amplitudes from $A=0$ to $1.1$. The red curves touch for the topological transition amplitude $A_c = 0.8944$ at $\omega=\epsilon_0=1.5$. The inset elaborates the gap-closing transition amplitude $A_c$ at which band inversion occurs. 
}\label{fig1}
\end{figure}

The considered model is a nonlinear SSH~\cite{PhysRevLett.42.1698} chain composed of $N$ classical dimer fields $\Psi_n = (\Psi_n^{(1)}, \Psi_n^{(2)})^\top$ ($\top$ is matrix transpose) coupled by nonlinear interactions, as represented
pictorially in Fig.\ref{fig1}(a). The chain dynamics is governed by the 1D generalized nonlinear Schr\"{o}dinger equations, 
\begin{eqnarray}\label{1}
 & {} & \mathrm{i}\partial_t\Psi^{(1)}_n = \epsilon_0\Psi^{(1)}_n+f_1(\Psi^{(1)}_n, \Psi_n^{(2)}) +f_2(\Psi^{(1)}_{n}, \Psi^{(2)}_{n-1}),  \nonumber \\
 & {} & \mathrm{i}\partial_t\Psi^{(2)}_n = \epsilon_0\Psi^{(2)}_n+f_1(\Psi^{(2)}_n, \Psi^{(1)}_n) +f_2(\Psi^{(2)}_{n}, \Psi^{(1)}_{n+1}) , 
\end{eqnarray}
subjected to periodic boundary condition (PBC), where $\epsilon_0\ge 0$ is the on-site potential, and $f_i(x,y)$ for $i=1$ and $i=2$ stand for intracell and intercell nonlinear couplings, respectively. $f_i(x,y)$ are real-coefficient general polynomials of $x$, $x^*$, $y$, and $y^*$ ($*$ represents complex conjugate), which offer time-reversal symmetry~\cite{RevModPhys.82.3045}. Given a nonlinear solution $\Psi_n(t)$, time-reversal symmetry demands a partner solution $\Psi_n^*(-t)$, as demonstrated in App.~B1. 
For systems such as those with Bose-Einstein condensates~\cite{RevModPhys.73.307}, 
$|\Psi(\vec r, t)|^2$ corresponds to a particle number density and third-order nonlinearities are thus limited to $|\Psi|^2 \Psi$ to enforce particle number conservation; in our case the fields do not correspond to particle densities and more general nonlinearities are thus permitted. 

In linear regime, the polynomials are approximated as $f_i(x,y)\approx c_i y$ ($c_{i=1,2}>0$) to have ``gapped" two-band models when $c_1\neq c_2$. The bulk mode eigenfunctions are sinusoidal in time, and Berry phase is quantized by reflection symmetry. In the ``strongly nonlinear regime" where nonlinear interactions become comparable to the linear ones, nonlinear bulk modes are significantly different from sinusoidal waves (e.g. figs.\ref{fig1}(b), \ref{SIfig8}(c)), and the frequencies naturally deviate from their linear counterparts. The nonlinearities become increasingly important as the bulk mode amplitude rises. Hence, the frequency of a nonlinear bulk mode is controlled both by wavenumber and amplitude. We thus define nonlinear band structure~\cite{hadad2018self, PhysRevB.93.155112} $\omega = \omega(q\in[-\pi,\pi], A)$ as the frequencies of nonlinear bulk modes for given amplitude $A$. We consider the simple case that nonlinear bulk modes are always non-degenerate (i.e., different modes at the same wavenumber have different frequencies) 
unless they reach the topological transition amplitude when the nonlinear bands merge at the band-touching frequency. Hence, given the amplitude, frequency, and wavenumber, a nonlinear bulk mode is uniquely defined. Extended from gapped linear models, the lattice is a ``gapped two-band nonlinear model''. In what follows, we define Berry phase for nonlinear bulk modes of the upper-band by adiabatically evolving the wavenumber across the Brillouin zone.

The considered nonlinear bulk mode is spatial-temporal periodic. It takes the traveling plane-wave ansatz, 
\begin{eqnarray}\label{1.2}
\Psi_{q} = (\Psi_q^{(1)}(\omega t-qn), \Psi_q^{(2)}(\omega t-qn+\phi_q))^\top,
\end{eqnarray}
where $\omega$ and $q$ are the frequency and wavenumber, respectively. $\Psi_q^{(j=1,2)}(\theta)$ are $2\pi$-periodic wave components, where the phase conditions are chosen by asking ${\rm Re\,}\Psi_{q}^{(j)}(\theta=0)=A$, and $A\overset{\rm def}{=}\max({\rm Re\,}\Psi_q^{(j)})$ is the amplitude. This is analogous to the phase condition ${\rm Re\,}\Psi(t=0)= \max({\rm Re\,}\Psi(t))$ adopted in Schr\"{o}dinger equation in order to have the eigenfunctions $\Psi(t) = |\Psi| e^{-\mathrm{i}\epsilon t/\hbar}$. 
Following this condition, $\phi_q$ in Eq.(\ref{1.2}) characterizes the relative phase between the two wave components. Nonlinear bulk modes are not sinusoidal. They fulfill $\mathrm{i} \partial_t \Psi_{q} = H(\Psi_{q})$, where $H(\Psi_{q})$ is the nonlinear function determined by Eqs.(\ref{1}) and is elaborated in App.~A. Given the band index and the amplitude $A$ of a nonlinear bulk mode, we find that $\omega$, $\phi_q$, and the waveform are determined by the wavenumber $q$.

We adopt the ansatz in Eq.(2) based on a number of reasons. First, typical studies on weakly nonlinear bulk modes~\cite{PhysRevB.99.125116, fronk2017higher, PhysRevE.97.032209, PhysRevB.101.104106, narisetti2010perturbation, zaera2018propagation, vakakis2001normal} reveal that the dynamics of all high-order harmonics are controlled by the single variable $\theta=\omega t-qn$: $\Psi_q^{(j)} = \sum_l \psi_{l,q}^{(j)} e^{-\mathrm{i}l(\omega t -qn)}$, where $\psi_{l,q}^{(j)} = (2\pi)^{-1}\int_0^{2\pi} e^{\mathrm{i} l\theta} \Psi_q^{(j)}d\theta$ is the $l$-th Fourier component of $\Psi_q^{(j)}$. Second, numerical experiments such as shooting method (see figs.\ref{fig1}(b), \ref{SIfig8}(a,c), and Refs.~\cite{renson2016numerical, ha2001nonlinear, peeters2008nonlinear}) manifest non-dispersive, plane-wave like bulk modes in the strongly nonlinear regime. Finally, it is demonstrated in App.~C3 that the analytic solutions of nonlinear bulk modes at high-symmetry wavenumbers are in perfect agreement with Eq.(2). 

We realize the adiabatic evolution of wavenumber $q(t')$ traversing the Brillouin zone from $q(0)=q$ to $q(t) = q+2\pi$, while the amplitude $A$ remains unchanged during this process. According to the nonlinear extension of the adiabatic theorem~\cite{RevModPhys.82.1959, PhysRevLett.90.170404, PhysRevLett.98.050406, PhysRevA.81.052112}, a system $H(\Psi_{q})$ initially in one of the nonlinear modes $\Psi_{q}$ will stay as an instantaneous nonlinear mode of $H(\Psi_{q(t)})$ throughout this procedure, provided that the nonlinear mode $\Psi_{q}$ is stable~\cite{PhysRevLett.98.050406} within the amplitude scope of this paper. The stability of nonlinear bulk modes is confirmed in App.~C via the algorithm of self-oscillation~\cite{fronk2017higher, PhysRevB.99.125116, PhysRevE.97.032209}. Therefore, the only degree of freedom is the phase of mode. At time $t$, the mode is $\Psi_{q(t)}(\int_0^t \omega(t', q(t'))dt' -\gamma(t))$, where $\gamma(t)$ defines the phase shift of the nonlinear bulk mode in the adiabatic evolution. The dynamics of $\gamma$ is depicted by $(d\gamma/dt) (\partial\Psi_q/\partial\theta)=(dq/dt) (\partial\Psi_q/\partial q)$. After $q$ traverses the Brillouin zone, the wave function acquires an extra phase $\gamma$ dubbed Berry phase of nonlinear bulk modes, 
\begin{eqnarray}\label{1.3}
\gamma
=
\oint_{\rm BZ}dq \frac{\sum_{l\in\mathcal{Z}} \left( l |\psi_{l,q}^{(2)} |^2 \frac{\partial\phi_q}{\partial q}+\mathrm{i}\sum_{j}\psi_{l,q}^{(j)*}\frac{\partial\psi_{l,q}^{(j)}}{\partial q}\right)}
{\sum_{l'\in\mathcal{Z}}  l'  \left(\sum_{j'}|\psi_{l',q}^{(j')}|^2\right)},\quad
\end{eqnarray}
where $j, j'=1,2$ denote the two wave components, and the mathematical derivations are in App.~A. In general, $\gamma$ is \emph{not quantized} unless additional symmetry properties are imposed on the model, which we will discuss below. We note that the eigenmodes of linear problems are sinusoidal in time, which reduces Eq.(\ref{1.3}) to the conventional form~\cite{RevModPhys.82.1959} $\gamma = \oint_{\rm BZ}\mathrm{d}q\,\mathrm{i}\langle\Psi_q|\partial_q|\Psi_q\rangle$.

Now we demonstrate that Berry phase defined in Eq.(\ref{1.3}) is quantized by reflection symmetry. The model in Eqs.(\ref{1}) respects reflection symmetry, which means that the nonlinear equations of motion are invariant under reflection transformation, 
\begin{eqnarray}\label{3.11}
(\Psi^{(1)}_n, \Psi^{(2)}_n) \to (\Psi^{(2)}_{-n}, \Psi^{(1)}_{-n}).
\end{eqnarray}
Given a nonlinear bulk mode $\Psi_q$ in Eq.(\ref{1.2}), reflection transformation demands a partner solution $\Psi_{-q}'  = (\Psi^{(2)}_q(\omega t +qn), \Psi^{(1)}_q(\omega t+qn-\phi_q))^\top$ that also satisfies the model. On the other hand, a nonlinear bulk mode of wavenumber $-q$ is by definition denoted as $
\Psi_{-q} = (\Psi_{-q}^{(1)}(\omega t+qn), \Psi_{-q}^{(2)}(\omega t+qn+\phi_{-q}))^\top
$. Since there is no degeneracy of nonlinear bulk modes, $\Psi_{-q}'$ and $\Psi_{-q}$ have to be identical, which imposes the constraints
\begin{eqnarray}\label{4}
\phi_{-q} = -\phi_q \mod 2\pi, \quad{\rm and}\quad 
 \Psi^{(2)}_q = \Psi^{(1)}_{-q}.
\end{eqnarray}
Thus, the Fourier components of nonlinear bulk modes satisfy $\psi_{l,q}^{(2)} = \psi_{l,-q}^{(1)}$. This relationship, together with Eqs.(\ref{4}), is the key to quantizing the Berry phase in Eq.(\ref{1.3}) (details in App.~B2),
\begin{eqnarray}\label{4.12}
\gamma = \frac{1}{2}\oint_{\rm BZ} \frac{d\phi_q}{dq} dq =\phi_\pi-\phi_0 =  0{\rm \,\, or\,\,}\pi\mod 2\pi, 
\end{eqnarray}
where $\phi_{q=0,\pi}$ are the relative phases of the upper-band nonlinear modes at high-symmetry points. They are determined by comparing the frequencies $\omega(\phi_q=0)$ and $\omega(\phi_q=\pi)$ for $q=0$ and $\pi$. $\gamma=\pi$ if $\omega(\phi_0=0)$ and $\omega(\phi_\pi = \pi)$ belong to the same band, whereas $\gamma=0$ if they are in different bands. Interestingly, $\gamma$ encounters a topological transition induced by the critical amplitude $A=A_c$ if frequencies merge at $\omega(\phi_\pi = 0, A_c) = \omega(\phi_\pi=\pi, A_c)$. This transition is exemplified by the minimal model of nonlinear topological lattice in Sec.\uppercase\expandafter{\romannumeral3}. It is worth emphasizing that despite all the discussions of nonlinear Schr\"{o}dinger equations and the quantization of Berry phase, the model is purely classical in the sense of $\hbar$ being zero.

Having established quantized Berry phase, we now search additional properties for vanishing on-site potential, $\epsilon_0=0$. The model's linear limit respects charge-conjugation symmetry~\cite{ryu2010topological, kane2014topological}, which demands that the states  appear in $\pm\omega$ pairs, and the topological mode  have zero-energy. To have $\pm\omega$ pairs of modes in the nonlinear problem, we require the parity of interactions to satisfy $f_i(-x, y) =-f_i(x, -y)= f_i(x,y)$. Consequently, the system is invariant under the transformation $(\Psi^{(1)}_n(\omega t), \Psi^{(2)}_n(\omega t)) \to (-\Psi^{(1)}_{n}(-\omega t), \Psi^{(2)}_{n}(-\omega t))$. Given a nonlinear mode $\Psi_{\omega}$ defined in Eq.(\ref{1.2}), this transformation demands a partner solution $\Psi_{-\omega} = (-\Psi^{(1)}_q(-\omega t -qn), \Psi^{(2)}_q(-\omega t-qn+\phi_q))^\top$. Therefore, nonlinear modes always appear in $\pm\omega$ pairs. Similar to charge-conjugation symmetric models in linear systems~\cite{ryu2010topological}, the frequencies of nonlinear topological modes are zero, which is illustrated in the following minimal model.

\section{Topological transition and bulk-boundary correspondence in the minimal model} 

We now clarify the nonlinear extension of bulk-boundary correspondence~\cite{PhysRevB.100.014302, PhysRevA.97.043602} by demonstrating topological edge modes in the minimal model that respects time-reversal symmetry, where the couplings are specified as  
\begin{eqnarray}\label{9}
f_i (x,y) = c_i y + d_i [({\rm Re\,}y)^3+\mathrm{i} ({\rm Im\,}y)^3],
\end{eqnarray}
with $c_i, d_i>0$ for $i= 1,2$. This interaction offers numerically stable nonlinear bulk and topological edge modes and can be realized in passive photonic and active electrical circuit metamaterials (Sec.\uppercase\expandafter{\romannumeral4} and App.~E). We are interested in attributes unique to nonlinear systems, in particular the topological phase transition induced by bulk mode amplitudes. Thus, the parameters yield $c_1<c_2$ and $d_1>d_2$ ($c_1>c_2$ and $d_1<d_2$) to induce topological-to-non-topological phase transition (non-topological-to-topological transition) as amplitudes increase. 
We abbreviate them as ``T-to-N" and ``N-to-T" transitions, and they are converted to one another by simply flipping intracell and intercell couplings. In the remainder of this paper, a semi-infinite lattice subjected to open boundary condition (OBC) is always considered whenever we refer to topological edge modes.

We first study the case $c_1<c_2$ and $d_1>d_2$, in which a T-to-N transition occurs. Fig.\ref{fig1}(d) numerically illustrates nonlinear band structures and topological transition by considering $\epsilon_0=1.5$, $c_1=0.25$, $c_2=0.37$, $d_1=0.22$, and $d_2=0.02$. Given that Berry phase $\gamma(A=0)=\pi$, the lattice is topologically nontrivial in the linear limit. As amplitudes rise, the topological invariant $\gamma(A<A_c)=\pi$ cannot change until it becomes ill-defined when the nonlinear bandgap closes at the transition amplitude $A_c$. The bandgap reopens above $A_c$, allowing the well-defined Berry phase to take the trivial value $\gamma(A>A_c)=0$, as depicted in the inset of Fig.\ref{fig1}(d). $A_c$ is numerically computed by solving the bandgap-closing equation $\omega(\phi_\pi = 0, A_c)=\omega(\phi_\pi=\pi, A_c)$. We propose a convenient approximation~\cite{detroux2014harmonic} $f(\Psi_{n'}^{(j')}, \Psi_n^{(j)})\approx (c_i+\frac{3}{4}d_i A^2)\Psi_n^{(j)}$ to estimate the transition amplitude $A_c\approx \sqrt{-4(c_2-c_1)/3(d_2-d_1)}$. The good agreement between this approximation and the numerical solutions is shown in App.~C. We highlight that $A_c^2\max(d_1,d_2) /\max(c_1, c_2) \approx 0.5$, which demonstrates the comparable nonlinear and linear interactions in the strongly nonlinear regime.

\widetext

\begin{figure}[htbp]
\includegraphics[scale=0.55]{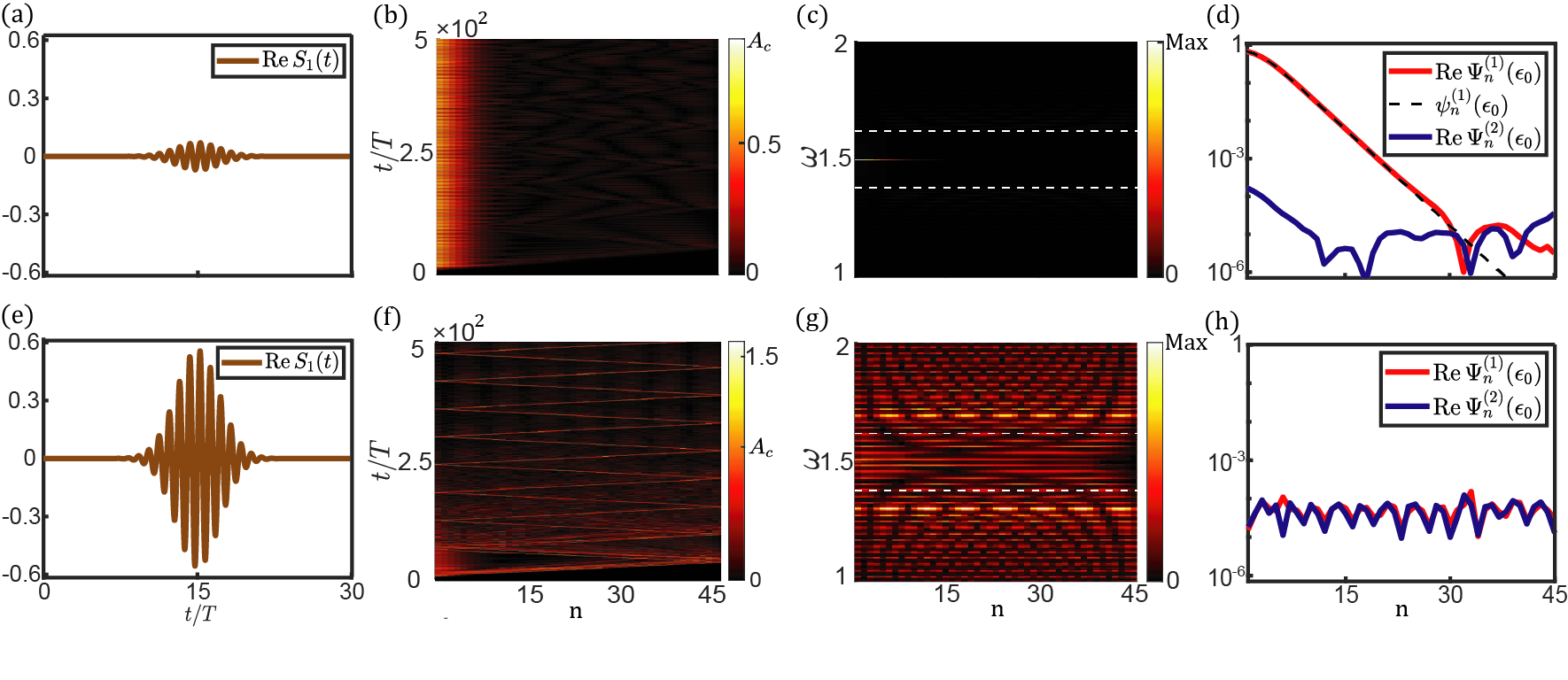}
\caption{Nonlinear edge excitations of the model subjected to T-to-N transition, where the parameters fulfill $c_1<c_2$ and $d_1>d_2$. (a-d) and (e-h) show lattice boundary responses in small-amplitude topological regime and large-amplitude non-topological regime, respectively. The magnitudes of Gaussian tone bursts are $S=7\times 10^{-2}$ in (a) and $S=56\times 10^{-2}$ in (e), respectively. (b) and (f), spatial-temporal profiles of $|{\rm Re\,}\Psi_n^{(1)}(t)|$ for all $45$ sites, where $|{\rm Re\,}\Psi_n^{(1)}(t)|$ denote the strength of the lattice excitations. (c) and (g), spatial profiles of the frequency spectra of the responding modes, where the time domain of performing Fourier analysis is from $250T$ to $500T$. White dashed lines mark the top and bottom of the linear bandgap. In (g), modes in the bandgap are triggered by energy absorption~\cite{fronk2017higher} from nonlinear bulk modes. (d) and (h), red and blue curves for the spatial profiles of the $\omega=\epsilon_0$ wave component of the excitations. The analytic prediction of the topological mode $\psi_n^{(1)}(\epsilon_0)$ is depicted by the black dashed curve in (d). 
}\label{fig2}
\end{figure}

\endwidetext

Bulk-boundary correspondence has been extended to weakly nonlinear Newtonian~\cite{PhysRevB.100.014302} and Schr\"{o}dinger~\cite{PhysRevA.97.043602} systems by showing topological boundary modes guaranteed by topologically non-trivial Berry phase. In the strongly nonlinear problem, we utilize analytic approximation and numerical experiment, to doubly confirm this correspondence by identifying nonlinear topological edge modes. In the former, the lattice is composed of $N=45$ unit cells with OBCs on both ends to mimic semi-infinite lattice, and the parameters are carried over from Fig.\ref{fig1}. The topological mode and frequency are denoted as $\Psi_n=(\Psi_n^{(1)}, \Psi_n^{(2)})^\top$ and $\omega_{\rm T}$, respectively. Analogous to linear SSH chain~\cite{PhysRevLett.42.1698}, the analytic scheme is to approximate $\Psi_{n}^{(1)}\gg\Psi_{n}^{(2)}$, which is numerically verified in Fig.\ref{fig2}(d). We make one further approximation to truncate the equations of motion to fundamental harmonics. Therefore, the nonlinear topological edge mode is approximated as $\Psi_n \approx (\psi_{1,n}^{(1)}, 0)^\top e^{-\mathrm{i}\epsilon_0 t}$, where $\psi_{1,n}^{(1)}$ are the fundamental harmonic components. By doing so, we find $\omega_{\rm T} = \epsilon_0$, and
\begin{eqnarray}\label{Recur}
\left(c_1+\frac{3}{4}d_1|\psi_{1,n}^{(1)}|^2\right) |\psi_{1,n}^{(1)}|=\left(c_2+\frac{3}{4}d_2|\psi_{1,n+1}^{(1)}|^2\right) |\psi_{1,n+1}^{(1)}|. \nonumber \\
\end{eqnarray}
From Eq.(\ref{Recur}), the semi-infinite lattice hosts topological evanescent modes when $|\Psi_{1}^{(1)}|<\sqrt{-4(c_2-c_1)/3(d_2-d_1)}\approx A_c$, whereas no such mode exists for $|\Psi_{1}^{(1)}|>\sqrt{-4(c_2-c_1)/3(d_2-d_1)}\approx A_c$. In App.~D, the frequency and analytic expression are applied in weakly nonlinear regime, and they are perfectly in line with method of multiple-scale~\cite{fronk2017higher, narisetti2010perturbation, zaera2018propagation, SNEE2019100487}. The numerical scenario is accomplished by applying a Gaussian profile signal $S_n =\delta_{n1} S e^{-\mathrm{i}\omega_{\rm ext} t-(t-t_0)^2/\tau^2} (1, 0)^\top$ on the first site, where the carrier frequency $\omega_{\rm ext} = \epsilon_0=1.5$, $T=2\pi/\omega_{\rm ext}$, $\tau=3T$ controls Gaussian spread, and $t_0 = 15T$ denotes trigger time. Figs.\ref{fig2}(b) and (f) together verify bulk-boundary correspondence~\cite{PhysRevB.100.014302, PhysRevA.97.043602} by identifying the presence and absence of topological boundary excitations below and above the critical amplitude $A_c$, respectively. In fig.\ref{fig2}(d), the flattened part near the lattice boundary is the manifestation of nonlinearities.

One may find it unusual that the frequencies of topological modes $\omega_{\rm T}=\epsilon_0$ are independent of amplitudes, although this result is in agreement with Ref.~\cite{hadad2018self, PhysRevB.93.155112, PhysRevB.100.014302} in weakly nonlinear regime. Here we propose an 
explanation for this intriguing result. Because the evanescent mode fades to zero in the bulk, the ``tail" of this mode eventually enters into small-amplitude regime where nonlinearities are negligible and linear analysis becomes effective. Linear topological theory~\cite{PhysRevLett.42.1698} demands the tail of the mode to be $\omega_{\rm T}=\epsilon_0$, which in turn requires the frequency of the nonlinear topological mode to be independent of the amplitude.

Topological protection is featured in multiple aspects. As visualized in Fig.\ref{fig1}(d), the frequencies of topological modes stay in the bandgap and are distinct from nonlinear bulk modes. The appearance and absence of these modes are captured by the topological invariant that cannot change continuously upon the variation of system parameters. Lastly, topological modes are insensitive to defects, which is numerically verified in App.~D.

When $\epsilon_0=0$, the model manifests nonlinear bulk modes in $\pm\omega$ pairs. 
Topologically protected nonlinear boundary modes do not oscillate in time, in contrast to the $\epsilon_0 \ne 0$ systems. Thus, we obtain exact solutions of nonlinear topological modes via the recursion relation, $f_1(\Psi_n^{(1)},\Psi^{(2)}_n) +f_2(\Psi_n^{(1)},\Psi^{(2)}_{n-1}) = f_1(\Psi_n^{(2)},\Psi^{(1)}_n) +f_2(\Psi_n^{(2)},\Psi^{(1)}_{n+1}) =0$. This is the nonlinear analog of charge-conjugation symmetric systems.

\widetext

\begin{figure}[htbp]
\includegraphics[scale=0.55]{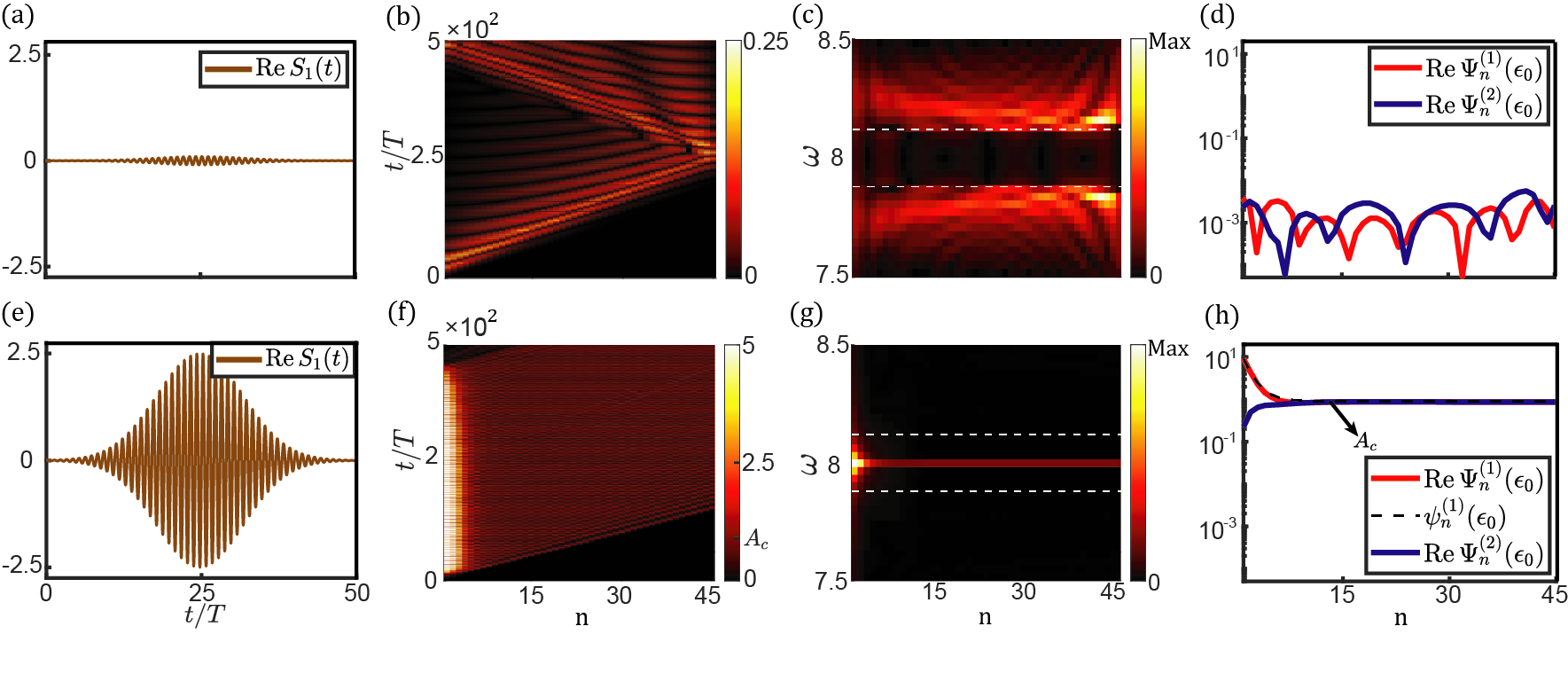}
\caption{Nonlinear boundary responses of the lattice subjected to N-to-T transition, where the parameters yield $c_1>c_2$ and $d_1<d_2$. (a-d) and (e-h) exhibit lattice boundary excitations in the small-amplitude non-topological regime and the large-amplitude topological regime, respectively. The magnitudes of Gaussian signals are $S=0.1$ in (a) and $S=2.5$ in (e), respectively. (b) and (f), Spatial-temporal profiles of $|{\rm Re\,}\Psi_n^{(1)}(t)|$ for $45$ sites. (c) and (g), frequency spectra of the lattice excitations for $45$ sites. Fourier analysis is executed from $250T$ to $500T$. White dashed lines encircle the linear bandgap. (d) and (h), red and blue curves for the spatial distributions of the $\omega=\epsilon_0$ mode component of the lattice excitations. The analytic result of the anomalous topological modes $\psi_n^{(1)}(\epsilon_0)$ is captured by the black dashed curve in (h). 
}\label{fig3}
\end{figure}

\endwidetext

In the second case of $c_1>c_2$ and $d_1<d_2$, N-to-T (non-topological-to-topological) transition occurs as amplitudes rise. We exemplify boundary excitations in Fig.\ref{fig3} by letting $\epsilon_0=8$, $c_1=0.37$, $c_2=0.25$, $d_1=0.02$, and $d_2=0.22$. A Gaussian signal is applied on the first site of the lattice, where the carrier frequency $\omega_{\rm ext} = \epsilon_0=8$, $T=2\pi/\omega_{\rm ext}$, Gaussian spread $\tau = 10T$, and trigger time $t_0 = 25T$. In the small-amplitude regime, we consider a chain of $N=45$ unit cells. As shown in Fig.\ref{fig3}(b), the lattice is free of topological modes for $|\Psi_{1}^{(1)}|<A_c=0.8944$. In the large-amplitude regime, the lattice is constructed from $N=120$ unit cells. Anomalous topological edge modes emerge when $|\Psi_{1}^{(1)}|>A_c$ (see figs.\ref{fig3}(f,h)). In contrast to conventional topological modes that shrink to zero over space, $\Psi_n^{(1)}$ decay to the plateau $A_c$ governed by the stable fixed point of Eq.(\ref{Recur}), whereas $\Psi_n^{(2)}$ increase to $A_c$ by absorbing energy~\cite{fronk2017higher} from $\Psi_n^{(1)}$. Theoretical analysis predicts that the plateau should extend to infinity, but the plateau is limited to reach site $60$ by the finite lifetime of topological modes due to the energy conversion to bulk modes, as elaborated in Fig.\ref{SIfig11}. Despite the huge nonlinearities ($|\Psi_1^{(1)}|/A_c\sim 10$, and $|\Psi_1^{(1)}|^2\max(d_1,d_2) /\max(c_1,c_2) \sim 10$), this mode is stable within the finite lifetime of more than 400 periods. This model serves as the combined prototype of long-lifetime, high-energy storage, long-distance transmission of topological modes, and efficient frequency converter from Gaussian inputs to monochromatic signals.

Although T-to-N and N-to-T transitions are converted to one another by choosing the unit cell, topological modes behave qualitatively different (Fig.\ref{fig2}(d) and \ref{fig3}(h)) due to the distinction in the fixed points of Eq.(\ref{Recur}). The modes converge to the stable fixed point $0$ in T-to-N transition ($A_c$ in N-to-T transition), but this fixed point becomes unstable in N-to-T transition (T-to-N transition).

\section{Proposals for experimental implementations} 
Upon establishing nonlinear topological band theory, it is natural to ask if any realistic physical systems enjoy these unconventional properties. Classical passive and active structures are proposed here to realize the minimal model of Eq.(\ref{9}), as detailed in App.~E.

\emph{Topological photonics~\cite{d2008ultraslow, Christodoulides:88}, passive system}: Our theoretical prototype is readily testified in 1D array of optic lattice. Each unit cell is composed of two waveguides to guide electro-magnetic modes along the axial direction, and the permittivity and permeability are nonlinearly modulated by the fields. Hence, the adjacent electro-magnetic fields are coupled nonlinearly. It can be shown that the propagation of electro-magnetic fields along the axial $z$-direction is depicted by 4-field extension of generalized nonlinear Schr\"{o}dinger equations, where the $z$-coordinate takes place of the time-like differential variable~\cite{d2008ultraslow, Christodoulides:88}. Consequently, this photonic system realizes the minimal model of Eq.(\ref{9}). 

\emph{Topoelectrical circuit~\cite{hadad2018self}, active system}: The second promising direction is to construct a ladder of cascaded diatomic unit cells composed of two LCR resonators and two capacitors $C_{j=1,2}\ll C$. The inductances are connected to external power sources which are nonlinear functions of $V_n^{(j=1,2)}$. The motion equation of the unit cell voltages $V_n^{(j=1,2)}$ are captured by Eqs.(\ref{1}) and Eq.(\ref{9}), and nonlinear topological attributes can be studied here.

\begin{figure}[htbp]
\includegraphics[scale=0.3]{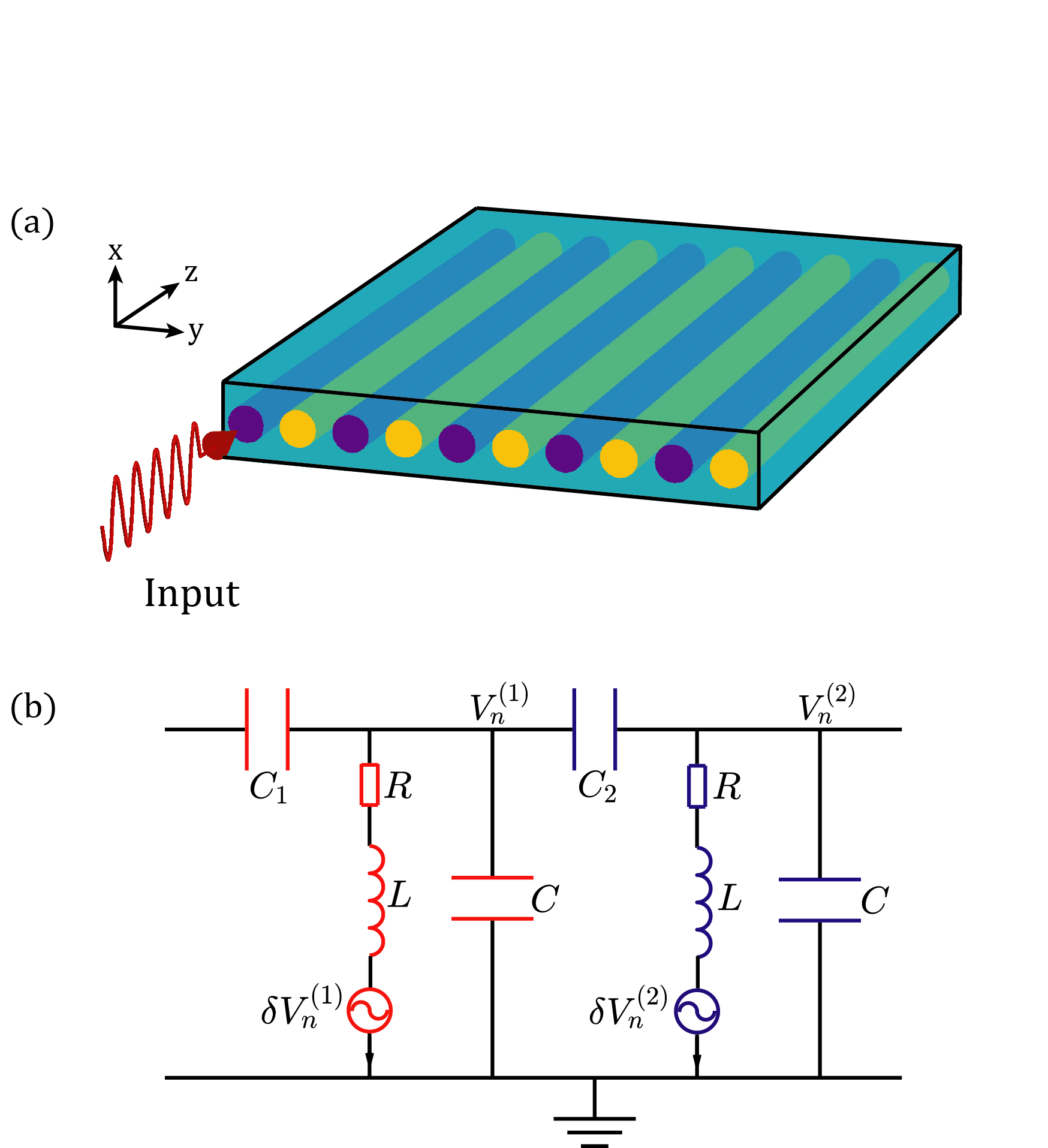}
\caption{Experimental proposals for passive and active nonlinear topological metamaterials. (a) 1D array of nonlinear optic lattice.  The nearest neighbor electro-magnetic fields are coupled nonlinearly. (b) The unit cell of nonlinear active topoelectrical circuit, where the inductances are connected by external alternating power sources nonlinearly controlled by voltage fields $V_n^{(j=1,2)}$. 
}\label{expProposal}
\end{figure}

\section{Conclusions}
In this paper, we extend topological band theory to strongly nonlinear Schr\"{o}dinger equations beyond Kerr-type nonlinearities. The proper definition of Berry phase is carried out for nonlinear bulk modes, and its quantization is demonstrated in reflection-symmetric models. The topological invariant experiences transitions induced by mode amplitudes. These results can be extended to higher dimensional systems with arbitrarily complex unit cells, but we leave the full proof for the future.

The advent (disappearance) of topological modes is associated with a change in the Berry phase to its topological (non-topological) value. As amplitudes increase, T-to-N (topological-to-non-topological) and N-to-T (non-topological-to-topological) transitions take place for different choices of unit cells. Anomalous topological modes decrease away from lattice boundaries to a plateau controlled by the stable fixed point of nonlinearities.

A rich variety of problems can be studied following this paper, such as the nonlinear extension of topological chiral edge modes in 2D systems~\cite{PhysRevLett.61.2015}, and higher-order topological states~\cite{benalcazar2017quantized}. Experimental characterizations of photonic, acoustic, and electrical metamaterials with built-in nonlinearities can also be studied in future. 

\begin{acknowledgements}
D. Z. would like to thank insightful discussions with Xueda Wen, Junyi Zhang, Feng Li, and Biao Wu. This work is supported by the National Key R$\&$D Program of China (Grant No. 2020YFA0308800), the NSF of China (Grants Nos. 11734003, 12061131002),  and the Strategic Priority Research Program of Chinese Academy of Sciences (Grant No. XDB30000000).

\end{acknowledgements}

\appendix






\section{Berry phase of nonlinear bulk modes}
In this section, we derive Berry phase of nonlinear bulk modes by adiabatically evolving the wave function as the wavenumber $q$ slowly traverses the Brillouin zone. We consider the nonlinear problem described by a classical two-field generalized nonlinear Schr\"{o}dinger equations presented in the main text, 
\begin{eqnarray}\label{1v2}
 & {} & \mathrm{i}\partial_t\Psi^{(1)}_n = \epsilon_0\Psi^{(1)}_n+f_1(\Psi^{(1)}_n, \Psi_n^{(2)}) +f_2(\Psi^{(1)}_{n}, \Psi^{(2)}_{n-1}),  \nonumber \\
 & {} & \mathrm{i}\partial_t\Psi^{(2)}_n = \epsilon_0\Psi^{(2)}_n+f_1(\Psi^{(2)}_n, \Psi^{(1)}_n) +f_2(\Psi^{(2)}_{n}, \Psi^{(1)}_{n+1}),\qquad 
\end{eqnarray}
where $f_i(x,y)$ for $i=1,2$ are real-coefficient general polynomials of $x$, $x^*$, $y$, and $y^*$. Berry phase is derived from this general model.

In the linear limit, the interactions are approximated as $f_i(x,y)\approx c_i y$ ($c_{i=1,2}>0$). The model is a $2\times 2$ matrix problem in which the bands are gapped. As the amplitude rises, nonlinearities become increasingly significant and the linear bulk modes evolve into nonlinear bulk modes. In this section, we study the simple case that the nonlinear bulk modes are non-degenerate and are stable. In other words, a nonlinear bulk mode is unique, provided that the amplitude $A$, the frequency $\omega$, and the wavenumber $q$ are given. In addition to these properties, we consider the simple case that the nonlinear bandgap~\cite{hadad2018self, PhysRevB.93.155112} never closes. As such, this system is a nonlinear extension of the linear two-band model.

We begin by defining the nonlinear periodic bulk mode of the system as follows,
\begin{eqnarray}\label{B0}
\Psi_n(t) = \Psi_q(\omega t-qn) = \left(\begin{array}{c}
\Psi_q^{(1)}(\omega t-qn)\\
 \Psi_q^{(2)}(\omega t-qn+\phi_q)\end{array}\right),\qquad
\end{eqnarray}
where $q$ is the wavenumber and $\omega$ is the frequency that belongs to the upper band of the nonlinear band structure. As such, the wave functions depend on the single variable $\theta = \omega t -qn$. We adopt this functional form based on a number of reasons. First, typical studies of weakly nonlinear bulk modes~\cite{PhysRevB.99.125116, fronk2017higher, PhysRevE.97.032209, PhysRevB.101.104106, narisetti2010perturbation, zaera2018propagation, vakakis2001normal} via the method of multiple-scale reveal that all Fourier components are captured by the single $\theta$ variable, $\Psi_n = \sum_l \psi_{l,n}e^{\mathrm{i}l\theta}$. Second, numerical experiments such as shooting method (see figs.\ref{fig1}(b), \ref{SIfig8}(a,c), and Refs.~\cite{renson2016numerical, ha2001nonlinear, peeters2008nonlinear}) manifests non-dispersive bulk modes in strongly nonlinear regime, which appear to be plane-wave like modes. Finally, analytic solutions for special wavenumbers $q=0,\pi$ demonstrate that strongly nonlinear bulk modes are in line with Eq.(\ref{B0}). It is at this point that we adopt Eq.(\ref{B0}) as the general form of nonlinear bulk modes.

In general, the waveforms of $\Psi_q^{(j)}(\theta)$ for $j=1,2$ are not sinusoidal in $\theta$. We note that because the wave function component $\Psi_q^{(j)}(\theta)$ is $2\pi$-periodic, it is defined up to an arbitrary phase condition. In this paper, the phase condition is chosen by asking that when $\theta=0$, the real part of wave component ${\rm Re\,}\Psi_q^{(j)}(\theta)$ reaches its amplitude/maximum,
\begin{eqnarray}\label{B1}
{\rm Re\,}\Psi_q^{(j)}(\theta=0) = \max({\rm Re\,}\Psi_q^{(j)}(\theta))\overset{\rm def}{=}A, \quad j = 1,2.\qquad
\end{eqnarray}
Note that the phase condition in Eq.(\ref{B1}) is similar to that in exponential functions, where ${\rm Re\,}e^{\mathrm{i}\theta=0}= \max({\rm Re\,}e^{\mathrm{i}\theta})$. Following this convention, $\phi_q$ in Eq.(\ref{B0}) characterizes the relative phase between $\Psi_q^{(1)}$ and $\Psi_q^{(2)}$. The nonlinear mode has to fulfill the differential equation parametrized by wavenumber $q$, 
\begin{eqnarray}\label{B2}
\mathrm{i} \partial_t \Psi_q(\theta) = H(\Psi_q),
\end{eqnarray}
where $\theta=\omega t$, and the nonlinear function $H(\Psi_q)$ is given by 
\begin{eqnarray}\label{B3}
H(\Psi_q)  & = &   \left(
\begin{array}{c}
\epsilon_0 \Psi_q^{(1)}(\theta) \\
\epsilon_0 \Psi_q^{(2)}(\theta+\phi_q) \\
\end{array}
\right)\nonumber \\
 & {} & +\left(
\begin{array}{c}
f_1(\Psi_q^{(1)}(\theta),\Psi_q^{(2)}(\theta+\phi_q))  \\
f_1(\Psi_q^{(2)}(\theta+\phi_q),\Psi_q^{(1)}(\theta))  \\
\end{array}
\right)
\nonumber\\
 & {} & +\left(
\begin{array}{c}
f_2(\Psi_q^{(1)}(\theta),\Psi_q^{(2)}(\theta+q+\phi_q)) \\
f_2(\Psi_q^{(2)}(\theta+\phi_q),\Psi_q^{(1)}(\theta-q)) \\
\end{array}
\right).
\end{eqnarray}
In what follows, we study the nonlinear bulk mode with the fixed amplitude $A$. Therefore, the mode frequency $\omega$, the relative phase $\phi_q$, and the waveform are controlled by the wavenumber $q$.

Next, we adiabatically evolve the wavenumber $q(t)$ traversing the Brillouin zone from $q(0)=q$ to $q(t) = q+2\pi$. According to the nonlinear extension of adiabatic theorem~\cite{RevModPhys.82.1959, PhysRevLett.90.170404, PhysRevLett.98.050406, PhysRevA.81.052112}, a system initially in one of its nonlinear mode $\Psi_{q}$ of the upper band will stay as an instantaneous upper-band nonlinear mode of $H(\Psi_{q(t)})$ throughout the process. This theorem is valid when the control parameter $q$ varies sufficiently slowly compared to the frequencies~\cite{PhysRevLett.90.170404}, and the nonlinear bulk modes are stable~\cite{PhysRevLett.98.050406} within the amplitude scope of this paper. In App.~C, we exploit the self-oscillation method~\cite{fronk2017higher} to confirm the stability of these nonlinear modes. Hence the only degree of freedom is the phase of the mode. At time $t$, the mode is 
\begin{eqnarray}\label{B4}
\Psi_{q(t)}\left(\int_0^t \omega(t', q(t'))dt' -\gamma(t)\right).
\end{eqnarray}
We are interested in the extra phase term $\gamma$, which will be carried out as follows. Substituting Eq.(\ref{B4}) into Eq.(\ref{B2}), we have
\begin{eqnarray}\label{B5}
\frac{d\gamma}{dt}\frac{\partial\Psi_q}{\partial\theta}=\frac{dq}{dt} \frac{\partial\Psi_q}{\partial q},
\end{eqnarray}
where $\theta=\omega t$ stands for the phase of the wave function $\Psi_q$. We bare in mind that the nonlinear bulk mode is $2\pi$-periodic in its phase, which grants Fourier transformation. We expand $\Psi_q^{(j)}(\theta)$, the component of periodic wave function, in terms of its Fourier series: 
\begin{eqnarray}\label{B6}
\Psi_q^{(j)}(\theta) = \sum_{l\in\mathcal{Z}} \psi_{l,q}^{(j)}e^{-\mathrm{i} l\theta} \qquad j = 1,2,
\end{eqnarray}
where $\psi_{l,q}^{(j)}$ is the $l$-th Fourier component of $\Psi_q^{(j)}$ ($l$ is integer). Inserting Eq.(\ref{B6}) into Eq.(\ref{B5}), we have
\begin{eqnarray}\label{B7}
 & {} & \frac{d\gamma}{dt} \sum_l \mathrm{i} l e^{-\mathrm{i} l\theta} 
\left(
\begin{array}{c}
\psi_{l,q}^{(1)}\\
\psi_{l,q}^{(2)}e^{-\mathrm{i} l\phi_q}\\
\end{array}
\right)=\nonumber \\
 & {} & 
- \frac{dq}{dt} \sum_l e^{-\mathrm{i} l \theta}   \left(
\begin{array}{c}
{\partial\psi_{l,q}^{(1)}}/{\partial q}\\
e^{-\mathrm{i} l\phi_q}[{\partial\psi_{l,q}^{(2)}}/{\partial q}-\mathrm{i} l \psi_{l,q}^{(2)}  ({\partial\phi_q}/{\partial q})]\\
\end{array}
\right).\qquad
\end{eqnarray}
We multiply Eq.(\ref{B7}) on both sides by $\Psi_q^\dag(\theta)$ and integrate $\theta$ from $0$ to $2\pi$, to obtain the following result, 
\begin{eqnarray}\label{B8}
 & {} & \frac{d\gamma}{dt} \sum_{l'} \mathrm{i} l'  \left(|\psi_{l',q}^{(1)}|^2+|\psi_{l',q}^{(2)}|^2\right)=\nonumber \\
 & {} & 
 \frac{dq}{dt} \sum_{l} \left(\mathrm{i} l |\psi_{l,q}^{(2)} |^2 \frac{\partial\phi_q}{\partial q}-\psi_{l,q}^{(1)*}\frac{\partial\psi_{l,q}^{(1)}}{\partial q}- \psi_{l,q}^{(2)*}\frac{\partial\psi_{l,q}^{(2)}}{\partial q}\right). \qquad\quad
\end{eqnarray}
Since the wavenumber $q$ traverses the Brillouin zone, by integrating over time $t$ we obtain the phase term $\gamma$ expressed in terms of a loop integration through the entire Brillouin zone, 
\begin{eqnarray}\label{B9}
\gamma
=
\oint_{\rm BZ}dq \frac{\sum_{l} \left( l |\psi_{l,q}^{(2)} |^2 \frac{\partial\phi_q}{\partial q}+\mathrm{i}\sum_j\psi_{l,q}^{(j)*}\frac{\partial\psi_{l,q}^{(j)}}{\partial q}\right)}
{\sum_{l'}  l'  \left(\sum_{j'}|\psi_{l',q}^{(j')}|^2\right)}.\qquad
\end{eqnarray}
Eq.(\ref{B9}) is Berry phase of the upper-band nonlinear bulk modes, which is the generalization of Berry phase in linear problems. 

Having established Berry phase of nonlinear bulk modes, we now build the connection between Eq.(\ref{B9}) and its conventional form in linear systems. In quantum mechanics, Schr\"{o}dinger equation $\mathrm{i}\partial_t\Psi(t) = H\Psi(t)$ is linear in $\Psi(t)$, where $H$ is the Hamiltonian as a linear operator, $\omega$ are the eigenvalues, and the eigenmodes $\Psi(t) = \Psi e^{-\mathrm{i}\omega t}$ are sinusoidal in time. Let us consider a 1D lattice of diatomic unit cells subjected to PBC. Translational symmetry allows plane-wave eigenmodes $\Psi(t) =\Psi_q e^{\mathrm{i}qn-\mathrm{i}\omega t} = (\Psi_{q}^{(1)}, \Psi_q^{(2)}e^{-\mathrm{i}\phi_q})^\top e^{\mathrm{i}qn-\mathrm{i}\omega t}$, where $q$ is the wave number, $\phi_q$ is the relative phase between the two parts of the wave function, and $\sum_{j=1,2}|\Psi_{q}^{(j)}|^2\equiv 1$ is the normalization condition. The phase condition is chosen such that both $\Psi_q^{(j=1,2)}$ are real, which is consistent with Eq.(\ref{B1}). Thus, $\Psi_q^{(j)}(\theta) = \Psi_q^{(j)}e^{-\mathrm{i}\theta}$, where $\theta = \omega t -qn$. According to Eq.(\ref{B6}), the Fourier components are that $\psi_{l,q}^{(j)} = \Psi_q^{(j)}\delta_{l,1}$, which greatly simplify Eq.(\ref{B9}) to the following form, 
\begin{eqnarray}\label{B9.1}
\gamma
 & = & 
\oint_{\rm BZ}dq \frac{\sum_{l} \left( l |\psi_{l,q}^{(2)} |^2 \frac{\partial\phi_q}{\partial q}+\mathrm{i}\sum_j\psi_{l,q}^{(j)*}\frac{\partial\psi_{l,q}^{(j)}}{\partial q}\right)\delta_{l1}}
{\sum_{l'}  l'  \left(\sum_{j'}|\psi_{l',q}^{(j')}|^2\right)\delta_{l'1}}\nonumber \\
 & = & \oint_{\rm BZ}dq \left( |\psi_{1,q}^{(2)} |^2 \frac{\partial\phi_q}{\partial q}+\mathrm{i}\sum_j\psi_{1,q}^{(j)*}\frac{\partial\psi_{1,q}^{(j)}}{\partial q}\right) \nonumber \\
 & = & 
 \oint_{\rm BZ}dq\,\mathrm{i}\left(\sum_j \Psi_q^{(j)*}\partial_q\Psi_q^{(j)}-\mathrm{i}|\Psi_q^{(2)}|^2\partial_q\phi_q\right) \nonumber \\
 & = & 
 \oint_{\rm BZ}dq\,\mathrm{i}\langle \Psi_q|\partial_q|\Psi_q\rangle = \gamma_{L},
\end{eqnarray}
where $\Psi_q = (\Psi_q^{(1)},\Psi_q^{(2)}e^{-\mathrm{i}\phi_q})^\top$ is the eigenvector of the Hamiltonian,  and $\gamma_L$ denotes the conventional form of Berry phase in linear systems.

Next, we briefly review a reflection-symmetric linear model and quantized $\gamma_L$, where the equations of motion
\begin{eqnarray}\label{B9.2}
 & {} & \mathrm{i}\partial_t\Psi^{(1)}_n = \epsilon_0\Psi^{(1)}_n+c_1 \Psi_n^{(2)} +c_2 \Psi^{(2)}_{n-1},  \nonumber \\
 & {} & \mathrm{i}\partial_t\Psi^{(2)}_n = \epsilon_0\Psi^{(2)}_n+c_1 \Psi^{(1)}_n +c_2 \Psi^{(1)}_{n+1}
\end{eqnarray}
are subjected to PBC, $\epsilon_0>0$ is the on-site potential, and $c_i>0$ for $i=1$ and $i=2$ stand for intracell and intercell couplings, respectively. In momentum space, the motion equations are reduced to $\mathrm{i}\partial_t \Psi_q = H_q \Psi_q$, where $H_q = \epsilon_0 I_2+(c_1+c_2\cos q)\sigma_x +(c_2\sin q)\sigma_y$. The eigenvalue of the upper band reads $\omega = \epsilon_0+|c_1+c_2e^{\mathrm{i}q}|$, and the associated eigenvector is $\Psi_q = (1 ,(c_1+c_2e^{\mathrm{i}q})/|c_1+c_2e^{\mathrm{i}q}|)^\top/\sqrt{2}$. We invoke Eq.(\ref{B9.1}) to reduce Berry phase to the form, $\gamma_L
=
\frac{\mathrm{i}}{2}\oint_{\rm BZ}dq
\left[\partial_q \ln (c_1+c_2e^{\mathrm{i}q})
-
\partial_q\ln |c_1+c_2e^{\mathrm{i}q}| 
\right]$, which can be interpreted in two ways. In the first way, we notice that the second part vanishes, because the length of $|c_1+c_2e^{\mathrm{i}q}|$ does not wind around the origin when integrated over the Brillouin zone. The second way is to denote $c_1+c_2e^{\mathrm{i}q}=\rho_q e^{-\mathrm{i}\phi_q}$, and then Berry phase simply represents how $\phi_q$ winds around the origin by 0 or $2\pi$ when $q$ traverses the Brillouin zone. Thus, $\gamma_L$ can be reduced to 
\begin{eqnarray}\label{B9.5}
\gamma_L
 & = & 
\frac{\mathrm{i}}{2}\oint_{\rm BZ}dq
\,\partial_q \ln (c_1+c_2e^{\mathrm{i}q}) = \frac{1}{2}\oint_{\rm BZ}dq
\,\partial_q \phi_q. \qquad
\end{eqnarray}
In Ref.~\cite{kane2014topological}, the topological index is captured by the winding number $\mathcal{N} =- (2\pi\mathrm{i})^{-1}\oint_{\rm BZ}dq \,\partial_q \ln\det C(q)=(2\pi)^{-1}\oint_{\rm BZ}dq\,\partial_q \phi_q$, where $C(q) = c_1+c_2 e^{\mathrm{i}q}$ is the compatibility matrix that describes floppy modes. Thus, Berry phase and winding number are related by $\gamma_L = \mathcal{N}/\pi$. In summary, Eq.(\ref{B9}) is the nonlinear extension of the topological index in Ref.~\cite{kane2014topological}.

\section{Symmetries of generalized nonlinear Schr\"{o}dinger equations and the quantization of Berry phase}

In this section, we study symmetry properties of the model in Eqs.(\ref{1v2}). We prove that the frequencies of nonlinear bulk modes are restricted to be real numbers due to the combined effect of time-reversal symmetry and spatial reflection symmetry. Then, we demonstrate that Berry phase in Eq.(\ref{B9}) is quantized by reflection symmetry.

\subsection{Time-reversal symmetry} 
Here, we demonstrate that the model in Eqs.(\ref{1v2}) is subjected to time-reversal symmetry, as long as the interactions yield the constraint
\begin{eqnarray}\label{C100.2}
f_i^*(x,y) = f_i(x^*,y^*),
\end{eqnarray}
which is met by any real-coefficient polynomials of $x,x^*,y,y^*$, including the minimal model of the main text. The considered nonlinear solution $\Psi_n(t)$ satisfies the equations of motion $\mathrm{i}\partial_{t}\Psi_n(t) = H(\Psi_n)$, where $H(\Psi_n)$ is the nonlinear function of $\Psi_n(t)$ elaborated by Eqs.(\ref{1v2}). Taking complex conjugation on both sides offers us a new equation 
\begin{eqnarray}\label{C100.1}
\mathrm{i}\partial_{t}\Psi_n^*(-t) = H^*(\Psi_n(-t)).
\end{eqnarray}
Substituting Eq.(\ref{C100.2}), we arrive at the new result,
\begin{eqnarray}\label{C100.3}
\mathrm{i}\partial_{t}\Psi_n^*(-t) = H(\Psi_n^*(-t)). 
\end{eqnarray}
Eq.(\ref{C100.3}) suggests that given a nonlinear solution $\Psi_n(t)$, we can always find a partner solution $\Psi_n^*(-t)$ for the same equations of motion. Consequently, the model respects time-reversal symmetry, in the sense that nonlinear solutions $\Psi_n(t)$ and $\Psi_n^*(-t)$ always come in pairs~\cite{RevModPhys.82.3045}.

For given amplitude $A$, time-reversal symmetry demands that the frequencies of nonlinear bulk modes are related by $\omega(q) = \omega^*(-q)$. To prove this, we consider a nonlinear bulk mode
\begin{eqnarray}\label{A1.15}
\Psi_{q} = (\Psi_q^{(1)}(\omega t-qn), \Psi_q^{(2)}(\omega t-qn+\phi_q))^\top,
\end{eqnarray}
where $q$ is the wavenumber, and $\omega=\omega(q)$. $\Psi_{q}$ is a solution of Eqs.(\ref{1v2}) only if it fulfills Eq.(\ref{B2}), which is equivalent to the following nonlinear differential equation, 
\begin{eqnarray}\label{A1}
 & {} & \Psi_{q}(\theta= \omega t-qn) = (\Psi^{(1)}_q(\theta), \Psi^{(2)}_q(\theta+\phi_q))^\top: \nonumber \\
 & {} & \mathcal{L}(\Psi_{q}) = 0,
\end{eqnarray}
where the nonlinear differential operator $\mathcal{L}(\Psi_{q})$ is defined as follows, 
\begin{eqnarray}\label{A1.3}
\mathcal{L}(\Psi_{q})  & = &  (\mathrm{i} \omega\partial_\theta -\epsilon_0) \left(
\begin{array}{c}
\Psi_q^{(1)}(\theta)\\
\Psi_q^{(2)}(\theta) \\
\end{array}
\right) \nonumber \\
 & {} &
 -\left(
\begin{array}{c}
f_1(\Psi_q^{(1)}(\theta),\Psi^{(2)}_q(\theta+\phi_q)) \\
f_1(\Psi_q^{(2)}(\theta),\Psi^{(1)}_q(\theta-\phi_q)) \\
\end{array}
\right)\nonumber \\
 & {} & 
 -\left(
\begin{array}{c}
f_2(\Psi_q^{(1)}(\theta),\Psi^{(2)}_q(\theta+q+\phi_q))\\
f_2(\Psi_q^{(2)}(\theta),\Psi^{(1)}_q(\theta-q-\phi_q)) \\
\end{array}
\right).
\end{eqnarray}
In general, the waveform of $\Psi_{q}$ is not sinusoidal, which is the natural result of nonlinearity. Time-reversal symmetry demands a partner solution
\begin{eqnarray}\label{A1.151}
\Psi_{q}^*(-t) = (\Psi_q^{(1)*}(-\omega t-qn), \Psi_q^{(2)*}(-\omega t-qn+\phi_q))^\top,\nonumber \\
\end{eqnarray}
where $\omega = \omega(q)$. This mode also renders the equations of motion to vanish, 
\begin{eqnarray}\label{A1.152}
 & {} & \Psi_{q}^*(\theta = -\omega t -qn) = (\Psi_q^{(1)*}(\theta), \Psi_q^{(2)*}(\theta+\phi_q))^\top: \nonumber \\
 & {} & \mathcal{L}(\Psi_{q}^*(-t)) = \left[\mathcal{L}(\Psi_{q})\right]^* = 0.
\end{eqnarray}
We note that the wavenumber and frequency of the mode are $-q$ and $\omega(-q)$, respectively. Hence, Eqs.(\ref{A1.152}) demonstrates the following relationship, 
\begin{eqnarray}\label{A1.153}
\omega(-q)= \omega^*(q).
\end{eqnarray}
In the following subsection, we will prove that together with reflection symmetry, the frequencies are constrained to be real numbers (i.e., $\omega^* = \omega$), and nonlinear bulk modes are periodic in time.

\subsection{Reflection symmetry and quantized Berry phase}
Before going into details of reflection symmetry in the nonlinear system, we briefly review this symmetry in the linearized model and demonstrate the quantization of Berry phase, when the coupling is linearized as $f_i(x,y)=c_i y$. We convert the wave function into momentum space $\Psi_n = \Psi_{q} e^{\mathrm{i}(qn-\omega t)}$, to reduce the equations of motion as $H_q \Psi_{q} = \omega \Psi_{q}$, where $H_q = \epsilon_0 I +(c_1+c_2\cos q)\sigma_x +(c_2\sin q)\sigma_y$, and $\sigma_{x,y,z}$ are Pauli matrices. $H_q$ is subjected to reflection symmetry, meaning that one can find a reflection symmetry operator $M_x = \sigma_x$, such that $M_x^2 = I$, and $M_xH_q M_x^{-1} = H_{-q}$. We notice $H_{-q}M_x \Psi_{q} = \omega M_x \Psi_{q}$. It demonstrates that $\Psi_{q}$ and $\Psi_{-q}$ are related by $M_x\Psi_{q} = e^{i\phi_q}\Psi_{-q}$, where $\phi_q$ is the phase factor connecting $\Psi_q$ and $\Psi_{-q}$. At high-symmetry points when $q_{\rm hs} = 0,\pi$ (``hs" is short for high symmetry), we find that $M_x$ and $H_q$ commute, which demands the phase factor $\phi_{\rm hs}=0$ or $\pi$. Finally, in the linear problem, we prove the quantization of Berry phase by showing that $\gamma = \phi_\pi - \phi_0 = 0$ or $\pi\mod 2\pi$.

We now proceed to investigate the nonlinear problem raised in Eqs.(\ref{1v2}). We notice that the nonlinear system is subjected to reflection symmetry: the equations of motion are invariant under the reflection transformation, 
\begin{eqnarray}\label{A3}
(\Psi^{(1)}_n, \Psi^{(2)}_n) \to (\Psi^{(2)}_{-n}, \Psi^{(1)}_{-n}). 
\end{eqnarray}
In Eq.(\ref{A1.15}), given a nonlinear bulk mode solution $\Psi_q$ that renders $\mathcal{L}(\Psi_q)$ to vanish, Eq.(\ref{A3}) demands a new nonlinear bulk mode solution $\Psi'_{-q}=(\Psi^{(2)}_q(\omega t+qn), \Psi^{(1)}_q(\omega t+qn-\phi_q))^\top$ that also renders $\mathcal{L}(\Psi_{-q}')$ to vanish, 
\begin{eqnarray}\label{A4}
 & {} &  \Psi_{-q}'(\theta = \omega t +qn) = (\Psi^{(2)}_q(\theta), \Psi^{(1)}_q(\theta-\phi_q))^\top:\nonumber \\
 & {} & \mathcal{L}(\Psi_{-q}') = \sigma_x \mathcal{L}(\Psi_{q}) = 0, 
\end{eqnarray}
where $\omega = \omega(q)$. Since the wavenumber and frequency of $\Psi'_{-q}$ are $-q$ and $\omega(-q)$, respectively, we reach the conclusion 
\begin{eqnarray}\label{A4.02}
\omega(-q) = \omega(q).
\end{eqnarray}
Together with Eq.(\ref{A1.153}), we show $\omega(q) = \omega^*(q)$ for all $q$, which means the frequencies of nonlinear bulk modes are real. From now on, we denote $\omega(q)$ as $\omega$ for simplicity, and $\Psi_{-q}'$ is a nonlinear mode with frequency $\omega$ and wavenumber $-q$.

On the other hand, following the notation of Eq.(\ref{A1.15}), the nonlinear bulk mode of frequency $\omega$ and wavenumber $-q$ is by definition denoted as 
\begin{eqnarray}\label{A2}
\Psi_{-q}(\theta=\omega t + qn) = (\Psi^{(1)}_{-q}(\theta), \Psi^{(2)}_{-q}(\theta+\phi_{-q}))^\top.\qquad
\end{eqnarray}
Due to the non-degenerate nature of nonlinear bulk modes, $\Psi_{-q}$ and $\Psi_{-q}'$ have to be the same solution, which in turn imposes the constraints
\begin{eqnarray}\label{A5}
\Psi^{(1)}_{-q}(\theta)
=
\Psi^{(2)}_q(\theta), 
\end{eqnarray}
and
\begin{eqnarray}\label{A5.2}
-\phi_{q} = \phi_{-q}\mod 2\pi. 
\end{eqnarray}

\renewcommand{\thefigure}{B1}
\begin{figure}[htb]
\includegraphics[scale=0.58]{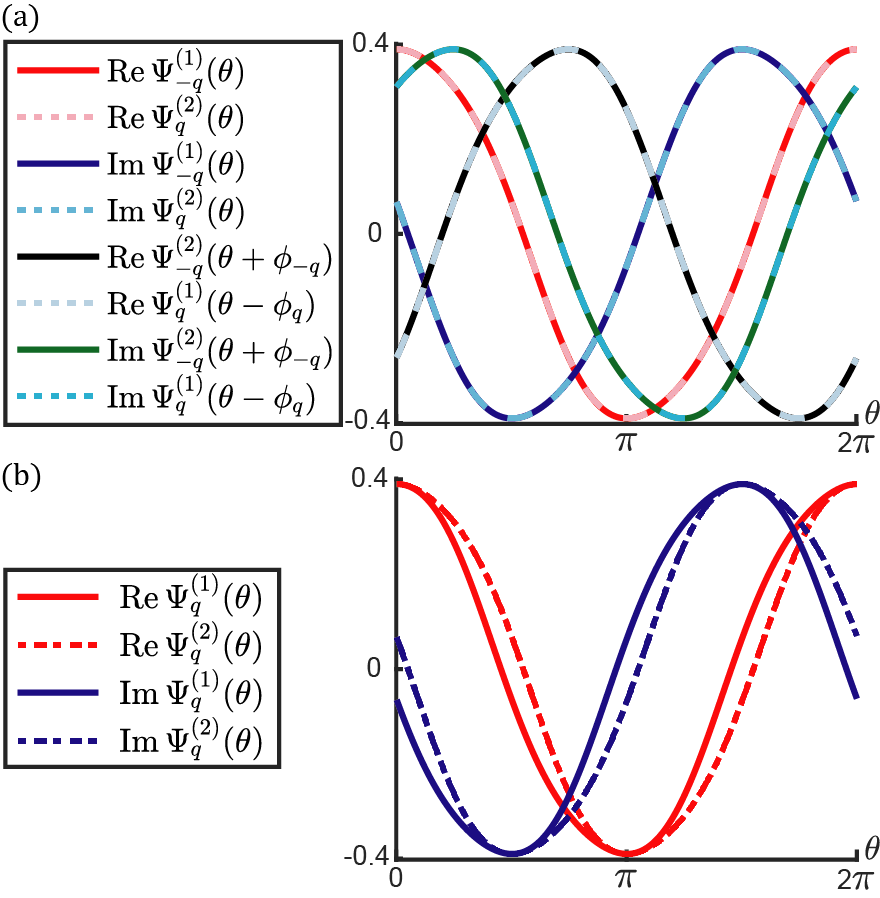}
\caption{Numerical verification of $\Psi^{(1)}_{-q}(\theta)=\Psi^{(2)}_q(\theta)$ and $-\phi_{q} = \phi_{-q}$ by computing nonlinear bulk modes. Here the nonlinear bulk modes are calculated in the lattice composed of classical dimer fields. The lattice is subjected to periodic boundary condition (PBC), and the interaction parameters are carried over from Fig.\ref{fig3}, where $\epsilon_0=0$, $c_1=0.25$, $c_2=0.37$, $d_1=0.22$, and $d_2=0.02$. (a) Numerical illustration of Eqs.(\ref{A5}, \ref{A5.2}) by comparing nonlinear bulk modes $\Psi_{-q}(t)$ in Eq.(\ref{A2}) and $\Psi_{-q}'(t)$ in Eq.(\ref{A4}), where the wavenumber $q=8\pi/9$ and the amplitude $A=0.3873$. $\Psi^{(1)}_{-q}(\theta)=\Psi^{(2)}_q(\theta)$ and $-\phi_{q} = \phi_{-q}$ are verified by the perfect overlap between the wave functions. (b) Numerical demonstration of ${\rm Re\,}\Psi_q^{(1)}(\theta)\neq {\rm Re\,}\Psi_q^{(2)}(\theta)$ and ${\rm Im\,}\Psi_q^{(1)}(\theta)\neq {\rm Im\,}\Psi_q^{(2)}(\theta)$. 
}\label{SIfig7}
\end{figure}

Having obtained Eqs.(\ref{A5}, \ref{A5.2}), we now attempt to prove the quantization of Berry phase defined in Eq.(\ref{B9}). To this end, we consider the Fourier components of $\Psi_{q}^{(1)}$ and $\Psi_{q}^{(2)}$, which are related to one another as follows,
\begin{eqnarray}\label{A9}
\psi^{(1)}_{l,-q}=\psi^{(2)}_{l,q}.
\end{eqnarray}
Employing Eq.(\ref{A5.2}) and Eq.(\ref{A9}), we compute Berry phase by separating it into two parts, $\gamma = \gamma_1+\gamma_2$, where 
\begin{eqnarray}\label{A10}
\gamma_1 = 
\mathrm{i} \oint_{\rm BZ}dq \frac{\sum_{l} \left(\psi_{l,q}^{(1)*}\frac{\partial\psi_{l,q}^{(1)}}{\partial q}+ \psi^{(1)*}_{l,-q}\frac{\partial\psi^{(1)}_{l,-q}}{\partial q}\right)}
{\sum_{l'}  l'  \left(|\psi_{l',q}^{(1)}|^2+|\psi^{(1)}_{l',-q}|^2\right)}
=
0,\qquad\quad
\end{eqnarray}
and 
\begin{eqnarray}\label{A11}
\gamma_2
 & = & 
\frac{1}{2}\oint_{\rm BZ}dq \frac{\sum_{l} l \left(|\psi^{(1)}_{l,q}|^2 +|\psi^{(1)}_{l,-q}|^2 \right)}
{\sum_{l'} l'  \left(|\psi_{l',q}^{(1)}|^2+|\psi^{(1)}_{l',-q}|^2\right)}\frac{\partial\phi_q}{\partial q}\nonumber \\
 & = & 
 \frac{1}{2}\oint_{\rm BZ}dq \frac{\partial\phi_q}{\partial q} = \phi_{\pi}-\phi_{0}.
\end{eqnarray}
Next, at high-symmetry points $q_{\rm hs} = 0,\pi$, we find that $\phi_{q_{\rm hs}} = \phi_{-q_{\rm hs}}$. Together with Eq.(\ref{A5.2}), we obtain $2\phi_{\rm hs} = 0\mod 2\pi$, meaning that 
\begin{eqnarray}\label{A8}
\phi_{\pi} -\phi_{0} = 0\,\,{\rm or}\,\,\pi \mod 2\pi.
\end{eqnarray}
Therefore, we demonstrate the quantization of Berry phase, 
\begin{eqnarray}\label{A12}
\gamma = 0 \,\,{\rm or}\,\, \pi\mod 2\pi.
\end{eqnarray}

\subsection{Additional properties when the nonlinear interactions yield $f_i(-x,y) = -f_i(x,-y)$} 
In the minimal model, the functional forms of nonlinear interactions yield $f_i(-x,y) = -f_i(x,-y)$ (or equivalently, $f_i(-x,-y) = -f_i(x,y)$). Given a nonlinear bulk mode $\Psi_q$, it is straightforward to prove that $-\Psi_q$ is a nonlinear solution as well. Hence, $\Psi_q$ and $-\Psi_q$ must differ by a phase $\Delta\theta$ only, such that $\Psi_q(\theta+\Delta \theta) = -\Psi_q(\theta)$. We perform the phase shift $\Delta\theta$ twice to have $\Psi_q(\theta+2\Delta \theta) = \Psi_q(\theta)$, which imposes $\Delta\theta=\pi$. Finally, we reach the conclusion 
\begin{eqnarray}\label{C1}
\Psi_q(\theta+\pi) = -\Psi_q(\theta).
\end{eqnarray}

\subsection{Symmetry properties of nonlinear bulk modes when $\epsilon_0=0$} 
When $\epsilon_0=0$, the linearized model of Eqs.(\ref{1v2}) is subjected to an additional symmetry called charge-conjugation symmetry~\cite{ryu2010topological, kane2014topological}. In the linear limit, the interactions are reduced to $f_i(\Psi_n^{(j)})=c_i \Psi_n^{(j)}$. We convert wave function to momentum space $\Psi_n^{(j)} = \Psi_{\omega}^{(j)}e^{\mathrm{i}(qn-\omega t)}$ to have the reduced equation of motion, $H \Psi_{\omega} = \omega \Psi_{\omega}$, where $H = (c_1+c_2\cos q)\sigma_x +(c_2\sin q)\sigma_y$. $H$ is subjected to charge-conjugation symmetry, meaning that one can find a symmetry operator $\Pi=\sigma_z$ such that $\Pi^2 =I$, and $\Pi H \Pi^{-1} = -H$. As a result, $H\Pi\Psi_{\omega} = -\omega \Pi\Psi_{\omega}$, meaning that the eigenvalues always come in $\pm\omega$ pairs, and the eigenmodes $\Psi_{\omega}$, $\Psi_{-\omega}$ are related by $\Pi\Psi_{\omega} = e^{i\phi_q}\Psi_{-\omega}$. This relationship demonstrates the quantization of Berry phase when we evaluate it in the upper band.

We then study the nonlinear model in Eqs.(\ref{1v2}) with $\epsilon_0=0$ and the associated nonlinear modes. In order to have the frequencies of nonlinear modes to appear in $\pm\omega$ pairs, we ask that the nonlinear interactions $f_i(x,y)$ to yield the following constraints:
\begin{eqnarray}\label{A14.1}
f_i(-x, y) =-f_i(x, -y)= f_i(x,y),
\end{eqnarray}
for $i=1,2$. In linear systems, $f_i(x,y)$ is reduced to $f_i(x,y)=c_i y$ and this property is naturally met. However, this property is not naturally satisfied by arbitrary nonlinear functions, and Eqs.(\ref{A14.1}) are the additional constraints for nonlinear interactions. As a result, the system is invariant under the transformation
\begin{eqnarray}\label{A14.3}
(\Psi^{(1)}_n(\omega t), \Psi^{(2)}_n(\omega t)) \to (-\Psi^{(1)}_{n}(-\omega t), \Psi^{(2)}_{n}(-\omega t)). \qquad
\end{eqnarray}
Let us consider a nonlinear bulk mode solution $\Psi_{\omega}$ of the upper band with the frequency $\omega>0$ and wavenumber $q$. It yields the following nonlinear differential equation, 
\begin{eqnarray}\label{A13}
 & {} & \Psi_{\omega}(\theta= \omega t-qn) = (\Psi^{(1)}_q(\theta), \Psi^{(2)}_q(\theta+\phi_q))^\top: \nonumber \\
 & {} &  \mathcal{L}(\Psi_{\omega}; \epsilon_0=0)=0.
\end{eqnarray}
Referring to Eq.(\ref{A14.3}), it is straightforward to find a ``partner solution $\Psi_{-\omega}$" of frequency $-\omega<0$ and wavenumber $q$, that satisfies the nonlinear differential equation, 
\begin{eqnarray}\label{A14}
 & {} & \Psi_{-\omega}(\theta= -\omega t-qn) = (-\Psi^{(1)}_q(\theta), \Psi^{(2)}_q(\theta+\phi_q))^\top:\nonumber \\
 & {} & \mathcal{L}(\Psi_{-\omega}; \epsilon_0=0)= \sigma_z\mathcal{L}(\Psi_{\omega}; \epsilon_0=0)=0. 
\end{eqnarray}
Eq.(\ref{A14}) demonstrates that the frequencies of nonlinear bulk modes always appear in $\pm\omega$ pairs. Consequently, $\Psi_{-\omega}$ is the nonlinear bulk mode solution that belongs to the lower band, and the nonlinear band structure is symmetric with respect to $\omega=0$ axis.

\section{Methods of computing nonlinear bulk modes}
In this section, we introduce the methods of computing nonlinear bulk modes, which are commonly used in solving nonlinear problems. We illustrate these methods by considering the model of Eqs.(\ref{1v2}) with the nonlinear interactions specified in Eq.(\ref{9}) of the main text, 
\begin{eqnarray}\label{9v2}
f_i (x,y) = c_i y + d_i [({\rm Re\,}y)^3+\mathrm{i} ({\rm Im\,}y)^3].
\end{eqnarray}
In the weakly nonlinear regime, the analytic and perturbative \emph{method of multiple-scale}~\cite{fronk2017higher, narisetti2010perturbation, zaera2018propagation, SNEE2019100487} finds nonlinear bulk modes asymptotically, which serves as the cornerstone of nonlinear modes for higher amplitudes. As the amplitude grows, the system enters into a region where this perturbative technique is unavailable. Instead, the numerical tactic called \emph{shooting method}~\cite{renson2016numerical, ha2001nonlinear, peeters2008nonlinear} finds nonlinear bulk modes for large amplitudes, and these modes are noticeably different from sinusoidal waves (figs.\ref{fig1}(b), \ref{fig3}(b), and \ref{fig3}(d)). In this paper, we combine these two methods, i.e., method of multiple-scale and shooting method, to obtain a series of nonlinear bulk modes for a wide range of amplitudes.

\subsection{Method of multiple-scale: bulk modes in weakly nonlinear regime}

First of all, we explore bulk modes in the weakly nonlinear regime. The perturbative approach, namely method of multiple-scale, is useful to solve the frequencies and waveforms of weakly nonlinear bulk modes.

This method is performed by introducing a small book-keeping parameter $\epsilon\ll 1$ that enforces small amplitudes for the bulk modes. Specifically, this parameter is introduced by rewriting $d_i$ as $\epsilon d_i$ in Eq.(\ref{9v2}). This method then expands the time derivatives in orders of slow-time derivatives, 
\begin{eqnarray}\label{MS1}
\frac{d}{dt} 
=\sum_{l=0}^{\infty} \epsilon^l D_{l},
\end{eqnarray}
where $T_{(l)} = \epsilon^l T_{(0)}$ is the $l$-th order slow time variable, and $D_{l} = \partial /\partial T_{(l)}$ is the corresponding slow time derivative.  Next, the wave function is also expanded in terms of the multiple-scale, 
\begin{eqnarray}\label{MS2}
\Psi_n = \sum_{l=0}^\infty \epsilon^l \Psi_{n,(l)},
\end{eqnarray}
where $\Psi_{n,(l)} = (\Psi_{n,(l)}^{(1)}, \Psi_{n,(l)}^{(2)})^\top$ is the $l$-th order wave function. In what follows, we calculate $\Psi_{n,(l=1)}$, which offers us the wave function correction and the frequency correction of the first order. Following Eqs.(\ref{MS1}, \ref{MS2}), we expand the equations of motion by matching all field variables with respect to the order of the book-keeping parameter $\epsilon$. To zeroth-order, the equations of motion are given as
\begin{eqnarray}\label{MS4}
L(\Psi_{n,(0)})=0,
\end{eqnarray}
where the Linear operator $L(\Psi_n)$ is specified below, 
\begin{eqnarray}\label{MS3}
L(\Psi_n) = \left(
\begin{array}{c}
\mathrm{i}D_0 \Psi^{(1)}_{n} - \epsilon_0 \Psi^{(1)}_{n}-c_1 \Psi^{(2)}_{n}-c_2 \Psi^{(2)}_{n-1}\\
\mathrm{i}D_0 \Psi^{(2)}_{n} - \epsilon_0 \Psi^{(2)}_{n}-c_1 \Psi^{(1)}_{n}-c_2 \Psi^{(1)}_{n+1} \\
\end{array}
\right).\qquad
\end{eqnarray}
The solution to the zeroth-order equations is 
\begin{eqnarray}\label{MS5}
\Psi_{n,(0)} & = & \mathrm{i} A(T_{(1)}) e^{\mathrm{i}qn-\mathrm{i}\omega_{(0)} T_{(0)}-\mathrm{i}\theta(T_{(1)})}   (e^{\mathrm{i}\phi_q^{(0)}}, 1)^\top,\qquad
\end{eqnarray}
where $\Delta\omega_{(0)}=\omega_{(0)} -\epsilon_0= \pm\sqrt{c_1^2+c_2^2+2c_1c_2\cos q}$, $\tan\phi_q^{(0)} = -{c_2\sin q}/{(c_1+c_2\cos q)}$, and $\theta = \theta(T_{(1)})$ is the arbitrary phase condition for the bulk modes. We note that this phase is a constant up to the fast time scale $T_{(0)}$ but can depend on the slow time scale $T_{(1)}$. It provides the frequency shift due to the nonlinearities. In what follows, we will focus on computing this frequency shift. To this end, we consider the first-order equations of motion,
\widetext
\begin{eqnarray}\label{MS6}
 L(\Psi_{n,(1)}) + 
\left(
\begin{array}{c}
{\rm i}D_1\Psi_{n,(0)}^{(1)}-d_1[({\rm Re\,}\Psi_{n,(0)}^{(2)})^3+{\rm i}({\rm Im\,}\Psi_{n,(0)}^{(2)})^3]
-d_2[({\rm Re\,}\Psi_{n-1,(0)}^{(2)})^3+{\rm i}({\rm Im\,}\Psi_{n-1,(0)}^{(2)})^3]\\
{\rm i}D_1\Psi_{n,(0)}^{(2)}-d_1[({\rm Re\,}\Psi_{n,(0)}^{(1)})^3+{\rm i}({\rm Im\,}\Psi_{n,(0)}^{(1)})^3] 
-d_2[({\rm Re\,}\Psi_{n+1,(0)}^{(1)})^3+{\rm i}({\rm Im\,}\Psi_{n+1,(0)}^{(1)})^3]\\
\end{array}
\right)
 =0.
\end{eqnarray}
The solution of Eq.(\ref{MS6}), namely the first-order correction of wave function $\Psi_{n,(1)}$, has two components $\Psi_{n,(1)} = \Psi_{n,(1)}(\omega)+\Psi_{n,(1)}(3\omega)$: a fundamental-harmonic part $\Psi_{n,(1)}(\omega)$ and a third-harmonic part $\Psi_{n,(1)}(3\omega)$. We are interested in how the nonlinearities modify the frequencies of the bulk modes, which stem from the secular term generated by the fundamental harmonics. On the other hand, the frequency-tripling part does not contribute to the secular term and the subsequent frequency shift. Hence, we consider the fundamental harmonic part only. The equations of the fundamental part $\Psi_{n,(1)}(\omega)$ are given as follows, 
\begin{eqnarray}\label{MS6.1}
 L(\Psi_{n,(1)}(\omega)) + 
e^{\mathrm{i}(qn-\omega_{(0)}T_{(0)}-\theta)}\left(
\begin{array}{c}
(-D_1 A+\mathrm{i}AD_1\theta)e^{\mathrm{i}\phi_q^{(0)}}-\frac{3}{4}\mathrm{i} A^3 (d_1+d_2 e^{-\mathrm{i}q})\\
(-D_1 A+\mathrm{i}AD_1\theta)-\frac{3}{4}\mathrm{i} A^3(d_1+d_2e^{\mathrm{i}q})e^{\mathrm{i}\phi_q^{(0)}}\\
\end{array}
\right)
 =0.
\end{eqnarray}
\endwidetext
We want to find $ \Psi_{n,(1)}(\omega)$ orthogonal to $\ker(L(\Psi_n))$, which is of the form 
\begin{eqnarray}\label{MS7}
 \Psi_{n,(1)}(\omega) =a (  e^{\mathrm{i}\phi_q^{(0)}},  -1 )^\top  e^{\mathrm{i}(qn-\omega_{(0)} T_{(0)} - \theta(T_{(1)}) )}, \qquad
\end{eqnarray}
where $a$ is a complex number. We use Eq.(\ref{MS6.1}) to solve $a$, $D_1 A$, and $D_1\theta$, 
\begin{eqnarray}\label{MS6.9}
 & {} & D_1\theta  =
\frac{3 A^2}{4}[d_1 \cos\phi_q^{(0)}+ d_2 \cos(\phi_q^{(0)}+q) ],
\nonumber \\
 & {} & D_1A = 0,
\nonumber \\
 & {} & 
a = \frac{3A^3}{8\Delta\omega_{(0)}}[d_1 \sin\phi_q^{(0)}+d_2 \sin(\phi_q^{(0)}+q)].\qquad
\end{eqnarray}
We note that the result $D_1A=0$ is natural for undamped systems. In Eqs.(\ref{MS6.9}), since $a\in\mathcal{R}$ is real, it is convenient to denote the real quantity $\phi_q^{(1)} =-2 a/A$. To the order $\mathcal{O}(\epsilon^1)$, the bulk mode solution can therefore be simplified as the following compact form, 
\begin{eqnarray}\label{MS6.2}
\Psi_n  =  \mathrm{i}A (e^{\mathrm{i}(\phi_q^{(0)}+\frac{1}{2}\epsilon\phi_q^{(1)})}, e^{-\frac{1}{2}\mathrm{i}\epsilon\phi_q^{(1)}})^\top   e^{\mathrm{i}qn-\mathrm{i} (\omega_{(0)}+\epsilon D_1\theta)T_{(0)} }.\nonumber \\
\end{eqnarray}
Hence, as the amplitude rises, the relative phase between two wave components changes from $\phi_q^{(0)}$ to $\phi_q^{(0)}+\epsilon \phi_q^{(1)}$.

Method of multiple-scale is a trustworthy technique in weakly nonlinear regime by allowing perturbative analysis. It provides nonlinear effects quantitatively, like the frequency shift $D_1\theta$. They help to verify the correctness of other numerical methods in strongly nonlinear regime. The good agreement of the frequency shift in weakly nonlinear regime between method of multiple-scale and shooting method is presented in Fig.\ref{SIfig6}.

\subsection{Shooting method: bulk modes in strongly nonlinear regime}
Secondly, we introduce shooting method which numerically computes nonlinear bulk modes of Eqs.(\ref{1v2}) in strongly nonlinear regime, where the nonlinearities are comparable to the linear interactions and perturbation theory breaks down. We define the $4N\times 1$ vector field $z(t)$ which describes the wave functions of all particles, 
\begin{eqnarray}
z(t)  & = &  \big({\rm Re\,}\Psi^{(1)}_1, {\rm Im\,}\Psi^{(1)}_1, {\rm Re\,}\Psi^{(2)}_1, {\rm Im\,}\Psi^{(2)}_1, \nonumber \\
 & {} & \ldots, {\rm Re\,}\Psi^{(1)}_N, {\rm Im\,}\Psi^{(1)}_N, {\rm Re\,}\Psi^{(2)}_N, {\rm Im\,}\Psi^{(2)}_N\big)^\top.\qquad
\end{eqnarray}
The equation of motion for $z(t)$ is $d{z}/dt = g(z)$, which in turn gives 
\begin{eqnarray}
z(t) = z(0)+\int_0^t g(z(t'))dt' ,
\end{eqnarray}
where $g(z)$ is a $4N\times 1$ vector derived from the nonlinear equations of motion. Each component is displayed as follows,
\begin{eqnarray}\label{D1.2}
 & {} & g_{4n-3} =+\epsilon_0 z_{4n-2}+F_1(z_{4n-0})+F_2(z_{4n-4}),\nonumber \\
 & {} & g_{4n-2} =-\epsilon_0 z_{4n-3}-F_1(z_{4n-1})-F_2(z_{4n-5}),\nonumber \\
 & {} & g_{4n-1} =+\epsilon_0 z_{4n-0}+F_1(z_{4n-2})+F_2(z_{4n+2}),\nonumber \\
 & {} & g_{4n-0} =-\epsilon_0 z_{4n-1}-F_1(z_{4n-3})-F_2(z_{4n+1}),
\end{eqnarray}
where $1\le n \le N$, and $F_i(x) = c_i x+d_i x^3$. 

The considered nonlinear wave function at time $t=0$ reads $z(t=0)$. It evolves forward in time for $T$, and then the wave function is given by $z(t=T)$. In general, $z(T)\neq z(0)$ since the considered wave may not be periodic in time. In the rest of this section, we denote the nonlinear mode that starts with $z(0)$ and evolves forward in time for $T$ as $\{z(0), T\}$. We further denote a periodic nonlinear solution as $\{z_{\rm p}(0), T_{\rm p}\}$, meaning that at $t=0$ the wave function is $z_{\rm p}(t=0)$ and the mode period is $T_{\rm p}$. Thus, it is straightforward to have $z_{\rm p}(T_{\rm p})-z_{\rm p}(0)=0$. In order to quantify ``how far away" $\{z(0), T\}$ is from $\{z_{\rm p}(0), T_{\rm p}\}$, we define the ``shooting function" $H(z(0), T)$ as follows,
\begin{eqnarray}\label{D2}
H(z(0), T) = z(T)-z(0) = \int_0^{T} g(z(t))dt.
\end{eqnarray}
$H(z(0), T)\neq 0$ for a temporal aperiodic mode $\{z(0), T\}$, and $H(z_{\rm p}(0), T_{\rm p})= 0$ for the periodic solution $\{z_{\rm p}(0), T_{\rm p}\}$. The smaller the shooting function $H(z(0), T)$ is, the closer $\{z(0), T\}$ is to the periodic solution.

From now on we attempt to find periodic solutions by lowering the shooting function in a recursive algorithm, which is known as shooting method. We start the algorithm with a guessing initial wave function $\{z_1(0), T_1\}$: at $t=0$, the imported guessing wave is $z_1(t=0)$ and the imported guessing period is $T_1$, which means we will evolve $z_1(t=0)$ forward in time for $T_1$ to evaluate the shooting function $H(z_1(0), T_1)$. Here, $z_1(t=0)$ and $T_1=2\pi/(\omega_{(0)}+\epsilon D_1\theta)$ are chosen from Eq.(\ref{MS6.2}), which implicitly determines the wavenumber $q$. $\{z_1(0), T_1\}$ is not a true periodic solution, and the subsequent shooting function $H(z_1(0), T_1)\neq 0$. In order to approach the true periodic solution $\{z_{\rm p}(0), T_{\rm p}\}$, we make corrections to $\{z_1(0), T_1\}$ to obtain the second guessing solution $\{z_2(0), T_2\}$. We repeat this process to obtain a series of guessing solutions $\{z_l(0), T_l\}$ ($l\ge 1$). The corresponding shooting functions $H(z_l(0), T_l)$ slowly converge to zero as $l$ increases.

In the $l$-th step, the guessing solution is denoted as $\{z_l(0), T_l\}$, and the associated shooting function is $H(z_l(0), T_l)\neq 0$. Therefore, we make corrections $\{\Delta z_l(0), \Delta T_l\}$ to the $l$-th step guessing solution to obtain the guessing solution of $(l+1)$-th step, 
\begin{eqnarray}\label{D1.1}
\{z_{l+1}(0), T_{l+1}\} = \{z_l(0)+\eta_A^{-1}\Delta z_l(0), T_l+\eta_T^{-1}\Delta T_l\},\nonumber \\
\end{eqnarray}
such that 
\begin{eqnarray}\label{D1.11}
|H(z_{l+1}(0), T_{l+1})|<|H(z_{l}(0), T_{l})|. 
\end{eqnarray}
In other words, $H(z_{l+1}(0), T_{l+1})$ is closer to zero than $H(z_{l}(0), T_{l})$. $\eta_A$ and $\eta_T$ in Eq.(\ref{D1.1}) are constants greater than 1, which slow down the evolution speed. In order to achieve Eq.(\ref{D1.11}), the correction $\{\Delta z_l(0), \Delta T_l\}$ is determined by the following matrix equation, 
\begin{eqnarray}\label{D3}
 & {} & H(z_{l+1}(0), T_{l+1}) \approx \nonumber \\
 & {} &  H(z_l(0), T_l)+\frac{\partial H}{\partial z_l(0)}\Delta z_l(0)+\frac{\partial H}{\partial T_l}\Delta T_l=0, 
\end{eqnarray}
where the matrices ${\partial H}/{\partial z_l(0)}$ and ${\partial H}/{\partial T_l}$ will be elaborated later. By repeating the above process, we generate a sequence of guessing solutions $\{z_l(0), T_l\}$, from which the shooting functions converge to zero, 
\begin{eqnarray}
 & {} & \lim_{l\to\infty}H(z_l(0), T_l) = 0\nonumber \\
 & \Rightarrow &  
\lim_{l\to\infty}\{z_l(0), T_l\} = \{z_{\rm p}(0), T_{\rm p}\}.
\end{eqnarray}
As the iteration step $l$ increases, we find the periodic nonlinear solution.

We now compute $\{\Delta z_l(0), \Delta T_l\}$. We first examine the number of variables in $\{\Delta z_l(0), \Delta T_l\}$ versus the number of constraints in Eq.(\ref{D3}). At first glance, there are $4N+1$ variables but $4N$ constraints in Eq.(\ref{D3}), which means that $\{\Delta z_l(0), \Delta T_l\}$ is indeterminate. However, we note that the solutions we seek are periodic in time. If $\{z_{\rm p}(0), T_{\rm p}\}$ is a periodic solution, so as $\{z_{\rm p}(t\neq 0), T_{\rm p}\}$ for an arbitrary initial time $t$. In other words, a phase condition has to be imposed to remove this arbitrariness. In our numerics, the phase condition is imposed by letting 
\begin{eqnarray}\label{D3.1}
\Delta z_l(0)|_{4N} = 0, \qquad  \forall l\ge 1,
\end{eqnarray}
and then in Eq.(\ref{D3}) the numbers of variables and constraints match. Next, we elaborate the matrices appeared in Eq.(\ref{D3}) as follows, 
\begin{eqnarray}\label{D4}
\frac{\partial H}{\partial T_l}  = g(z_l(T_l)),
\end{eqnarray}
and
\begin{eqnarray}\label{D5}
\frac{\partial H}{\partial z_l(0)}=\zeta(T_l)-I,
\end{eqnarray}
where $\zeta(t) \overset{\rm def}{=} \partial z_l(t) /\partial z_l(0)$, and it is obvious that $\zeta(0)=I$. $\zeta(T_l)$ can be computed in the following way. We find that ${d\ln\zeta}/{dt} = M_l$, which in turn gives 
\begin{eqnarray}\label{D6}
\zeta(T_l) = \exp \int_0^{T_l} M_l(t) dt,
\end{eqnarray}
where the monodromy matrix $M$ is defined as below
\begin{eqnarray}\label{D7}
M = {\partial g(z)}/{\partial z}.
\end{eqnarray}
In our problem, each element of the monodromy matrix $M$ is elucidated as follows, 
\begin{eqnarray}
 & {} & M_{4n-3,4n-4} = +dF_2(x)/dx|_{z_{4n-4}}, \nonumber \\
 & {} & M_{4n-3,4n-2} = +\epsilon_0, \nonumber \\
 & {} & M_{4n-3,4n-0} = +dF_1(x)/dx|_{z_{4n-0}},\nonumber \\
 & {} & M_{4n-2,4n-5} = -dF_2(x)/dx|_{z_{4n-5}}, \nonumber \\
 & {} & M_{4n-2,4n-3} = -\epsilon_0, \nonumber \\
 & {} & M_{4n-2,4n-1} = -dF_1(x)/dx|_{z_{4n-1}},\nonumber \\
 & {} & M_{4n-1,4n-2} = +dF_1(x)/dx|_{z_{4n-2}}, \nonumber \\
 & {} & M_{4n-1,4n-0} = +\epsilon_0, \nonumber \\
 & {} & M_{4n-1,4n+2} = +dF_2(x)/dx|_{z_{4n+2}},\nonumber \\
 & {} & M_{4n-0,4n-3} = -dF_1(x)/dx|_{z_{4n-3}}, \nonumber \\
 & {} & M_{4n-0,4n-1} = -\epsilon_0, \nonumber \\
 & {} & M_{4n-0,4n+1} = -dF_2(x)/dx|_{z_{4n+1}}.
\end{eqnarray}
In summary, we employ Eqs.(\ref{D3.1}, \ref{D4}, \ref{D5}), to solve $\{\Delta z_l(0), \Delta T_l\}$ in Eq.(\ref{D3}) in every iteration step of shooting method.

So far, we have evolved a guessing solution into a periodic solution of certain small amplitude $A_1$. Our next goal is to find nonlinear periodic solutions of the amplitudes greater than $A_1$. Let us denote the above well-established nonlinear bulk mode as $\{z_{\rm p}(0; A_1), T_{\rm p}(A_1)\}$. We find nonlinear bulk modes of highers amplitudes by using the following strategy. We rescale the wave function by a uniform factor $1+\xi$ ($\xi\ll 1$), to initialize shooting method with the new guessing solution, 
\begin{eqnarray}\label{D7.1}
\{z_1(0), T_1\} = \{(1+\xi)z_{\rm p}(0; A_1), T_{\rm p}(A_1)\}. 
\end{eqnarray}
Shooting method morphs it into a new periodic solution of the amplitude $A_2$, which we denote $\{z_{\rm p}(0; A_2), T_{\rm p}(A_2)\}$. We note that $A_2$ is slightly greater than $A_1$, but $A_2\neq (1+\xi)A_1$ because the trial wave function in Eq.(\ref{D7.1}) is not a periodic solution. By repeating this strategy, we get nonlinear bulk modes for a wide range of amplitudes.

Having established the algorithm of shooting method, we now elaborate the numerical details of all parameters. Two sets of parameters of 1D generalized nonlinear Schr\"{o}dinger equations are considered in this paper. 

In the first set of parameters, the nonlinear model is subjected to reflection symmetry only. The on-site potential $\epsilon_0$ adopted in Eqs.(\ref{1v2}) are $\epsilon_0=1.5$ for Fig.\ref{fig1} and Fig.\ref{fig2} and $\epsilon_0=8$ for Fig.\ref{fig3}, and the parameters of nonlinear interactions in Eq.(\ref{9v2}) are specified as $c_1=0.25$, $c_2=0.37$, and $d_1 = 0.22$, $d_2 = 0.02$. We note that the  topological attributes are not sensitive to the parameters. These parameters are randomly chosen. In order to numerically solve a nonlinear mode of wavenumber $q = 2\pi m/N$ ($m,N\in\mathcal{Z}$), we construct a chain of $N$ unit cells composed of classical dimer fields, subjected to PBC. Consequently, the wavenumbers are rational numbers multiple of $2\pi$. 
By constructing lattices with different unit cell numbers $N$, we initialize nonlinear modes with different wavenumbers. Since the wavenumbers $q = 2\pi\times (2m/2N)= 2\pi (m+N)/N=2\pi m/N\mod 2\pi$, we further restrict $0\le m \le N-1$, and $\gcd(m,N)=1$. We begin shooting method by employing the guessing perturbative solution $\{z_1(0), T_1\}$ in Eq.(\ref{MS6.2}), with the period $T_1$, the wavenumber $q = 2\pi m/N$, and the small amplitude $A_1\lesssim 10^{-1} \min(\sqrt{|c_1/d_1|}, \sqrt{|c_2/d_2|}, \sqrt{|(c_1-c_2)/(d_1-d_2)|})$. We simulate the differential equation by executing Runge-Kutta 6th-order~\cite{RK6} (RK6) and converting the time-differential operator $\partial/\partial t$ to the time-step $\Delta t = T_1/N_T$, where $N_T = 1000$. After $N_T$ steps of the motion equations, the wave function should go back to the beginning state if it is a periodic solution. Thus, we compute the shooting function $H(z_{1}(0), T_{1})$ to quantify how far away the wave function is from the periodic solution, and then slowly evolve the wave function towards it. $\eta_A = 300$ and $\eta_T = 10$ are adopted in Eq.(\ref{D1.1}) to slow down the evolution process. In the $l$-th step of shooting method, the period is evolved to $T_l$, which in turn asks the time step to be $\Delta t = T_l/N_T$. In other words, we adjust the time difference $\Delta t$ while keep the number of time steps $N_T$ unchanged throughout the evolution procedure of shooting method. We keep evolving a nonlinear mode before the error of shooting function $e$ reaches the numerical tolerance $e_{\rm max}$, 
\begin{eqnarray}\label{E179}
e \overset{\rm def}{=} \frac{1}{4N}\sum_{i=1}^{4N}|H_i(z_l(0), T_l)| < e_{\rm max}=3\times 10^{-3},\qquad
\end{eqnarray}
where $H_i$ is the $i$th component of the $4N\times 1$ vector of shooting function, and $e_{\rm max}$ is the numerical tolerance. In later discussions of this section, we will demonstrate the correspondence between the condition of $e<e_{\rm max}$ and the stability of nonlinear traveling waves by illustrating a stable mode ($e\ll e_{\rm max}$), a mode on the verge of stability ($e\lesssim e_{\rm max}$), and an unstable mode ($e>e_{\rm max}$) in Fig.\ref{SIfig1}. It is at this point that shooting method returns a periodic nonlinear traveling wave of amplitude $A_1$ and wavenumber $q=2\pi m/N$. The next goal is to find periodic bulk modes of higher amplitudes. To this end, we uniformly rescale the aforementioned wave function by a factor of $1+\xi$ ($\xi=3\times 10^{-3}$), to establish a new shooting procedure. Again, shooting method morphs the trial wave function into a traveling solution of amplitude $A_2$. We repeat this strategy to obtain a series of nonlinear bulk modes with the given wavenumber $q=2\pi m/N$ and a wide range of amplitudes.

Given the wave amplitude $A$, the nonlinear band structure $\omega = \omega(q\in[0,2\pi],A)$ is plotted by selecting the frequencies of nonlinear bulk modes when the mode amplitudes $A'$ are within the numerical tolerance, 
\begin{eqnarray}\label{E1.3}
|(A-A')/A| < \xi=3\times 10^{-3}.
\end{eqnarray}

\renewcommand{\thefigure}{C1}
\begin{figure}[htb]
\includegraphics[scale=0.58]{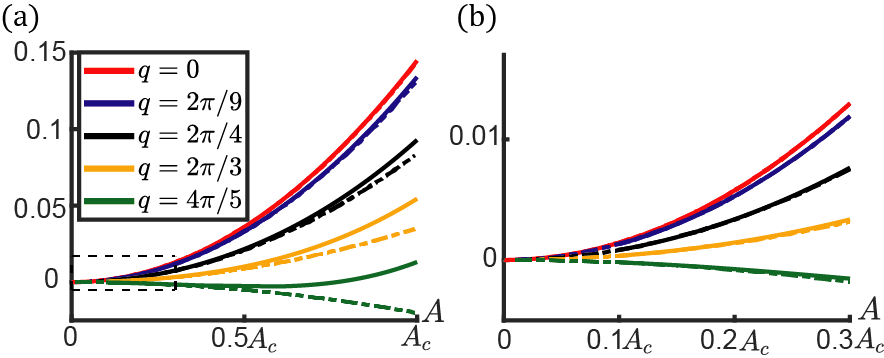}
\caption{Comparing shooting method (solid curves) and method of multiple-scale (dashed curves) on the frequency shift of nonlinear bulk waves. These nonlinear bulk modes start from $A=0$ in the upper nonlinear band to $A_c$ for a list of wavenumbers from $q=0$ to $4\pi/5$. The model and interaction parameters are depicted by Fig.\ref{fig1}. Frequency shift computed by shooting method is $\delta\omega(q,A) = \omega(q,A)-\omega(q,A=0)$, where $\omega(q,A)$ is the frequency of nonlinear bulk mode. Frequency shift obtained by method of multiple-scale is given by $\delta \omega(q,A)=D_1\theta$ in Eqs.(\ref{MS6.9}). (a) These two methods agree quite well in weakly nonlinear regime when $A\ll A_c$, while for $A\gtrsim 0.5A_c$, the large deviations demonstrate the breaking down of perturbation theory. (b) Enlarged data for $A\le 0.3A_c$ encircled by the black dashed box in (a).
}\label{SIfig6}
\end{figure}

\renewcommand{\thefigure}{C2}
\begin{figure}[htb]
\includegraphics[scale=0.58]{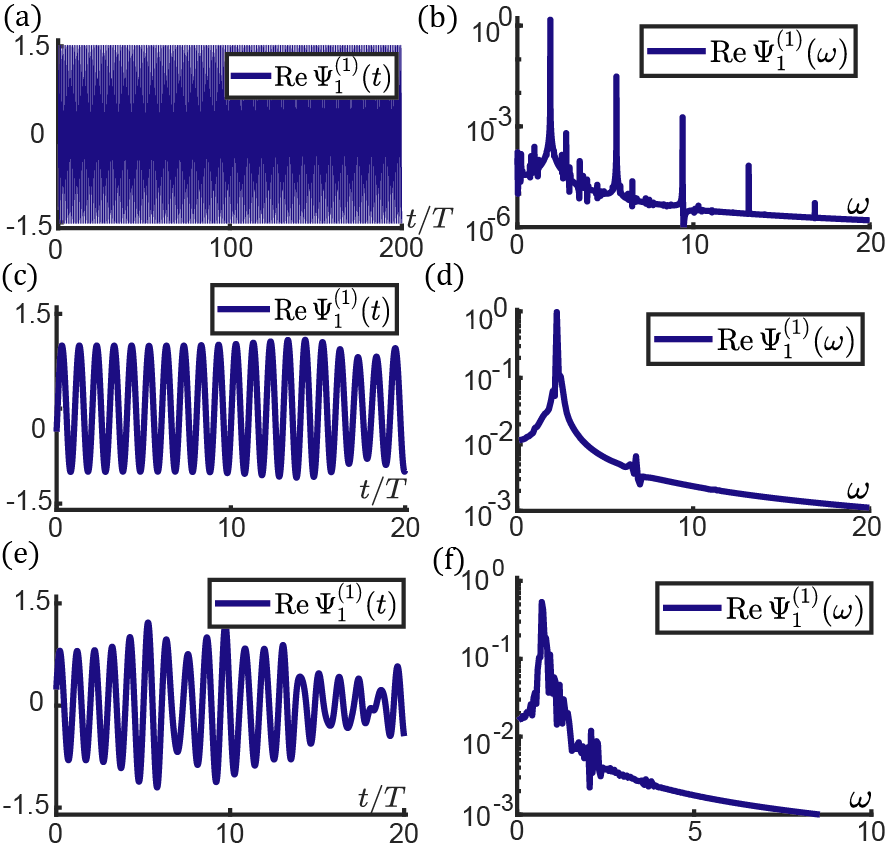}
\caption{Stability analysis of nonlinear bulk modes by performing the algorithm of self-oscillation. (a) A nonlinear bulk mode with the amplitude $A=1.515$ and wavenumber $q=4\pi/5$. The error of shooting function is $e=10^{-6}\ll e_{\max} = 3\times 10^{-3}$ (see Eq.(\ref{E1})), which suggests that the mode is stable. To verify our expectation, we initialize the mode from shooting method, and additionally impose a random perturbation $\delta\Psi_n^{(i)}$ on the wave function, where ${\rm Re\,}\delta\Psi_n^{(i)}$ and ${\rm Im\,}\delta\Psi_n^{(i)}$ are random numbers within $10^{-3}$. The mode persists for more than 200 periods without generating other nonlinear modes, which demonstrates mode stability. We note that $A^2\max(d_1, d_2)/\max(c_1, c_2)=1.364$, which means the nonlinearities are larger than the linear parts of interactions. (b) The Fourier analysis of (a) after 200 periods provides additional evidence of mode stability. (c) A nonlinear bulk mode with $A=0.9971$ and $q=2\pi/9$. The error of shooting function is $e=2.1\times 10^{-3}< e_{\max}$, which suggests that the mode is on the verge of stability. We initialize the mode from shooting method without imposing any wave function perturbation. The mode persists for 10 periods with the change of amplitude~\cite{fronk2017higher} smaller than $2.3\%$ and is therefore on the verge of stability. (d) The frequency spectrum of the mode in (c) manifests fundamental harmonic and frequency-tripling components. (e) A nonlinear bulk mode with $A=0.8037$ and $q=2\pi/21$. The error of shooting function is $e=7.9\times 10^{-3}>e_{\max}$, which indicates that the mode is unstable. The mode initialized by shooting method persists in 5 periods of oscillation, and it quickly exhibits mode instability by producing other nonlinear modes. (f) The frequency profile of (e) demonstrates the emergence of other Fourier components, which identifies mode instability. 
}\label{SIfig1}
\end{figure}

We now turn to discuss the stability analysis of nonlinear bulk modes. The stability analysis of nonlinear modes~\cite{fronk2017higher, peeters2008nonlinear} is to measure how many periods they persist in an undriven, undamped lattice before falling apart. According to Ref.~\cite{fronk2017higher}, the mode is considered stable if an instability does not occur within 10 periods of oscillation. In order to perform the stability analysis for a nonlinear bulk mode with the wavenumber $q=2\pi m/N$, we construct a lattice that comprises $N$ dimer unit cells and is subjected to PBC. We establish a nonlinear bulk mode obtained from shooting method. After letting the mode to oscillate by itself for more than 10 periods, Fourier analysis is applied to characterize whether the mode experiences instability and falls apart to other nonlinear modes. In Fig.\ref{SIfig1}, we exemplify three different bulk modes to verify the correspondence between Eq.(\ref{E179}) and the mode stability. Hence, all nonlinear bulk modes depicted in the nonlinear band structures of figs.\ref{fig1}(e, f) and Fig.\ref{fig3}(a) are considered stable, and they fulfill the criteria of the nonlinear extension of adiabatic theorem~\cite{RevModPhys.82.1959, PhysRevLett.90.170404, PhysRevLett.98.050406, PhysRevA.81.052112}.

In the second set of parameters, $c_1 = 0.25$, $c_2 = 0.37$, $d_1=0.22$, $d_2=0.02$ are carried over, while $\epsilon_0$ is now set to zero. Nonlinear bulk modes always appear in $\pm\omega$ pairs. Similar to the linear counterpart in which charge-conjugation symmetry~\cite{ryu2010topological} is present, the frequencies of nonlinear topological edge modes are zero in the second case.

\subsection{Topological transition amplitude $A_c$: calculating nonlinear bulk modes at high-symmetry points}
In this subsection, we solve nonlinear bulk modes at high-symmetry points when $q_{\rm hs} = 0,\pi$. This allows us to numerically find the topological transition amplitude $A_c$ as well as the band-touching frequency $\omega$.

We denote the nonlinear bulk modes at high-symmetry points as $\Psi_{\rm hs}$. According to Eq.(\ref{A5.2}), the relative phase at high-symmetry points are $\phi_{\rm hs}=0$ or $\pi$. The motion equation of $\Psi_{\rm hs}$ is greatly simplified by employing Eqs.(\ref{A5}, \ref{C1}), 
\begin{eqnarray}\label{C2}
  & {} & (\mathrm{i} \omega\partial_\theta-\epsilon_0) \Psi^{(j)}_{\rm hs} = \nonumber \\ 
 & {} & e^{\mathrm{i}\phi_{\rm hs}}f_1(\Psi^{(j')}_{\rm hs},\Psi^{(j)}_{\rm hs}) +e^{\mathrm{i}(q_{\rm hs}+\phi_{\rm hs})}f_2(\Psi^{(j')}_{\rm hs},\Psi^{(j)}_{\rm hs}), \qquad
\end{eqnarray}
for $j=1,2$. The nonlinear interactions are adopted from Eq.(\ref{9v2}). By solving Eq.(\ref{C2}), $({\rm Re\,}\Psi^{(j)}_{\rm hs},{\rm Im\,}\Psi^{(j)}_{\rm hs})$ yield the trajectory, 
\begin{eqnarray}\label{C5}
 [({\rm Re\,}\Psi^{(j)}_{\rm hs})^2-x_0]^2+[({\rm Im\,}\Psi^{(j)}_{\rm hs})^2 -x_0]^2= R^2,\qquad
\end{eqnarray}
where $R^2$ is the constant of integration which quantifies the ``radius" of the trajectory, and 
\begin{eqnarray}\label{C6}
x_0 = -\frac{\epsilon_0 + e^{\mathrm{i}\phi_{\rm hs}}c_1 +e^{\mathrm{i}(q_{\rm hs}+\phi_{\rm hs})}c_2}{e^{\mathrm{i}\phi_{\rm hs}}d_1 +e^{\mathrm{i}(q_{\rm hs}+\phi_{\rm hs})}d_2}.
\end{eqnarray}
In the linear limit, the trajectory simply reduces to a circle, which is in perfect agreement with linear models. Based on Eq.(\ref{C5}), we further obtain the mode frequencies:
\begin{eqnarray}\label{C7}
 & {} & \omega(q_{\rm hs}, \phi_{\rm hs}) =  
\frac{\pi }{2}
\left[\int_0^{A}\frac{\mathrm{d}u/|d_1 +e^{\mathrm{i}q_{\rm hs}}d_2|}{y(u) \sqrt{|y(u)+e^{\mathrm{i}\phi_{\rm hs}}x_0 |}}\right]^{-1},\qquad
\end{eqnarray}
where $A$ is the mode amplitude, and 
\begin{eqnarray}\label{C7.1}
y(u) = \sqrt{x_0^2+(A^2-x_0)^2-(u^2-x_0)^2}.
\end{eqnarray}
A quick check of the above result is to perform the intergration in the weakly nonlinear regime when $A\ll \sqrt{|x_0|}$. Eq.(\ref{C7}) reduces to $\omega = |\epsilon_0 + e^{\mathrm{i}\phi_{\rm hs}}c_1 +e^{\mathrm{i}(q_{\rm hs}+\phi_{\rm hs})}c_2|$, which is in line with the high-symmetry eigenfrequencies in the linear models. In this paper, the numerical parameters we adopt yield $\epsilon_0, c_1, c_2, d_1, d_2>0$, and $c_1<c_2$, $d_1>d_2$. Thus, the topological phase transition occurs when the frequencies of nonlinear modes merge at the critical amplitude $A_c$ when 
\begin{eqnarray}\label{C7.2}
\omega(\phi_{\pi} = 0, A_c) = \omega(\phi_\pi = \pi, A_c). 
\end{eqnarray}
This transition amplitude $A_c$ can be obtained by numerically solving the above equation, which is shown in Fig.\ref{SIfig3}(b). In linear SSH model, the topological transition point occurs at the frequency $\omega(\phi_{\pi} = 0) = \omega(\phi_\pi = \pi)=\epsilon_0$, which is in perfect agreement with the frequency of topological boundary modes, $\omega_{\rm T} = \epsilon_0$. Thus, the frequency of topological modes is always separated from the bulk bands unless topological transition is reached. Unlike linear models, the topological transition of the nonlinear system occurs at the frequency $\omega(\phi_{\pi} = 0, A_c) = \omega(\phi_\pi = \pi, A_c) = (1+3\times 10^{-4})\epsilon_0$, which is slightly different from $\epsilon_0$ (Fig.\ref{fig1}(d) of the main text). The small rectification of band-touching frequency stems from the coupling between higher-order and fundamental bulk mode components. On the other hand, the frequencies of nonlinear topological modes $\omega_{\rm T} = \epsilon_0$ are approximately solved by truncating the motion equations to the fundamental harmonics. Thus, if we consider all couplings among higher-order harmonics, the frequencies of topological modes are rectified as well to stay in the bandgap and are thus separated from nonlinear bulk bands. In addition to the distinguishable frequencies, the amplitudes of each site are remarkably different between nonlinear bulk and edge modes. In nonlinear bulk modes, the amplitudes are equal for A and B-sites, whereas the amplitudes of B-sites are negligible compared to A-sites for topological edge modes. When nonlinear topological modes reach the critical amplitude and penetrate infinitely into the lattice, this \emph{nonlinear} mode cannot be decomposed as the superposition of two nonlinear bulk modes. This is in sharp contrast to linear systems in which at the transition point, topological modes can be represented as the superposition of two bulk modes. Thus, nonlinear topological boundary modes are separated from bulk modes, in the sense that they cannot be continuously deformed into one another.

\section{Nonlinear topological edge modes}

In this section, we study nonlinear topological edge modes based on the model of Eqs.(\ref{1v2}) with the interactions specified in Eq.(\ref{9v2}). To have topological edge modes, we consider a semi-infinite lattice subjected to the open boundary condition (OBC)
\begin{eqnarray}\label{E1}
 & {} & \mathrm{i}\partial_t\Psi^{(1)}_n = \epsilon_0\Psi^{(1)}_n+f_1(\Psi^{(1)}_n,\Psi^{(2)}_n) +f_2(\Psi^{(1)}_n,\Psi^{(2)}_{n-1}),  \nonumber \\
 & {} & \mathrm{i}\partial_t\Psi^{(2)}_n = \epsilon_0\Psi^{(2)}_n+f_1(\Psi^{(2)}_n,\Psi^{(1)}_n) +f_2(\Psi^{(2)}_n,\Psi^{(1)}_{n+1}), \nonumber \\
 & {} & {\rm for\quad} n\ge 1, \quad {\rm and \quad} \Psi_0^{(2)} = 0.
\end{eqnarray}
In subsections 1 and 2, we investigate topological edge modes for the model with $\epsilon_0\neq 0$. In subsection 3, we explore topological modes for the vanishing on-site potential $\epsilon_0= 0$. The parameters we consider yield $0<c_1<c_2$, $d_1>d_2>0$.

\subsection{Method of multiple-scale: topological edge modes for the $\epsilon_0\neq 0$ case in weakly nonlinear regime}

Based on the numerical simulation and qualitative analysis presented in the main text, it is demonstrated that the frequency of topological edge mode is $\omega_{\rm T}=\epsilon_0$ and is independent of the mode amplitude $A$. This result is in sharp contrast to the amplitude-dependent frequencies of nonlinear bulk modes. Here in weakly nonlinear regime, we quantitatively exhibit this result by employing the method of multiple-scale.

\widetext

\renewcommand{\thefigure}{D1}
\begin{figure}[htb]
\includegraphics[scale=0.6]{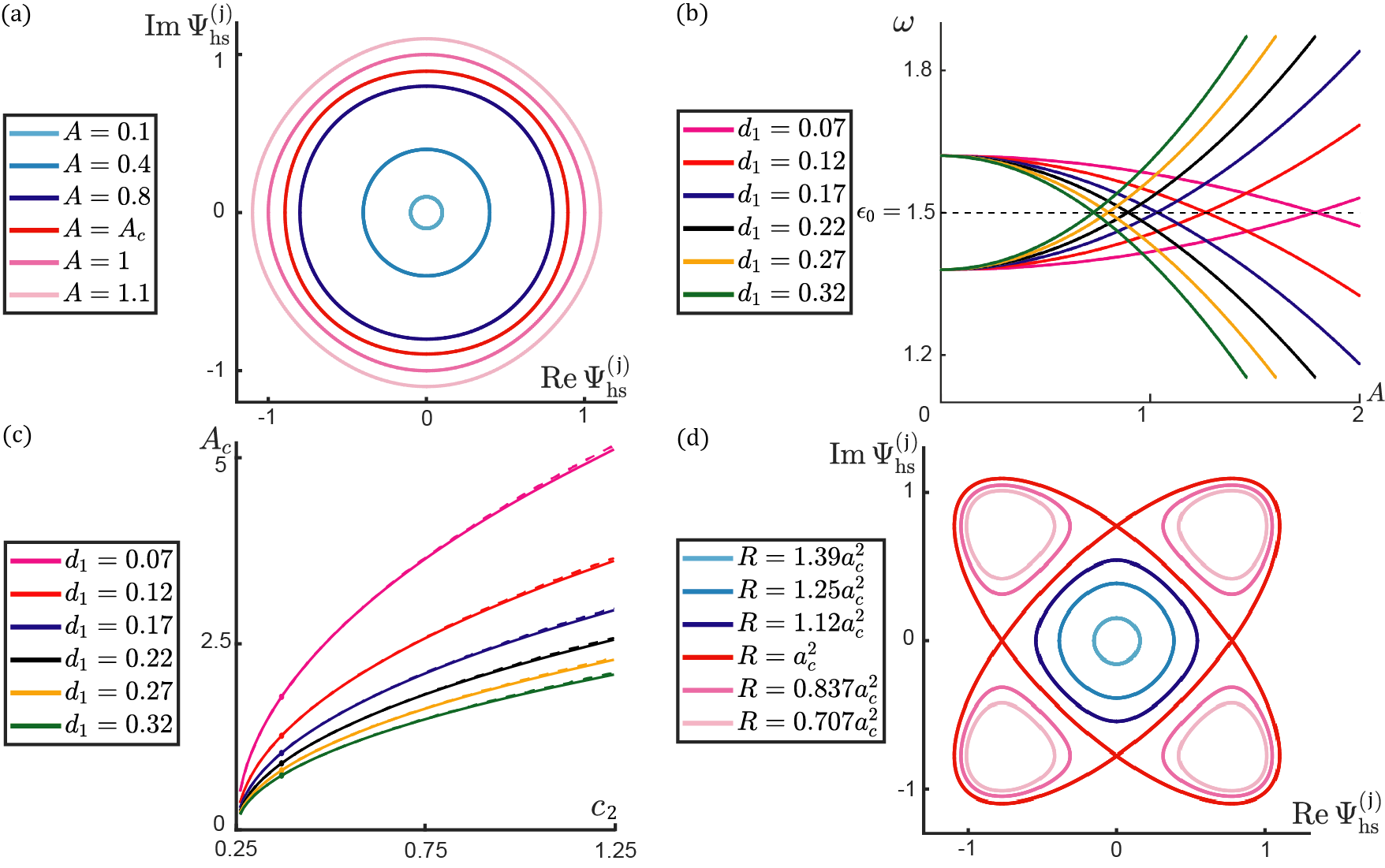}
\caption{(a) In the model with the parameters enumerated in Fig.\ref{fig1}, we plot the $({\rm Re\,}\Psi_{\rm hs}^{(j)}, {\rm Im\,}\Psi_{\rm hs}^{(j)})$ trajectories for nonlinear bulk modes at high-symmetry points with a set of amplitudes ranging from $A=0.1$ to $1.1$. The trajectories are noticeably different from regular circles for $A\gtrsim A_c$. (b) Transition amplitude $A_c$ is numerically solved by Eq.(\ref{C7.2}). Here, we exemplify these numerical solutions by varying $d_1$ from $0.07$ to $0.32$, where the transition amplitude $A_c=0.8944$ for $d_1=0.22$ is depicted by the intersection of black curves. All illustrated transition amplitudes occur at the merging frequency $\omega(\phi_{\pi} = 0, A_c) = \omega(\phi_\pi = \pi, A_c)=\epsilon_0$. (c) The nice agreement between numerically solved $A_c$ (solid curves) and its estimation $A_c\approx \sqrt{4/3}a_c$ (dashed curves), where $c_2$ varies from $0.26$ to $1.25$, and $d_1$ varies from $0.07$ to $0.32$.  The transition amplitudes in (b) are marked by colored dots here. We note that the estimations of $A_c$ are worse for $c_2\gtrsim 1.25$, which is much greater than $c_2=0.37$ in our model. (d) In the second case, all interaction parameters are the same as (a) except that $\epsilon_0=0$. We plot multiple $({\rm Re\,}\Psi_{\rm hs}^{(j)}, {\rm Im\,}\Psi_{\rm hs}^{(j)})$ trajectories with the ``constant of integration" $R$ that varies from $1.39a_c^2$ to $0.707a_c^2$ ($R$ is defined in Eq.(\ref{C5})). Blue and red curves describe nonlinear modes before and after instability occurs, respectively. The instability happens at $R=a_c^2$ which corresponds to mode amplitude $\max|{\rm Re}\,\Psi_1^{(1)}|=a_c$. Above the instability point (i.e., $R< a_c^2$ and $\max|{\rm Re}\,\Psi_1^{(1)}|>a_c$), wave functions oscillate around new equilibrium positions. 
}\label{SIfig3}
\end{figure}

\endwidetext

Method of multiple-scale introduces a book-keeping small parameter $\epsilon\ll 1$ that enforces small amplitudes for the edge modes, which is practically realized by rewriting $d_i$ as $\epsilon d_i$ in the nonlinear interactions. The time derivative and the wave function are expanded in orders of $\epsilon$ (see Eqs.(\ref{MS1}, \ref{MS2})). We expand the equations of motion and match them in orders of $\epsilon$. The zeroth-order equations of motion are presented by Eq.(\ref{MS4}) respecting the OBC $\Psi^{(2)}_{0,(0)} = 0$. The zeroth-order solution reads
\begin{eqnarray}
\Psi_{n,(0)} =  -(-\kappa_0)^{n-1}A(T_{(1)}) e^{-\mathrm{i}\omega_{\rm T(0)} T_{(0)}-\mathrm{i}\theta(T_{(1)})} (1, 0)^\top,\nonumber \\
\end{eqnarray}
where $\kappa_0 = c_1/c_2$ and $\omega_{\rm T(0)} = \epsilon_0$. The first-order equations of motion are given by Eq.(\ref{MS6}) subjected to the open boundary $\Psi^{(2)}_{0,(1)}=0$. There are two parts in this first-order correction of the wave function, namely the fundamental harmonic part $\Psi_{n,(1)}(\omega_{\rm T})$ and the frequency-tripling part $\Psi_{n,(1)}(3\omega_{\rm T})$: $\Psi_{n,(1)} = \Psi_{n,(1)}(\omega_{\rm T})+\Psi_{n,(1)}(3\omega_{\rm T})$. We are interested in the frequency correction due to the nonlinearities, which stems from the secular term generated by the fundamental harmonic part. The fundamental harmonics $\Psi_{n,(1)}(\omega_{\rm T})$ obey the following recursive equations, 
\begin{eqnarray}\label{MS22}
 & {} & 
 \Psi_{n,(1)}^{(2)} (\omega_{\rm T}) + \frac{ \Psi_{n-1,(1)}^{(2)}(\omega_{\rm T})}{\kappa_0}  +\frac{\mathrm{i}D_1A+AD_1\theta}{(-\kappa_0)^{1-n} c_1}    e^{-\mathrm{i}\Phi} = 0,\nonumber \\
 & {} & 
\Psi_{n,(1)}^{(1)}(\omega_{\rm T})+  \frac{\Psi_{n+1,(1)}^{(1)}(\omega_{\rm T})}{\kappa_0}
- \frac{3(d_1- d_2 \kappa_0^{3} ) A^3 e^{-\mathrm{i}\Phi }}{4 (-\kappa_0)^{3-3n} c_1} = 0,\nonumber \\
\end{eqnarray}
subjected to the OBC $\Psi_{0,(1)}^{(2)}=0$, where the phase factor $\Phi = \omega_{\rm T(0)} T_{(0)} + \theta(T_{(1)})$. If $D_1A\neq 0$ or $D_1\theta\neq 0$, Eqs.(\ref{MS22}) lead to the unphysical result that $\lim_{n\to\infty} |{\Psi}_{n,(1)}^{(2)}(\omega_{\rm T})|\to\infty$. Hence Eqs.(\ref{MS22}) demand that $D_1A = D_1\theta= 0$. The result $D_1\theta= 0$ demonstrates that the first-order correction of the frequency of the topological mode is zero. Up to the first-order correction, the frequency is 
\begin{eqnarray}\label{MS23}
\omega_{\rm T} = \omega_{\rm T(0)}+\epsilon D_1\theta = \epsilon_0,
\end{eqnarray}
which is independent of the mode amplitude. This conclusion is in line with the qualitative analysis and the numerical computation of nonlinear topological modes carried out in the main text. We note that $D_1A=0$ is the natural result of undamped systems. The total wave function up to the first-order correction is summarized as follows, 
\begin{eqnarray}\label{MS23.5}
 & {} & \Psi_{n} = \Psi_{n,(0)}+\epsilon (\Psi_{n,(1)}^{(1)}, 0)^\top, \nonumber \\
 & {} & \Psi_{2n,(1)}^{(1)} = \kappa_0^{2n-1} \frac{1-\kappa_0^{4n-2}}{1-\kappa_0^2} \frac{3(d_1-d_2\kappa_0^3)A^3 e^{-\mathrm{i}\Phi}}{4c_1} , \nonumber \\
 & {} & \Psi_{2n+1,(1)}^{(1)} = -\kappa_0^{2n} \frac{1-\kappa_0^{4n}}{1-\kappa_0^2} \frac{3(d_1-d_2\kappa_0^3)A^3 e^{-\mathrm{i}\Phi}}{4c_1}.
\end{eqnarray}
It is notable that the first-order correction of wave function exponentially decays in space and it does not diverge to infinity. In addition to this, the wave function in Eqs.(\ref{MS23.5}) fulfills Eq.(\ref{MS35}), which is the recursion relation of topological edge modes in strongly nonlinear regime. In summary, these results derived from the perturbative method of multiple-scale are in perfect agreement with the methods in strongly nonlinear regime discussed in subsection 2.

\subsection{Harmonic balance method: topological edge modes for the $\epsilon_0\neq 0$ case in strongly nonlinear regime}

\renewcommand{\thefigure}{D2}
\begin{figure}[htb]
\includegraphics[scale=0.55]{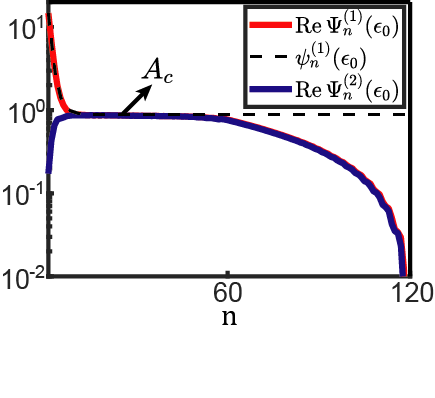}
\caption{Here we plot the entire spatial profile of the $\omega=\epsilon_0$ Fourier component of the nonlinear topological excitation in Fig.\ref{fig3} to complement the results. The plateau reaches site $n\sim 60$ before falling apart to other nonlinear modes. Fourier analysis is performed by considering the excitations from $200T$ to $400T$.
}\label{SIfig11}
\end{figure}

We now employ the harmonic balance method~\cite{detroux2014harmonic} to study topological edge modes in strongly nonlinear regime. Since the mode is periodic in time, it can be expressed as the Fourier series $\Psi_n =\sum_l \psi_{l,n} e^{-\mathrm{i}l\omega_{\rm T} t}$. We take the approximation by truncating the wave function to the fundamental harmonics, 
\begin{eqnarray}\label{MS30}
 & {} & \Psi_n \approx \psi_{1,n} e^{-\mathrm{i}\omega_{\rm T} t}+\psi_{-1,n} e^{\mathrm{i}\omega_{\rm T} t}
 = \nonumber \\
 & {} & 
\frac{1}{2}
\left(
\begin{array}{c}
\alpha^{(1)}_n+\mathrm{i}\alpha^{(2)}_n\\
\beta^{(1)}_n+\mathrm{i}\beta^{(2)}_n\\
\end{array}
\right)e^{-\mathrm{i}\omega_{\rm T} t}+
\frac{1}{2}\left(
\begin{array}{c}
\alpha^{(1)*}_n+\mathrm{i}\alpha^{(2)*}_n\\
\beta^{(1)*}_n+\mathrm{i}\beta^{(2)*}_n\\
\end{array}
\right)e^{\mathrm{i}\omega_{\rm T} t},\nonumber \\
\end{eqnarray}
where $\alpha_n = (\alpha_n^{(1)}, \alpha_n^{(2)})^\top$ and $\beta_n = (\beta_n^{(1)}, \beta_n^{(2)})^\top$ are $2\times 1$ complex vectors parametrizing $\psi_{\pm1, n}$. Hence, the real and imaginary parts of the wave functions can be expressed as 
\begin{eqnarray}\label{MS30.1}
 & {} & {\rm Re\,}\Psi_n^{(1)} = \frac{1}{2}\left(\alpha_n^{(1)}e^{-\mathrm{i}\omega_{\rm T}t}+\alpha_n^{(1)*}e^{\mathrm{i}\omega_{\rm T}t}\right),\nonumber \\
 & {} & {\rm Im\,}\Psi_n^{(1)} = \frac{1}{2}\left(\alpha_n^{(2)}e^{-\mathrm{i}\omega_{\rm T}t}+\alpha_n^{(2)*}e^{\mathrm{i}\omega_{\rm T}t}\right),\nonumber \\
 & {} & {\rm Re\,}\Psi_n^{(2)} = \frac{1}{2}\left(\beta_n^{(1)}e^{-\mathrm{i}\omega_{\rm T}t}+\beta_n^{(1)*}e^{\mathrm{i}\omega_{\rm T}t}\right),\nonumber \\
 & {} & {\rm Im\,}\Psi_n^{(2)} = \frac{1}{2}\left(\beta_n^{(2)}e^{-\mathrm{i}\omega_{\rm T}t}+\beta_n^{(2)*}e^{\mathrm{i}\omega_{\rm T}t}\right).
\end{eqnarray}
We further truncate the equations of motion to the fundamental harmonics to find 
\begin{eqnarray}\label{MS40}
 & {} & 
 (\epsilon_0 I + \omega_{\rm T} \sigma_y) \alpha_n+\nonumber \\
 & {} & \left(\begin{array}{c}
c_1(\sqrt{3}\beta_n^{(1)}/2)\beta_n^{(1)}+c_2(\sqrt{3}\beta_{n-1}^{(1)}/2)\beta_{n-1}^{(1)}\\
c_1(\sqrt{3}\beta_n^{(2)}/2) \beta_n^{(2)}+c_2(\sqrt{3}\beta_{n-1}^{(2)}/2) \beta_{n-1}^{(2)}
\end{array}\right) = 0,\nonumber \\
 & {} & 
 (\epsilon_0 I + \omega_{\rm T} \sigma_y) \beta_n+\nonumber \\
 & {} & 
\left(\begin{array}{c}c_1(\sqrt{3}\alpha_n^{(1)}/2)\alpha_n^{(1)}+c_2(\sqrt{3}\alpha_{n+1}^{(1)}/2)\alpha_{n+1}^{(1)}\\
c_1(\sqrt{3}\alpha_n^{(2)}/2) \alpha_n^{(2)}+c_2(\sqrt{3}\alpha_{n+1}^{(2)}/2) \alpha_{n+1}^{(2)}\end{array}\right)
 = 0,\qquad
\end{eqnarray}
where $\beta_0 = 0$, and $c_i(x) = c_i + d_i |x|^2$, $i = 1,2$. We solve Eqs.(\ref{MS40}) by exploiting the approximation $\alpha_n\gg \beta_n$. By doing so, we obtain $\omega_{\rm T} = \epsilon_0$, $\alpha_n^{(1)} = \mathrm{i}\alpha_n^{(2)}$, $\arg\alpha_{n}^{(1)} = \arg \alpha_{1}^{(1)}+(n-1)\pi$, and 
\begin{eqnarray}\label{MS35}
c_1(\sqrt{3}\alpha_n^{(j)}/2)\alpha_n^{(j)}
+c_2(\sqrt{3}\alpha_{n+1}^{(j)}/2)\alpha_{n+1}^{(j)}=0
\end{eqnarray}
for $j=1,2$, which in turn demands that 
\begin{eqnarray}\label{MS35.1}
c_1(\sqrt{3}\psi_{1,n}^{(1)}/2)|\psi_{1,n}^{(1)}|
=c_2(\sqrt{3}\psi_{1,n+1}^{(1)}/2)|\psi_{1,n+1}^{(1)}|.\qquad
\end{eqnarray}
Consequently, the analytic waveform of nonlinear topological edge mode is approximately solved as $\Psi_n \approx (\psi_{1,n}^{(1)}, 0)^\top e^{-\mathrm{i}\epsilon_0 t}$. Let us denote $a_c = \sqrt{-(c_1-c_2)/(d_1-d_2)}$. 
If $|\psi_{1,1}^{(1)}|>\sqrt{4/3}a_c \approx A_c$, the mode keeps increasing and there is no topological edge mode, whereas for $|\psi_{1,1}^{(1)}|<\sqrt{4/3}a_c \approx A_c$, a topological evanescent mode fades away from the boundary. On the other hand, Berry phase of nonlinear bulk modes changes at the critical amplitude $A_c$. Above this critical amplitude, Berry phase $\gamma(A>A_c)=0$. Below the transition point, Berry phase $\gamma(A<A_c)=\pi$. The relationship between the emergence of topological edge modes and Berry phase is the manifestation of the nonlinear extension of bulk-boundary correspondence.

\widetext

\renewcommand{\thefigure}{D3}
\begin{figure}[htb]
\includegraphics[scale=0.4]{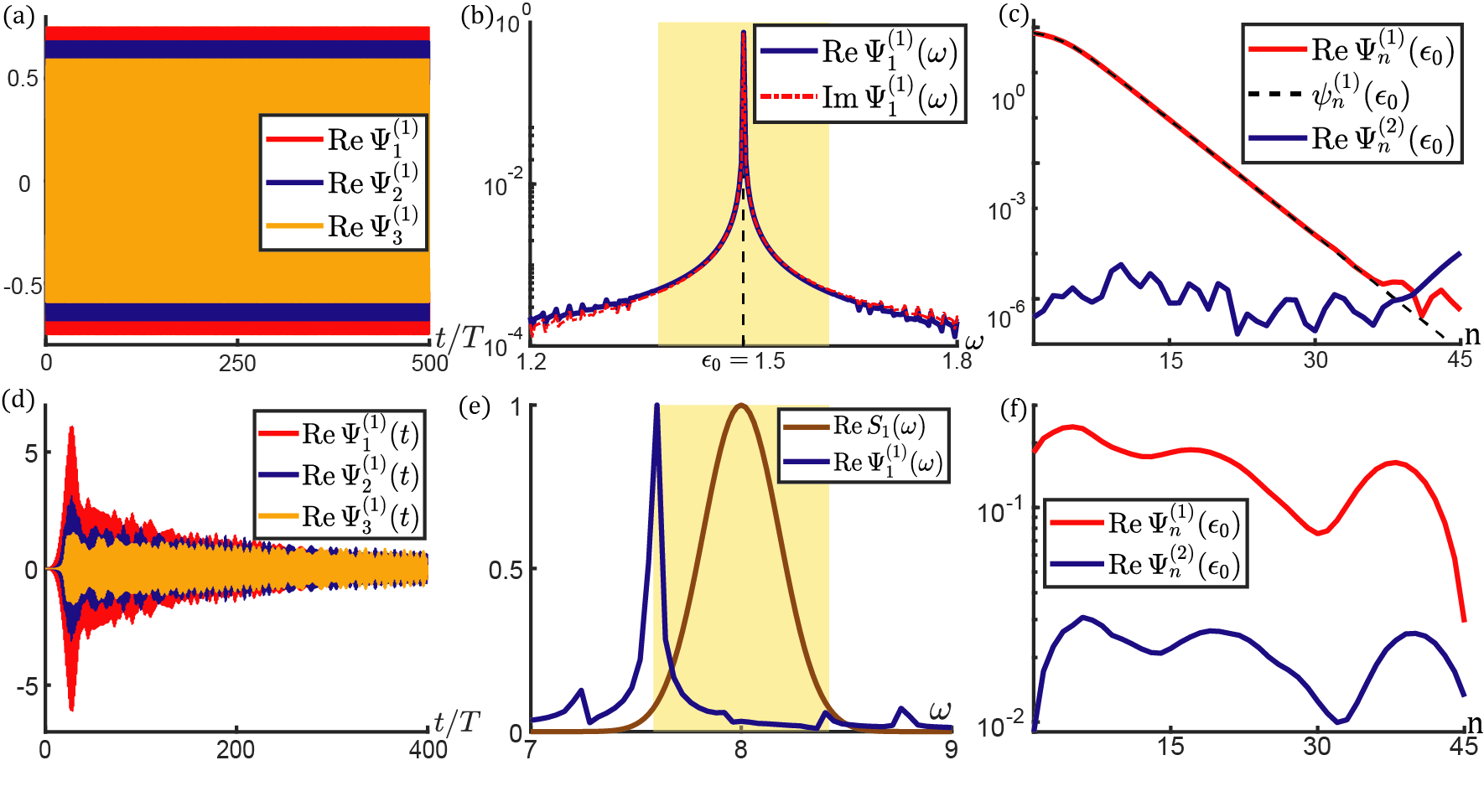}
\caption{(a-c) Stability analysis of nonlinear topological evanescent modes by performing the algorithm of self-oscillation in an undamped, undriven lattice. The lattice is constructed from $N=45$ unit cells subjected to OBC on both ends to mimic a semi-infinite lattice. The parameters of interactions are carried over from Fig.\ref{fig1} of the main text, namely $\epsilon_0=1.5$, $c_1=0.25$, $c_2=0.37$, $d_1=0.22$, and $d_2=0.02$. (a) A nonlinear topological edge mode with amplitude ${\rm Re\,}\Psi_1^{(1)} = 0.75 < A_c$. The mode is initialized by its analytic approximating form $\Psi_n \approx (\psi_{1,n}^{(1)}, 0)^\top e^{-\mathrm{i}\epsilon_0 t}$ derived from Eq.(\ref{MS35.1}), and is truncated in the finite lattice. The mode is allowed to self-oscillate in the lattice for more than $500T$, where $T=2\pi/\epsilon_0$ is the theoretical prediction of the period. (b) Fourier analysis of the topological mode in frequency space, where the peak is in perfect agreement with $\omega_{\rm T} = \epsilon_0$, our theoretical anticipation of the mode frequency. The yellow shaded area is the linear band structure $|\epsilon_0+c_1-c_2|<\omega<|\epsilon_0-c_1+c_2|$. (c) Red and blue curves stand for the spatial profile of the peaks at $\omega = \epsilon_0$ of the Fourier components of the unit cells. Black dashed line is the analytic approximating solution $\Psi_n \approx (\psi_{1,n}^{(1)}, 0)^\top e^{-\mathrm{i}\epsilon_0 t}$ derived from Eq.(\ref{MS35.1}). (d-f) Nonlinear response of the open boundary of a semi-infinite lattice, where reflection symmetry is broken by replacing $\epsilon_0$ with $\epsilon_A = (1+5\%)\epsilon_0$ on A-sites and $\epsilon_B = (1-5\%)\epsilon_0$ on B-sites, respectively. (d) The boundary manifest bulk-mode excitations in response to external Gaussian shaking signal. (e) Frequency spectrum of boundary response is composed of bulk mode components and is remarkably different from (b). (f) The spatial profile of the $\omega = \epsilon_0$ components is in strong contrast to (c). 
}\label{SIfig2}
\end{figure}

\endwidetext

We now present the numerical details of exciting nonlinear topological edge modes, given the parameters $\epsilon_0=1.5$, $c_1=0.25$, $c_2=0.37$, and $d_1 = 0.22$, $d_2 = 0.02$. We construct a lattice subjected to OBCs on both ends. The lattice consists of $N=45$ unit cells to mimic a semi-infinite lattice. According to our theory, the lattice is in the topological phase when the bulk wave amplitude $A<A_c\approx \sqrt{4/3}a_c$. Bulk-boundary correspondence demands that an evanescent mode should appear on lattice boundary, if the edge mode amplitude $\max({\rm Re}\,\Psi_1^{(1)})<A_c$. Theoretical analysis indicates that the spatial profile of this edge mode shall obey Eq.(\ref{MS35.1}). We now attempt to numerically verify this result by exciting a topological edge mode with amplitude $A<A_c$. To this end, a Gaussian tone burst 
\begin{eqnarray}\label{E1.4}
S_n =\delta_{n1} S e^{-\mathrm{i}\omega_{\rm ext} t-(t-t_0)^2/\tau^2} (1, 0)^\top
\end{eqnarray}
is applied on the open boundary at site $n=1$, where the driving amplitude $S = 7\times 10^{-2}$, the carrier frequency $\omega_{\rm ext} = \epsilon_0$, the mode period $T=2\pi/\omega_{\rm ext}$, the half height width $\tau = 3T$, and $t_0 = 15T$. In order to confirm the steady-state conditions, we wait $5000T$ before making any wave function measurements. We compute the frequency spectrum ${\rm Re\,}\Psi_1^{(1)}(\omega)$ by performing fast Fourier transformation (FFT) for the time interval $t\in [10, 5000] T$ in Fig.\ref{fig2}(d). In Fig.\ref{fig2}(e), we plot the spatial profile of the amplitude of the boundary excitation, $\max({\rm Re\,}\Psi_n^{(1,2)}(t))$.  In Fig.\ref{fig2}(f), we plot the spatial profile of the Fourier component ${\rm Re\,}\Psi_n^{(1,2)}(\omega=\epsilon_0)$. The curves are in perfect agreement with the theoretical predictions of nonlinear topological mode $\Psi_n \approx (\psi_{1,n}^{(1)}, 0)^\top e^{-\mathrm{i}\epsilon_0 t}$, where $\psi_{1,n}^{(1)}$ are computed by Eq.(\ref{MS35.1}).

The stability analysis of nonlinear topological edge modes is similar to what has been done in nonlinear bulk modes. We construct a lattice that is composed of $N=45$ unit cells and is subjected to the OBCs on both ends, to mimic a semi-infinite lattice. We initialize the topological mode by employing the analytic approximating solution $\Psi_n \approx (\psi_{1,n}^{(1)}, 0)^\top e^{-\mathrm{i}\epsilon_0 t}$ derived from Eq.(\ref{MS35.1}). After more than 10 periods of self-oscillation in the undamped, undriven lattice, we perform Fourier analysis to characterize if the mode has fallen apart to other nonlinear modes. As shown in Fig.\ref{SIfig2}, the mode remains intact for more than $500T$, which demonstrates mode stability. What is more, all features of this nonlinear topological mode, including the frequency and the spatial profile of mode amplitude, are in perfect alignment with the approximated theoretical solution of Eq.(\ref{MS35.1}).

According to our theory, nonlinear topological modes do not exist if $\max({\rm Re\,}\Psi_1^{(1)})>A_c$ in the T-to-N transition (topological-to-non-topological transition). We numerically verify this by driving the lattice boundary with a Gaussian tone burst (Eq.(\ref{E1.4})), where the stimulation amplitude is $S = 53\times 10^{-2}$. As shown in figs.\ref{fig2}(f), the amplitude of the responding signal is nearly the same for all sites, and the frequency spectrum comprises bulk modes. We note that due to the large amplitude of excitation, the responding nonlinear mode quickly shows instability~\cite{RevModPhys.63.991} and falls apart to other nonlinear modes. To have a stable responding signal, we introduce small damping $\eta = 10^{-3}$ for this large-amplitude driven case. Damping is ubiquitous in dissipative classical systems (see Eqs.(\ref{G2.10}) for example). 

\widetext 

\renewcommand{\thefigure}{D4}
\begin{figure}[htb]
\includegraphics[scale=0.55]{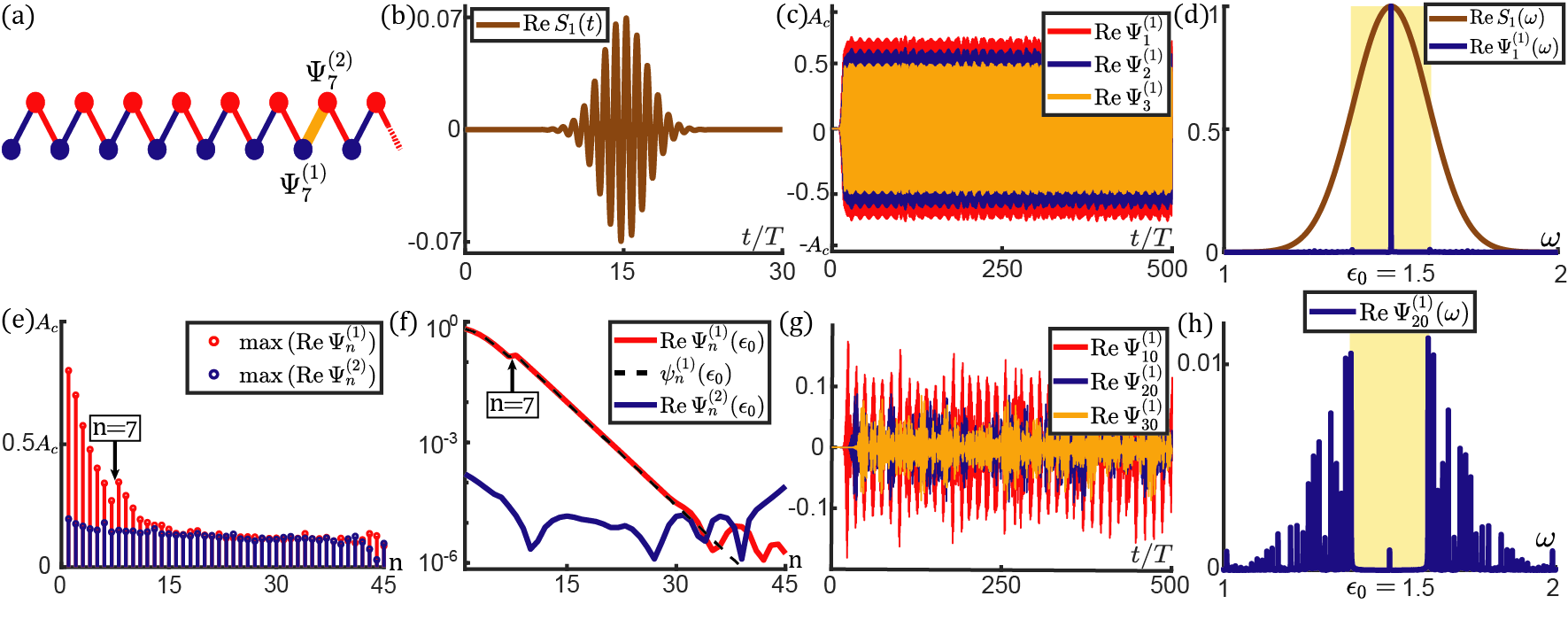}
\caption{Exciting topological edge modes in nonlinear SSH lattice with a defect. The interaction parameters $\epsilon_0, c_1, c_2, d_1$ and $d_2$ are the same as Fig.\ref{fig1}. (a) We construct a long chain that consists of $N=45$ unit cells and is subjected to the OBCs on both ends to mimic a semi-infinite lattice. Red and blue bonds stand for nonlinear interactions between nearest neighbors. The defect is introduced by replacing the blue bond with an orange one connecting $\Psi_7^{(1)}$ and $\Psi_7^{(2)}$, where the interaction parameters are replaced by $c_1'=0.4$ and $d_1'=0.15$. (b) A Gaussian tone burst is employed on the first site to excite topological edge mode, where all parameters of this driving signal are carried over from Fig.\ref{fig2}(b). (c) Wave functions of $n=1,2,3$ sites exhibit the localization of topological mode, where the amplitude $\max({\rm Re\,}\Psi_1^{(1)})<A_c$. (d) Brown and blue curves represent the frequency profiles of Gaussian shaking and responding mode of site $n=1$, respectively. Yellow shaded area is the linear bandgap. Despite the defect, the frequency of topological mode is still $\omega_{\rm T}=\epsilon_0=1.5$. (e) The spatial profile of mode amplitude captures a noticeable jump at site $n=7$ which stems from the defect. (f) Red and blue curves are the spatial profiles of the $\omega=\epsilon_0$ wave component, where the noticeable jump is presented in the ${\rm Re\,}\Psi_n^{(1)}(\epsilon_0)$ curve at the $7$th site. The analytic prediction of the topological mode $\psi_n^{(1)}(\epsilon_0)$ is described by the black dashed line, which is in perfect agreement with numerical results. (g) The wave functions of $n=10, 20, 30$ sites exhibit echo-like shapes indicating multiple reflections at the boundaries, which in turn show the bulk mode excitations. These bulk mode components are excited by the input Gaussian tone burst in (b) which contains all frequencies. (h) The frequency spectrum indicates that the mode at site $n=20$ is mainly composed of bulk modes. 
}\label{SIfig4}
\end{figure}

\endwidetext

In figs.\ref{SIfig2}(d-f), we study the nonlinear boundary response of the semi-infinite lattice, where reflection symmetry is broken by replacing the on-site potentials $\epsilon_0$ with $\epsilon_A = (1+\delta)\epsilon_0$ on A-sites and $\epsilon_B = (1-\delta)\epsilon_0$ on B-sites, respectively. We drive the lattice with the same external Gaussian shaking presented in Eq.\ref{E1.4}. Different from the reflection-symmetric models, the aforementioned symmetry-protected topological boundary modes quickly disappear due to the violation of reflection symmetry that quantizes Berry phase.

Fig.\ref{SIfig4} studies nonlinear topological edge modes in a lattice where the bond connecting $\Psi_7^{(1)}$ and $\Psi_7^{(2)}$ is replaced by the interaction $f_1'(\Psi_7^{(1)}, \Psi_7^{(2)}) = c_1' \Psi_7^{(2)} + d_1' [({\rm Re\,}\Psi_7^{(2)})^3+\mathrm{i}({\rm Im\,}\Psi_7^{(2)})^3]$. The topologically protected boundary mode is insensitive to the defect, in the sense that there is no change of frequency (i.e., $\omega_{\rm T} = \epsilon_0$), and the excitation is still robust.

\widetext

\renewcommand{\thefigure}{D5}
\begin{figure}[htb]
\includegraphics[scale=0.55]{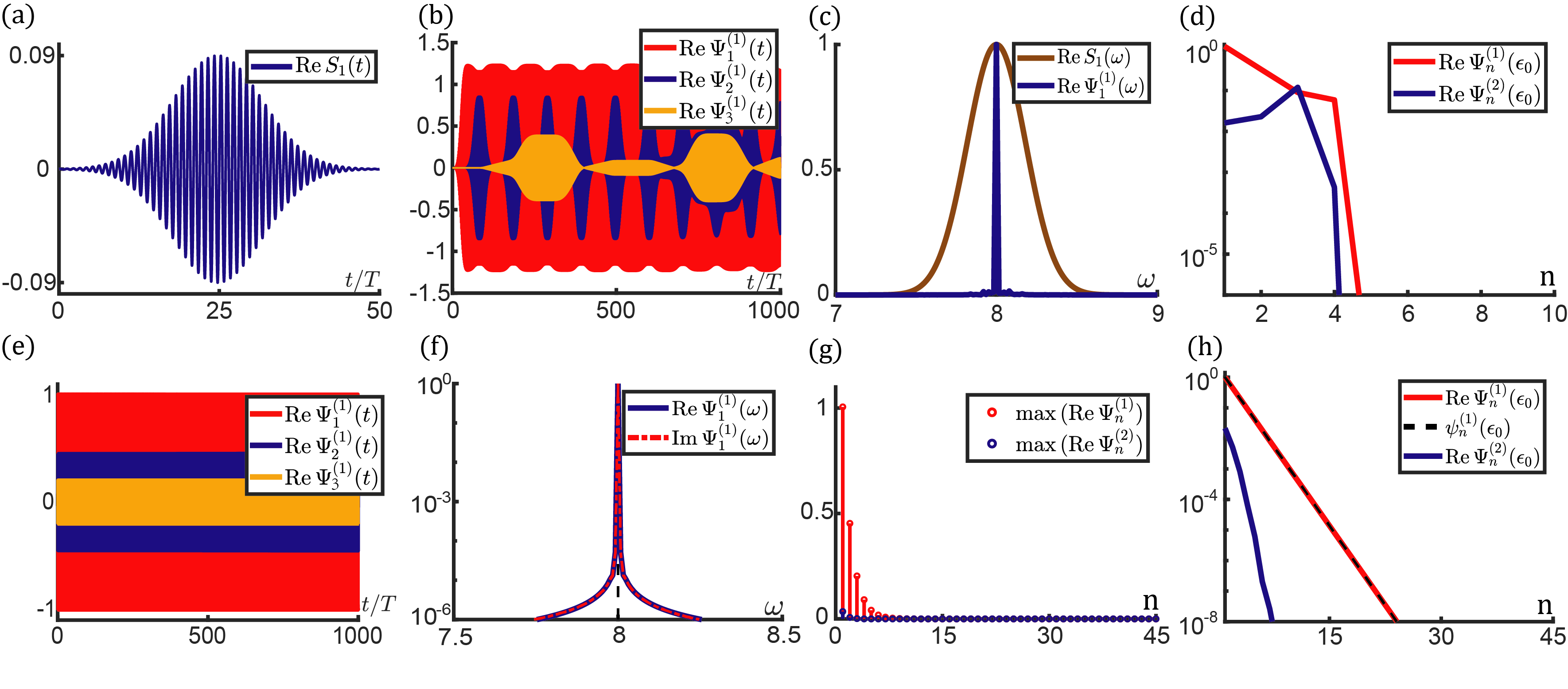}
\caption{Topological evanescent modes in the ``purely nonlinear" model, where $\epsilon_0=8$, $c_1=c_2=0$, $d_1=0.02$ and $d_2=0.22$. A chain composed of $N=45$ unit cells is considered, where OBCs are adopted on both ends. (a) A Gaussian shaking signal is applied on the $n=1$ site to excite nonlinear topological modes, where $S=9\times 10^{-2}$, $\omega_{\rm ext} = \epsilon_0=8$, $T=2\pi/\omega_{\rm ext}$, $\tau = 3T$, and $t_0 = 15T$. (b) Responding mode on $n=1,2,3$ sites exhibits mode localization. (c) Brown and blue curves stand for the frequency spectra of external Gaussian signal and the responding wave function of the $n=1$ site, respectively. The mode frequency is in in perfect alignment with theoretical predictions. (d) Red and blue curves are the spatial profiles of the $\omega=\epsilon_0$ Fourier component of the boundary mode, which manifest the evanescent nature of topological modes. (e) We execute the stability analysis by initializing the mode via Eq.(\ref{MS50.2}) with $\psi_1^{(1)}=1$, and impose an additional perturbation by multiplying a random factor $1+\xi_n$ ($\xi_n\le 10^{-2}$) on the wave function of each site $n$. We let the mode to self-oscillate in an undamped, undriven lattice for more than $1000$ periods before measuring the wave functions. The mode remains intact without generating other components, which demonstrates the mode stability. (f) Frequency profile of the topological mode. (g) Spatial profile of the amplitude of this mode. (h) Spatial profile of the $\omega=\epsilon_0$ mode component is captured by red and blue curves. Theoretical analysis is depicted by the black dashed curve, which is in perfect agreement with numerical computations. 
}\label{SIfig10}
\end{figure}

\endwidetext

\subsection{Analytic solution of topological edge modes in the ``purely nonlinear" model}

An analytic solution of topological edge modes can be carried out when the linear components of interactions vanish, i.e., $c_1=c_2=0$. The topological edge mode is expressed as $\Psi_n = (\psi_n^{(1)}, \psi_n^{(2)})^\top e^{-\mathrm{i}\epsilon_0 t}$, where the mode amplitudes $\psi_n^{(1)}$ and $\psi_n^{(2)}$ satisfy the recursion relations, 
\begin{eqnarray}\label{MS50.1}
{\psi_{n+1}^{(1)}}/{\psi_n^{(1)}} = {\psi_{n}^{(2)}}/{\psi_{n+1}^{(2)}} = \left(-{d_1}/{d_2}\right)^{1/3}.
\end{eqnarray}
Given that $0<d_1<d_2$, the lattice is topologically nontrivial for all amplitudes because Berry phase always takes the non-trivial value, $\gamma(A)\equiv \pi$. As a result, an evanescent mode exponentially localizes on the open boundary, where the waveform is given by 
\begin{eqnarray}\label{MS50.2}
\Psi_n =(\psi_1^{(1)} ,0)^\top \left(-{d_1}/{d_2}\right)^{(n-1)/3}  e^{-\mathrm{i}\epsilon_0 t}.
\end{eqnarray}
In figs.\ref{SIfig10}(a--d), we study the nonlinear topological evanescent mode by driving an undamped lattice with a Gaussian shaking signal elaborated in Eq.(\ref{E1.4}). In figs.\ref{SIfig10}(e--h), we perform the algorithm of self-oscillation for the stability analysis. Both numerical methods suggest that the topological mode is stable in the ``purely nonlinear regime" where the linear parts of hopping terms vanish.

\subsection{Exact solution of static nonlinear topological edge modes for the $\epsilon_0=0$ case}
In contrast to the $\epsilon_0\neq 0$ model, this $\epsilon_0=0$ model features two qualitatively different properties.

The first property lies in the lattice under PBC. At the critical amplitude $a_c$, the nonlinear bands merge at zero-frequency. When the mode amplitude goes beyond this critical amplitude, the lattice experiences instability to reach new ground states. There are eight new ground states described by the equilibrium wave functions, 
\begin{eqnarray}\label{MS50}
\bar{\Psi}_n = (-1)^n \sqrt{2} a_c(e^{\mathrm{i}s_1\pi/4}, s_2 e^{\mathrm{i}s_3\pi/4})^\top,
\end{eqnarray}
where $s_1, s_2, s_3 = \pm 1$. Without loss of generality, we pick one of the eight equilibrium ground states, $\bar{\Psi}_n = (-1)^n e^{\mathrm{i}\pi/4} \sqrt{2} a_c(1, 1)^\top$, to study small fluctuations $\delta\Psi_n = \Psi_n - \bar{\Psi}_n$ around it. By expanding the equations to the linear order in $\delta\Psi_n$, we obtain 
\begin{eqnarray}\label{MS39}
H_q\delta \Psi_q = \mathrm{i}\partial_t \delta\Psi_q,
\end{eqnarray}
where $\delta\Psi_q = \sum_q \delta \Psi_n e^{-\mathrm{i}qn}$ is the momentum-space wave function, and the new ground state Hamiltonian $H_q$ reads
\begin{eqnarray}\label{MS51}
H_q = [c_1(\sqrt{3}a_c)+c_2(\sqrt{3}a_c)\cos q ]\sigma_x +[c_2(\sqrt{3}a_c)\sin q]\sigma_y. \nonumber \\
\end{eqnarray}

The second peculiar property is that the nonlinear topological edge modes are static in time, which allows for analytic solutions governed by the following nonlinear recursion relations, 
\begin{eqnarray}\label{MS38}
 & {} & c_1({\rm Re\,}\Psi_{n}^{(1)})|{\rm Re\,}\Psi_{n}^{(1)}|
=c_2({\rm Re\,}\Psi_{n+1}^{(1)})|{\rm Re\,}\Psi_{n+1}^{(1)}|, \nonumber \\
 & {} & c_1({\rm Im\,}\Psi_{n}^{(1)})|{\rm Im\,}\Psi_{n}^{(1)}|
=c_2({\rm Im\,}\Psi_{n+1}^{(1)})|{\rm Im\,}\Psi_{n+1}^{(1)}|, \nonumber \\
 & {} & c_1({\rm Re\,}\Psi_{n}^{(2)})|{\rm Re\,}\Psi_{n}^{(2)}|
=c_2({\rm Re\,}\Psi_{n-1}^{(2)})|{\rm Re\,}\Psi_{n-1}^{(2)}|, \nonumber \\
 & {} & c_1({\rm Im\,}\Psi_{n}^{(2)})|{\rm Im\,}\Psi_{n}^{(2)}|
=c_2({\rm Im\,}\Psi_{n-1}^{(2)})|{\rm Im\,}\Psi_{n-1}^{(2)}|, 
\qquad
\end{eqnarray}
subjected to the OBC $\Psi_0^{(2)}=0$. 

\renewcommand{\thefigure}{D6}
\begin{figure}[htbp]
\includegraphics[scale=0.52]{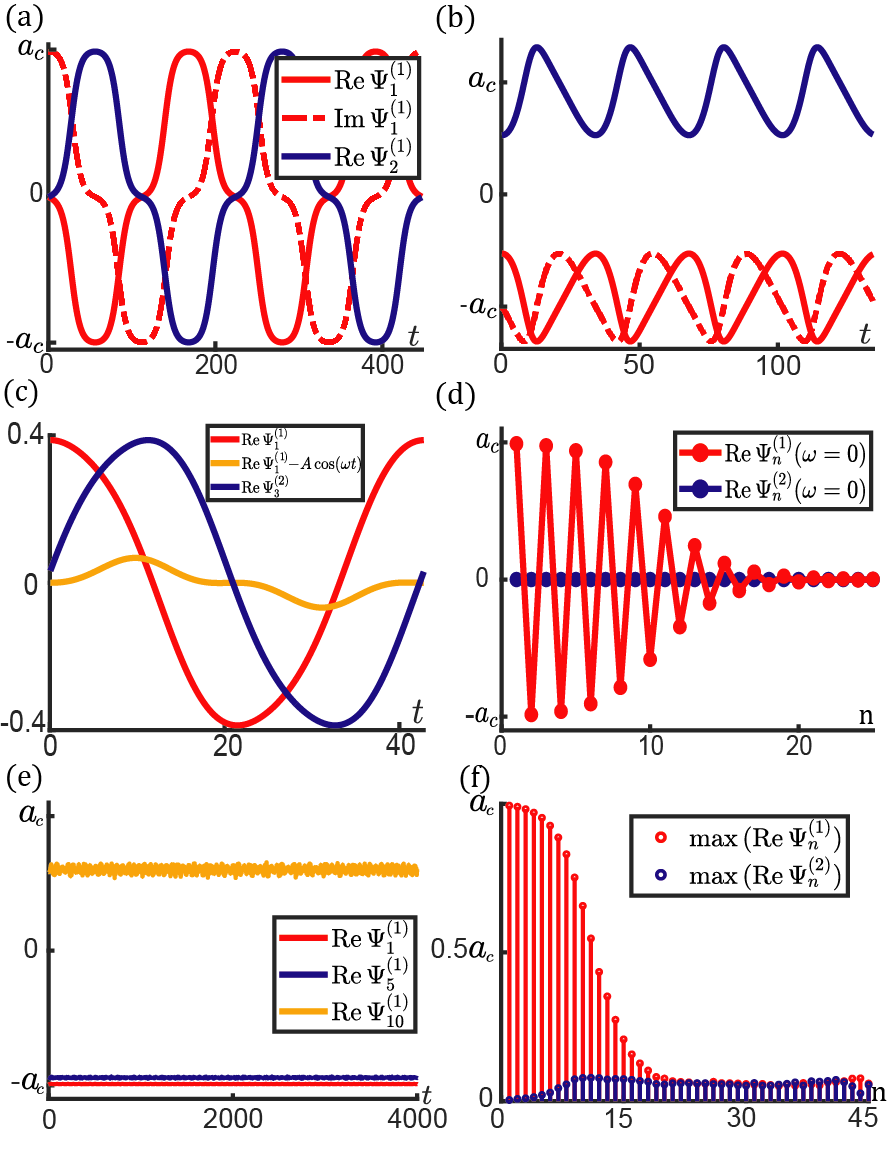}
\caption{We now consider the $\epsilon_0=0$ model, while all other parameters keep the same as Fig.\ref{fig1}. (a) A nonlinear bulk mode on the verge of instability, where the amplitude $A=0.9877 a_c\lesssim a_c$. (b) A nonlinear bulk mode with $\max{|{\rm Re\,}\Psi_n^{(1)}|}=1.312 a_c>a_c$ experiences instability and oscillates around new ground states. (c) A nonlinear bulk mode with $A=0.5a_c$ and $q=8\pi/9$ is obtained via shooting method. Red and blue curves stand for the wave functions of $n=1,3$ sites. Orange curve indicates that the nonlinear mode is noticeably different from sinusoidal function. (d) Spatial profile of static topological mode with amplitude ${\rm Re\,}\Psi_1^{(1)} = 0.99a_c$. (e) Stability analysis of nonlinear topological edge modes. Temporal profile of the perturbed topological mode for the time interval $t\in[0,75\times 2\pi/(c_2-c_1)]$. The mode remains intact without generating other wave components, which demonstrates mode stability. (f) Spatial profile of the amplitude of the mode in (c) on each site. 
}\label{SIfig8}
\end{figure}

Stability analysis of topological modes is elaborated as follows. We construct a lattice with $N=45$ unit cells subjected to the OBCs on both ends, and initialize the mode via the following procedure. We establish the analytic solution of Eqs.(\ref{MS38}) with the amplitude ${\rm Re\,}\Psi_1^{(1)}={\rm Im\,}\Psi_1^{(1)}=0.99a_c$. Next, we perturb the aforementioned mode by multiplying a random factor $1+\xi_n$ ($\xi_n\le 10^{-2}$) on the wave function of each site $n$. Finally, we let the initialized mode to self-oscillate in the undamped, undriven lattice. We wait $t=75\times 2\pi/(c_2-c_1)$ before making any wave function measurements. As shown in Fig.\ref{SIfig8}, the mode remains intact without producing other wave components, which demonstrates the stability of nonlinear topological edge modes.

\section{Deriving generalized nonlinear Schr\"{o}dinger equations for classical models}

In this section, we derive the nonlinear equations of motion for classical systems that exhibit topological properties, and realize the minimal model of nonlinear interactions in Eq.(\ref{9v2}). Both passive and active structures are discussed here.

\subsection{Topological photonics (passive system)}

Extended from Refs.~\cite{d2008ultraslow, Christodoulides:88}, the first model is a nonlinear optic metamaterial that serves as a passive system to emerge topological edge modes. As shown in Fig.\ref{expProposal}, it is a 1D array of waveguides to propagate electro-magnetic waves along the axial $z$-direction without backscattering. The unit cell comprises two waveguides to host electro-magnetic fields $\vec E_n^{(j)} = \sum_{k=1,2}\hat{e}_k {E}_{k,n}^{(j)}(z,t)$ and $\vec {H}_n^{(j)} = \sum_{k=1,2}\hat{e}_k {H}_{k,n}^{(j)}$ for $j=1,2$, where $\hat{e}_1$ and $\hat{e}_2$ represent the unit vectors of $x$ and $y$ directions, respectively, and $E_{1,n}^{(j)}$, $E_{2,n}^{(j)}$ ($H_{1,n}^{(j)}$, $H_{2,n}^{(j)}$) are the projections of the fields. We now show that the motion equations for the field variables are 4-field generalized nonlinear Schr\"{o}dinger equations. We demonstrate the 4-field extension of quantized Berry phase due to reflection symmetry, and indicate the physical realization of the minimal nonlinear interaction in this passive system.

Maxwell equations demand the field variables to obey $\nabla\times \vec {E} = -\partial_t \vec {B}$ and $\nabla\times \vec {H} = \partial_t \vec {D}$, which in turn are converted to $\partial_z {E}_1 = -\partial_t {B}_2$, $\partial_z {E}_2 = \partial_t {B}_1$ and $\partial_z {H}_1 = \partial_t {D}_2$, $-\partial_z {H}_2 = \partial_t {D}_1$. For the $j$th waveguide of the $n$th unit cell, the motions of electro-magnetic fields are
\begin{eqnarray}\label{G2.52}
\partial_z {E}_{k,n}^{(j)} & = & (-1)^k \sum_{n'} \sum_{j'=1,2} \partial_t {B}_{k'}( r_{nn'}^{(jj')}; \vec {H}_{n'}^{(j')}),\nonumber \\
-\partial_z {H}_{k,n}^{(j)} & = & (-1)^{k} \sum_{n'} \sum_{j'=1,2} \partial_t {D}_{k'}( r_{nn'}^{(jj')}; \vec {E}_{n'}^{(j')}),
\end{eqnarray}
where $k'\neq k$ represent the $x$ ($k=1$) and $y$ ($k=2$) components of the field variables. $ r_{nn'}^{(jj')} =| \vec r_{n'}^{(j')}-\vec r_n^{(j)}|$ is the distance between waveguides, $\vec {D}( r_{nn'}^{(jj')}; \vec {E}_{n'}^{(j')})$ is the electric displacement on the target waveguide induced by the $j'$th waveguide of the $n'$th unit cell, and $\vec {B}( r_{nn'}^{(jj')}; \vec {H}_{n'}^{(j')})$ is the induced magnetic field on the target. It is worth of emphasizing that as long as the geometric structure of the 1D array yields reflection symmetry, Eqs.(\ref{G2.52}) respect reflection symmetry and therefore potentially host quantized Berry phase. Here, reflection symmetry means that the equations of motion are invariant under the change of variables, 
\begin{eqnarray}\label{G2.56}
(\vec {E}_n^{(1)}, \vec {H}_n^{(1)}, \vec {E}_n^{(2)}, \vec {H}_n^{(2)}) \to (\vec {E}_{-n}^{(2)}, \vec {H}_{-n}^{(2)}, \vec {E}_{-n}^{(1)}, \vec {H}_{-n}^{(1)}).\qquad
\end{eqnarray}

\widetext

\noindent We simply Eqs.(\ref{G2.52}) by considering nearest neighbor interactions only, 
\begin{eqnarray}\label{G2.53}
(-1)^k\partial_z {E}_{k,n}^{(j)} & = &  \partial_t \left[{B}_{k'}(\vec{ H}_{n}^{(j)}) + {B}_{k'}( r_{nn}^{(jj')},\vec {H}_n^{(j')}) + {B}_{k'}( r_{n,n+(-1)^j}^{(jj')},\vec {H}_{n+(-1)^j}^{(j')})\right],\nonumber \\
-(-1)^{k}\partial_z {H}_{k,n}^{(j)} 
 & = &  \partial_t \bigg[ {D}_{k'}(\vec {E}_{n}^{(j)})+ {D}_{k'}( r_{nn}^{(jj')}, \vec {E}_n^{(j')}) + {D}_{k'}( r_{n,n+(-1)^j}^{(jj')},\vec {E}_{n+(-1)^j}^{(j')})\bigg],
 \end{eqnarray}
where $j\neq j'=1,2$ labels the waveguides within a unit cell, and $k\neq k'=1,2$ denotes the $x$ and $y$ field components. Next, we write the fields as the product of an envelope function $\hat{E}_n^{(j)} = \sum_{k=1,2}\hat{e}_k \hat{E}_{k,n}^{(j)}$ and $\hat{H}_n^{(j)} = \sum_{k=1,2}\hat{e}_k \hat{H}_{k,n}^{(j)}$ multiplied by a harmonic oscillation at frequency $\omega_0$:
\begin{eqnarray}\label{G2.54}
\vec {E}_n^{(j)}(z,t) = \frac{1}{2} \left[\hat{E}_n^{(j)}(z,t)e^{-\mathrm{i}\omega_0 t}+{\rm c.c.}\right],
\qquad\qquad \vec {H}_n^{(j)}(z,t) = \frac{1}{2} \left[\hat{H}_n^{(j)}(z,t)e^{-\mathrm{i}\omega_0 t}+{\rm c.c.}\right].
\end{eqnarray}
In the hypothesis that the time modulation of the fields are mostly captured by the carrier frequency $\omega_0$, and the envelope slowly varies in time, we adopt the following approximations by assuming $\partial_t \hat{E}_n^{(j)}\ll \omega_0 \hat{E}_n^{(j)}$ and $\partial_t \hat{H}_n^{(j)}\ll \omega_0 \hat{H}_n^{(j)}$,
\begin{eqnarray}\label{G2.57}
 & {} & \partial_t \vec {D}(\vec {E}_{n}^{(j')}) \approx -\frac{\mathrm{i}}{2}e^{-\mathrm{i}\omega_0 t} \omega_0 \hat{D}_0(\hat{E}_n^{(j')})+{\rm c.c.}, \qquad\qquad \qquad\quad\,
 \partial_t \vec {B}(\vec {H}_{n}^{(j')}) \approx -\frac{\mathrm{i}}{2}e^{-\mathrm{i}\omega_0 t} \omega_0 \hat{B}_0(\hat{H}_n^{(j')})+{\rm c.c.} ,\nonumber \\
 & {} & \partial_t \vec {D}(r_{nn}^{(j\neq j')},\vec {E}_{n}^{(j')}) \approx -\frac{\mathrm{i}}{2}e^{-\mathrm{i}\omega_0 t} \omega_0 \hat{D}_1(\hat{E}_n^{(j')})+{\rm c.c.}, \qquad\qquad
 \partial_t \vec {B}(r_{nn}^{(j\neq j')},\vec {H}_n^{(j')}) \approx -\frac{\mathrm{i}}{2}e^{-\mathrm{i}\omega_0 t} \omega_0 \hat{B}_1(\hat{H}_n^{(j')})+{\rm c.c.} ,\nonumber \\
 & {} & \partial_t \vec {D}(r_{n,n+(-1)^j}^{(j\neq j')},\vec {E}_{n+(-1)^j}^{(j')}) \approx -\frac{\mathrm{i}}{2}e^{-\mathrm{i}\omega_0 t} \omega_0 \hat{D}_2(\hat{E}_{n+(-1)^j}^{(j')})+{\rm c.c.}, \nonumber \\
 & {} & \partial_t \vec {B}(r_{n,n+(-1)^j}^{(j\neq j')},\vec {H}_{n+(-1)^j}^{(j')}) \approx -\frac{\mathrm{i}}{2}e^{-\mathrm{i}\omega_0 t} \omega_0 \hat{B}_2(\hat{H}_{n+(-1)^j}^{(j')})+{\rm c.c.}, 
\end{eqnarray}
where $\hat{D}_i= \sum_{k=1,2}\hat{e}_k \hat{D}_{k,i}$ and $\hat{B}_i= \sum_{k=1,2}\hat{e}_k \hat{B}_{k,i}$ ($i=0,1,2$) are the envelope functions of electric displacement and magnetic fields, respectively. Here, these nonlinear functions have taken the distance $r_{nn'}^{(jj')}$ and the shapes of waveguides into consideration. The equations of motion are now reduced to 
\begin{eqnarray}\label{G2.55}
(-1)^k \mathrm{i}\omega_0^{-1}\partial_z \hat{E}_{k,n}^{(j)} & = & \hat{B}_{k',0}(\hat{H}_n^{(j)})+\hat{B}_{k',1}(\hat{H}_n^{(j')})+\hat{B}_{k',2}(\hat{H}_{n+(-1)^j}^{(j')}),\nonumber \\
-(-1)^k\mathrm{i}\omega_0^{-1}\partial_z \hat{H}_{k,n}^{(j)} & = & \hat{D}_{k',0}(\hat{E}_n^{(j)})+\hat{D}_{k',1}(\hat{E}_n^{(j')})+\hat{D}_{k',2}(\hat{E}_{n+(-1)^j}^{(j')}).
 \end{eqnarray}
In linear regime, the electric displacement and magnetic fields are simply given by $\hat{D}_0(\hat{E})=\epsilon_0 \hat{E}$ and $\hat{B}_0 (\hat{H}) = \mu_0 \hat{H}$, where $\epsilon_0$ and $\mu_0$ are linear permittivity and permeability of the waveguide, respectively. Thus, we introduce the constant $\alpha = (\mu_0/\epsilon_0)^{1/2}$ to construct new field variables and nonlinear functions as follows, 
\begin{eqnarray}\label{G2.70}
\Psi_n^{(2j-2+k)}=\hat{E}_{k,n}^{(j)}+\alpha \hat{H}_{k',n}^{(j)}
\qquad\qquad
f_i^{(k)}(\Psi_n^{(2j-1)}, \Psi_n^{(2j)}) = \hat{B}_{k,i}(\hat{H}_n^{(j)})+\alpha \hat{D}_{k',i}(\hat{E}_n^{(j)}).
\end{eqnarray}
where $k'\neq k =1,2$. The equations of motion of electro-magnetic fields are converted as 4-field generalized nonlinear Schr\"{o}dinger equations,
\begin{eqnarray}\label{G2.171}
-\mathrm{i}\omega_0^{-1}\partial_z \Psi_n^{(1)} & = &  f_0^{(2)}(\Psi_n^{(1)}, \Psi_n^{(2)})+f_1^{(2)}(\Psi_n^{(3)}, \Psi_n^{(4)})+f_2^{(2)}(\Psi_{n-1}^{(3)}, \Psi_{n-1}^{(4)}),\nonumber \\
\mathrm{i}\omega_0^{-1}\partial_z \Psi_n^{(2)} & = & f_0^{(1)}(\Psi_n^{(1)}, \Psi_n^{(2)})+f_1^{(1)}(\Psi_n^{(3)}, \Psi_n^{(4)})+f_2^{(1)}(\Psi_{n-1}^{(3)}, \Psi_{n-1}^{(4)}),\nonumber \\
-\mathrm{i}\omega_0^{-1}\partial_z \Psi_n^{(3)} & = & f_0^{(2)}(\Psi_n^{(3)}, \Psi_n^{(4)})+f_1^{(2)}(\Psi_n^{(1)}, \Psi_n^{(2)})+f_2^{(2)}(\Psi_{n+1}^{(1)}, \Psi_{n+1}^{(2)}),\nonumber \\
\mathrm{i}\omega_0^{-1}\partial_z \Psi_n^{(4)} & = & f_0^{(1)}(\Psi_n^{(3)}, \Psi_n^{(4)})+f_1^{(1)}(\Psi_n^{(1)}, \Psi_n^{(2)})+f_2^{(1)}(\Psi_{n+1}^{(1)}, \Psi_{n+1}^{(2)}),
 \end{eqnarray}
\endwidetext

\noindent which are invariant under the reflection transformation, 
\begin{eqnarray}\label{G2.57}
(\Psi_n^{(1)}, \Psi_n^{(2)}, \Psi_n^{(3)}, \Psi_n^{(4)}) \to (\Psi_{-n}^{(3)}, \Psi_{-n}^{(4)}, \Psi_{-n}^{(1)}, \Psi_{-n}^{(2)}).\qquad
\end{eqnarray}
Given a nonlinear bulk mode of the form 
\begin{eqnarray}\label{G2.59}
\Psi_q = \left(
\begin{array}{c}
\Psi_q^{(1)}(\omega t -qn)\\
\Psi_q^{(2)}(\omega t -qn+\phi_q^{(2)})\\
\Psi_q^{(3)}(\omega t -qn+\phi_q^{(3)})\\
\Psi_q^{(4)}(\omega t -qn+\phi_q^{(4)})\\
\end{array}
\right),
\end{eqnarray}
we repeat the adiabatic evolution in App.~A to have Berry phase 
\begin{eqnarray}\label{G2.58}
\gamma
=
\oint_{\rm BZ}dq \frac{\sum_{l}\sum_{j} \left( l |\psi_{l,q}^{(j)} |^2 \frac{\partial\phi_q^{(j)}}{\partial q}+\mathrm{i}\psi_{l,q}^{(j)*}\frac{\partial\psi_{l,q}^{(j)}}{\partial q}\right)}
{\sum_{l'}\sum_{j'}  l'  |\psi_{l',q}^{(j')}|^2},\qquad
\end{eqnarray}
where $\sum_j$ is summed over the four field components, and $\phi_q^{(1)}\equiv 0$. Based on Eq.(\ref{G2.57}) and (\ref{G2.59}), reflection symmetry demands a partner solution 
\begin{eqnarray}\label{G2.60}
\Psi_{-q}' = \left(
\begin{array}{c}
\Psi_q^{(3)}(\omega t +qn)\\
\Psi_q^{(4)}(\omega t +qn+\phi_q^{(4)}-\phi_q^{(3)})\\
\Psi_q^{(1)}(\omega t +qn-\phi_q^{(3)})\\
\Psi_q^{(2)}(\omega t +qn+\phi_q^{(2)}-\phi_q^{(3)})\\
\end{array}
\right).
\end{eqnarray}
On the other hand, a nonlinear bulk mode of the wavenumber $-q$ is by definition written as  
\begin{eqnarray}\label{G2.61}
\Psi_{-q} = \left(
\begin{array}{c}
\Psi_{-q}^{(1)}(\omega t +qn)\\
\Psi_{-q}^{(2)}(\omega t +qn+\phi_{-q}^{(2)})\\
\Psi_{-q}^{(3)}(\omega t +qn+\phi_{-q}^{(3)})\\
\Psi_{-q}^{(4)}(\omega t +qn+\phi_{-q}^{(4)})\\
\end{array}
\right).
\end{eqnarray}
Since we assume that there is no degeneracy of nonlinear bulk modes, $\Psi_{-q}$ and $\Psi_{-q}'$ have to be the same solution, which in turn demands
\begin{eqnarray}\label{G2.63}
\phi_{-q}^{(4)}-\phi_{q}^{(2)}=\phi_{-q}^{(3)}, \qquad\qquad \phi_{-q}^{(3)} = -\phi_q^{(3)}
\end{eqnarray}
and
\begin{eqnarray}\label{G2.62}
\Psi_{q}^{(1)} = \Psi_{-q}^{(3)},\qquad\qquad \Psi_{q}^{(2)} = \Psi_{-q}^{(4)}.
\end{eqnarray}
Employing Eqs.(\ref{G2.63}) and (\ref{G2.62}), we demonstrate the quantization of $\gamma$ by separating it into two parts $\gamma_1$ and $\gamma_2$. The first part $\gamma_1$ is given as follows, 
\begin{eqnarray}\label{G2.64}
\gamma_1
 & = & 
\oint_{\rm BZ}dq \frac{\sum_{l}\sum_{j}  l |\psi_{l,q}^{(j)} |^2 \frac{\partial\phi_q^{(j)}}{\partial q}}
{\sum_{l'}\sum_{j'}  l'  |\psi_{l',q}^{(j')}|^2}\nonumber \\
 & = & 
\oint_{\rm BZ}dq \frac{\sum_{l}l\left(|\psi_{l,q}^{(3)} |^2 \frac{\partial\phi_q^{(3)}}{\partial q}-|\psi_{l,q}^{(2)} |^2 \frac{\partial(\phi_{-q}^{(4)}-\phi_q^{(2)})}{\partial q}\right)}
{\sum_{l'}\sum_{j'=2,3}  l'  \left(|\psi_{l',q}^{(j')}|^2+|\psi_{l',-q}^{(j')}|^2\right)}
 \nonumber \\
 & = & 
\frac{1}{2}\oint_{\rm BZ}dq\frac{\partial\phi_q^{(3)}}{\partial q}\nonumber \\
 & = & \phi_\pi^{(3)}-\phi_0^{(3)} = 0\,\,{\rm or\,\,}\pi\mod 2\pi.
\end{eqnarray}
The second part, 
\begin{eqnarray}\label{G2.65}
\gamma_2
=
\oint_{\rm BZ}dq \frac{\mathrm{i} \sum_{l}  \sum_{j=1,2}\left(\psi_{l,q}^{(j)*}\frac{\partial\psi_{l,q}^{(j)}}{\partial q}+\psi_{l,-q}^{(j)*}\frac{\partial\psi_{l,-q}^{(j)}}{\partial q}\right)}
{\sum_{l'}\sum_{j'=1,2}  l' \left( |\psi_{l',q}^{(j')}|^2+|\psi_{l',-q}^{(j')}|^2\right)}=0.\nonumber \\
\end{eqnarray}
Thus, we have proved the quantization of Berry phase in this 4-field generalized nonlinear Schr\"{o}dinger equations, $\gamma = \gamma_1+\gamma_2 = 0$ or $\pi\mod 2\pi$.

Finally, we realize the nonlinear interaction specified in Eq.(\ref{9}) of the main text by considering a linearly polarized incident light. As a result, $\hat{E}_{2,n}^{(j)}=\hat{H}_{1,n}^{(j)}=0$ and $\hat{H}_{2,n}^{(j)}$ is delayed by a phase of $\pi/2$ compared to $\hat{E}_{1,n}^{(j)}$. Thus, it is convenient to re-write $\hat{H}_{2,n}^{(j)}\to \mathrm{i}\hat{H}_{2,n}^{(j)}$ such that both $\hat{E}_{1,n}^{(j)}$ and $\hat{H}_{2,n}^{(j)}$ are real quantities that represent the real and imaginary parts of the field variables, respectively. The induced fields of the inversion-symmetric material are given by 
\begin{eqnarray}\label{G2.73}
 & {} & \hat{D}_{1,i}(\hat{E}_1) = \epsilon_i \hat{E}_1 + \epsilon_i^{(3)} \hat{E}_1^3, \nonumber \\
 & {} & \hat{B}_{2,i}(\hat{H}_2) = \mu_i \hat{H}_2 + \mu_i^{(3)} \hat{H}_2^3.
\end{eqnarray}
We demand the parameters to yield 
\begin{eqnarray}\label{G2.76}
{\mu_i}/{\epsilon_i}=\alpha^2,\qquad {\mu_i^{(3)}}/{\epsilon_i^{(3)}} = -\alpha^4,\qquad{\rm for}\qquad i=0,1,2,\nonumber \\
\end{eqnarray}
which reduces Eqs.(\ref{G2.171}) to 
\begin{eqnarray}\label{G2.75}
-\mathrm{i}(\alpha\omega_0)^{-1}\partial_z \Psi_n^{(1)} & = &  f_0(\Psi_n^{(1)})+f_1(\Psi_n^{(3)})+f_2(\Psi_{n-1}^{(3)}),\nonumber \\
-\mathrm{i}(\alpha\omega_0)^{-1}\partial_z \Psi_n^{(3)} & = & f_0(\Psi_n^{(3)})+f_1(\Psi_n^{(1)})+f_2(\Psi_{n+1}^{(1)}),\nonumber \\
 \end{eqnarray}
where $f_i(y) = \epsilon_i y+  \epsilon_i^{(3)} [({\rm Re\,}y)^3+ \mathrm{i}({\rm Im\,}y)^3]$ for $i=0,1,2$. Finally, in the parameter regime 
\begin{eqnarray}\label{G2.67}
 & {} & \epsilon_0^{(3)} |\Psi_n^{(j=1,3)}|^2/\epsilon_0 \ll 1, \nonumber \\
 & {} & (\epsilon_1^{(3)}-\epsilon_2^{(3)})|\Psi_n^{(j=1,3)}|^2/(\epsilon_1-\epsilon_2)\sim\mathcal{O}(1),
\end{eqnarray}
Eqs.(\ref{G2.75}) can be finally simplified as the minimal model proposed in Eqs.(\ref{1v2}) and (\ref{9v2}).

\subsection{Topoelectrical circuit (active system)}

\renewcommand{\thefigure}{E1}
\begin{figure}[htb]
\includegraphics[scale=0.28]{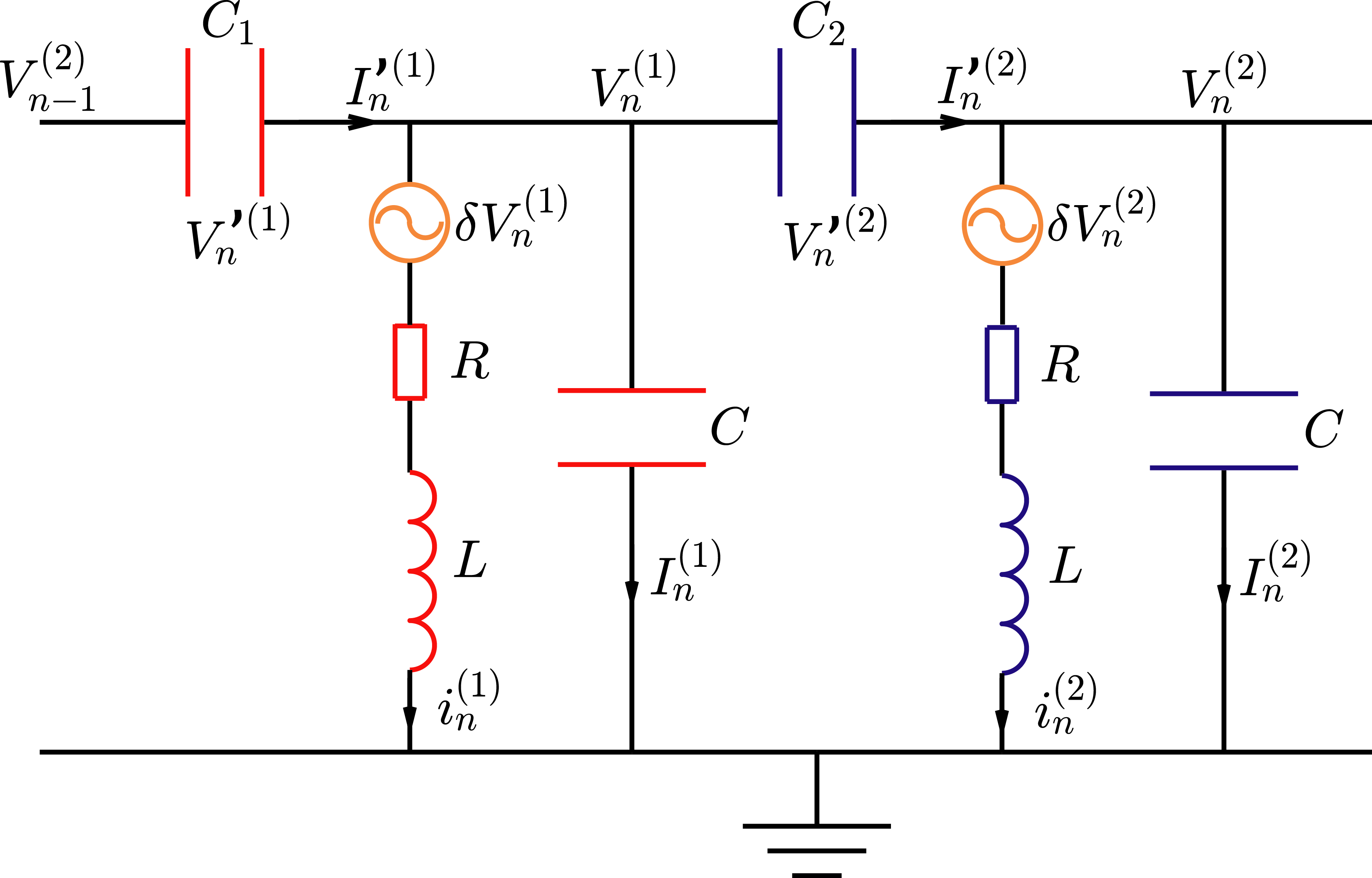}
\caption{The unit cell of nonlinear topoelectric circuit subjected to external active drivings. It is composed of two pairs of LCR resonators of natural frequency $\omega_0$, which are connected by two small capacitors $C_1$ and $C_2$. The inductances are connected to alternating power sources  $\delta V_n^{(1)}(V_n^{(2)}, V_{n-1}^{(2)})$ and $\delta V_n^{(2)}(V_n^{(1)}, V_{n+1}^{(1)})$ controlled by the nearest neighbor voltages, which in turn serve as the nonlinear couplings between dimer voltage fields.   
}\label{eCircuit2}
\end{figure}

Here, we demonstrate that the equations of motion of 1D nonlinear topoelectrical LCR circuit can be converted to generalized nonlinear Schr\"{o}dinger equations with the specific nonlinear interactions in Eq.(\ref{9v2}). As shown in Fig.\ref{eCircuit2}, the unit cell of the ladder circuit is composed of two resonators of natural frequency $\omega_0 = 1/\sqrt{LC}$, where $L$ is the inductance and $C$ is the capacitance. The resonators are connected by small capacitors $C_{j=1,2}\ll C$. We denote the voltages of the resonators as $V_n^{(j)}$, the currents of the inductances as $i_n^{(j)}$, and the currents of the capacitors as $I_n^{(j)}$. We further denote the voltages and currents of the nonlinear capacitors as $V_n'^{(j)}$ and $I_n'^{(j)}$, respectively. Finally, external active sources $\delta V_n^{(j)}$ as the nonlinear functions of $V_n^{(j)}$ are internally installed in resonators. Kirchhoff's law tells us
\begin{eqnarray}\label{G2.3}
 & {} & V_n'^{(1)} = V_{n-1}^{(2)}-V_n^{(1)}, \nonumber \\
 & {} & V_n'^{(2)} = V_n^{(1)} - V_n^{(2)}, \nonumber \\
 & {} & I_n'^{(1)} = i_n^{(1)} + I_n^{(1)} + I_n'^{(2)}, \nonumber \\
 & {} & I_n'^{(2)} = i_n^{(2)}+I_n^{(2)}+I_{n+1}'^{(1)}.
\end{eqnarray}
We also have 
\begin{eqnarray}\label{G2.4}
 & {} & I_n^{(j)} = C\dot{V}_n^{(j)},\nonumber \\
 & {} & I_n'^{(j)} = C_{j} \dot{V}_n'^{(j)},
\end{eqnarray}
for $j=1,2$. By adopting the limit of small capacitances~\cite{hadad2018self} $C_{j}\ll C$, from Eqs.(\ref{G2.3}, \ref{G2.4}) we obtain
\begin{eqnarray}\label{G2.6}
i_n^{(1)} 
 & \approx &  C_{1} \dot{V}_{n-1}^{(2)}+ C_{2} \dot{V}_n^{(2)}  - C\dot{V}_n^{(1)},\nonumber \\
i_n^{(2)}
 & \approx & C_{1} \dot{V}_{n+1}^{(1)}+C_{2}\dot{V}_n^{(1)}-C\dot{V}_n^{(2)}. 
\end{eqnarray}
The equations of motion for the inductances are 
\begin{eqnarray}\label{G2.12}
L\dot{i}_n^{(j)}+R i_n^{(j)}  = V_n^{(j)}-\delta V_n^{(j)} \overset{\rm def}{=} U_n^{(j)}.
\end{eqnarray}
where $\delta V_n^{(j)}\ll V_n^{(j)}$ is assumed here. Finally, we employ the approximation $C_{j}\ll C$ again to simplify Eq.(\ref{G2.12}) as follows, 
\begin{eqnarray}\label{G2.7}
 & {} & \frac{\ddot{V}_n^{(1)}}{\omega_0^{2}}+RC \dot{V}_n^{(1)} = -U_n^{(1)}-\frac{C_{1}}{C}U_{n-1}^{(2)} - \frac{C_{2}}{C}U_n^{(2)},\nonumber \\
 & {} & \frac{\ddot{V}_n^{(2)}}{\omega_0^2}+RC \dot{V}_n^{(2)} = -U_n^{(2)}-\frac{C_{1}}{C}U_{n+1}^{(1)}-\frac{C_{2}}{C}U_n^{(1)}. \qquad
\end{eqnarray}
We note that $C_j/C\ll 1$, which means 
\begin{eqnarray}\label{G2.71}
\frac{C_j}{C}U_n^{(j')} \approx \frac{C_j}{C}V_n^{(j')}.
\end{eqnarray}
Thus, Eqs.(\ref{G2.7}) further reduce to 
\begin{eqnarray}\label{G2.72}
 & {} & \frac{\ddot{V}_n^{(1)}}{\omega_0^{2}}+RC \dot{V}_n^{(1)} = -V_n^{(1)}+\delta V_n^{(1)}-\frac{C_{1}}{C}V_{n-1}^{(2)} - \frac{C_{2}}{C}V_n^{(2)},\nonumber \\
 & {} & \frac{\ddot{V}_n^{(2)}}{\omega_0^2}+RC \dot{V}_n^{(2)} = -V_n^{(2)}+\delta V_n^{(2)}-\frac{C_{1}}{C}V_{n+1}^{(1)}-\frac{C_{2}}{C}V_n^{(1)}. \nonumber \\
\end{eqnarray}
We then express the voltages as the envelope function 
\begin{eqnarray}\label{G2.8}
V_n^{(j)} = \Psi_n^{(j)}e^{-\mathrm{i}\omega_0 t}
\end{eqnarray}
which in turn gives us 
\begin{eqnarray}\label{G2.9}
\ddot{V}_n^{(j)} & = &  (\ddot{\Psi}_n^{(j)}-2\mathrm{i}\omega_0\dot{\Psi}_n^{(j)}-\omega_0^2\Psi_n^{(j)} )e^{-\mathrm{i}\omega_0 t} \nonumber \\
 & \approx & 
 (-2\mathrm{i}\omega_0\dot{\Psi}_n^{(j)}-\omega_0^2\Psi_n^{(j)}) e^{-\mathrm{i}\omega_0 t},
\end{eqnarray}
where in the second step we assume that the time-modulation of $V_n^{(j)}$ is mostly captured by the factor $e^{-\mathrm{i}\omega_0 t}$ and hence $\Psi_n^{(j)}$ varies slowly in time, giving $\ddot{\Psi}_n^{(j)}\ll \omega_0 \dot{\Psi}_n^{(j)}$. We denote the damping coefficient $\eta = RC\omega_0/2\ll 1$ for simplicity. It is at this point that we obtain the equations of motion which is expressed as generalized nonlinear Schr\"{o}dinger equations with small damping
\begin{eqnarray}\label{G2.10}
 & {} & \left(\mathrm{i}-\eta\right)\dot{\Psi}_n^{(1)}+\mathrm{i}\eta\omega_0\Psi_n^{(1)} = \nonumber \\
 & {} & \frac{\omega_0 C_{1}}{2C} \Psi_{n-1}^{(2)} + \frac{\omega_0 C_{2}}{2C} \Psi_n^{(2)}-\frac{\omega_0}{2}e^{\mathrm{i}\omega_0 t}\delta V_n^{(1)},\nonumber \\
 & {} & \left(\mathrm{i}-\eta\right)\dot{\Psi}_n^{(2)}+\mathrm{i}\eta\omega_0\Psi_n^{(2)} = \nonumber \\
 & {} & \frac{\omega_0 C_{1}}{2C} \Psi_{n+1}^{(1)}+\frac{\omega_0 C_{2}}{2C} \Psi_n^{(1)}-\frac{\omega_0}{2}e^{\mathrm{i}\omega_0 t}\delta V_n^{(2)}. 
\end{eqnarray}
Having established the nonlinear Schr\"{o}dinger equations, we now discuss how to realize the nonlinear interactions specified in Eq.(\ref{9v2}). This can be achieved by asking 
\begin{eqnarray}\label{G2.101}
\delta V^{(1)}_n & = & -\delta c_1 V_{n-1}^{(2)}-\delta c_2 V_{n}^{(2)} \nonumber \\
 & {} & -e^{-\mathrm{i}\omega_0 t}\bigg\{d_1\left[ ({\rm Re\,}\Psi_{n-1}^{(2)})^3+\mathrm{i}({\rm Im\,}\Psi_{n-1}^{(2)})^3\right] \nonumber \\
 & {} & +d_2\left[ ({\rm Re\,}\Psi_{n}^{(2)})^3+\mathrm{i}({\rm Im\,}\Psi_{n}^{(2)})^3\right]\bigg\}, \nonumber \\
\delta V^{(2)}_n & = & -\delta c_1 V_{n+1}^{(1)}-\delta c_2 V_{n}^{(1)} \nonumber \\
 & {} & 
-e^{-\mathrm{i}\omega_0 t}\bigg\{d_1\left[ ({\rm Re\,}\Psi_{n+1}^{(1)})^3+\mathrm{i}({\rm Im\,}\Psi_{n+1}^{(1)})^3\right] \nonumber \\
 & {} & +d_2\left[ ({\rm Re\,}\Psi_{n}^{(1)})^3+\mathrm{i}({\rm Im\,}\Psi_{n}^{(1)})^3\right]\bigg\}.
\end{eqnarray}

\widetext

\renewcommand{\thefigure}{F1}
\begin{figure}[htb]
\includegraphics[scale=0.55]{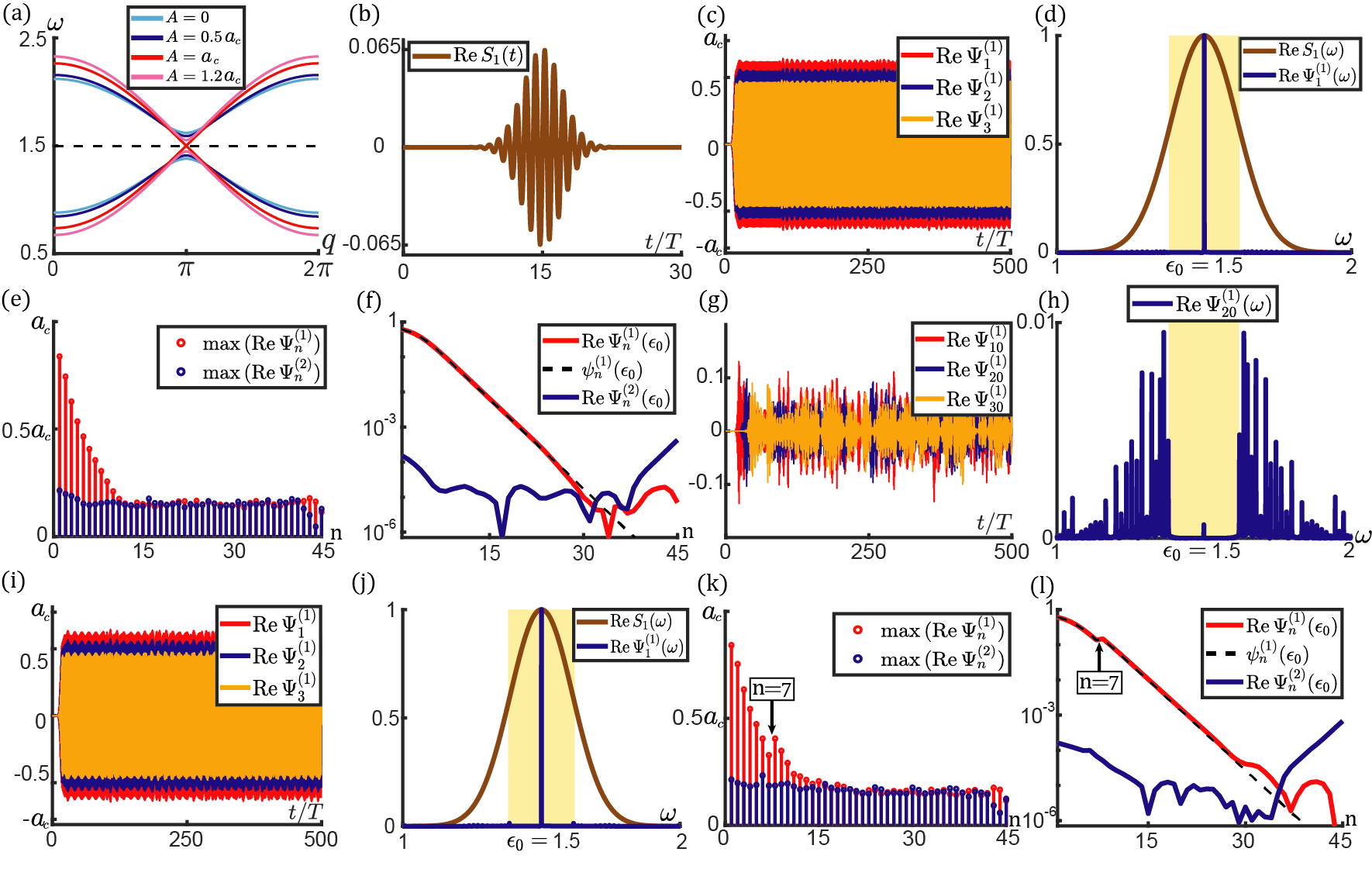}
\caption{Topological properties of nonlinear SSH lattice subjected to Kerr-type nonlinearities. The interaction parameters, including $\epsilon_0$, $c_1$, $c_2$, $d_1$ and $d_2$, are the same as those in Fig.\ref{fig1}. (a) Nonlinear band structures for various amplitudes ranging from $A=0$ to $1.2a_c$, where $a_c=\sqrt{-(c_2-c_1)/(d_2-d_1)}=0.7746$. The nonlinear bands touch at the critical amplitude $A=a_c$, and Berry phase changes abruptly from $\gamma(A<a_c) = \pi$ to $\gamma(A>a_c)=0$. (b) By constructing a finite lattice shown in Fig.\ref{fig2}(a) with open boundary conditions, we shake the boundary on site $n=1$ by imposing a Gaussian tone burst in Eq.(\ref{E1.4}) to excite nonlinear topological modes, where $S=6.5\times 10^{-2}$, $\omega_{\rm ext} = \epsilon_0=1.5$, $T=2\pi/\omega_{\rm ext}$, $\tau = 3T$, and $t_0 = 15T$. (c) Responding mode on $n=1,2,3$ sites indicates the mode localization, where the mode amplitude $\max({\rm Re\,}\Psi_1^{(1)})<a_c$. (d) Brown and blue curves stand for the frequency spectra of external Gaussian signal and $n=1$ site responding wave function. Yellow shaded area is the linear bandgap. (e) Spatial profile of the boundary excitation amplitude. We note that the bulk mode components reflected here are excited by Gaussian signal, which contains all frequencies. (f) Red and blue curves are the spatial profiles of the $\omega=\epsilon_0$ Fourier component of the boundary mode. The analytic result of the $\omega=\epsilon_0$ Fourier component is depicted by the black dashed curve. (g) The echo-like wave functions of sites $n=10,20,30$ manifest bulk mode components excited by the external Gaussian shaking signal. (h) The spectrum of $n=20$ site contains a wide range of frequency components of bulk modes. (i) We now study topological edge modes in the lattice depicted by Fig.\ref{SIfig4}(a), where the interaction parameters $c_1'$ and $d_1'$ of the defect bond at site $n=7$ are carried over from that figure. Responding modes are plotted for $n=1,2,3$ sites which exhibit the feature of mode localization. (j) Fourier analysis of frequency space for wave function at site $n=1$. (k) Spatial profile of the responding amplitudes. A noticeable bump at site $n=7$ is induced by the defect. (l) Spatial profile of the $\omega=\epsilon_0$ frequency component is captured by red and blue curves, and the theoretical analysis of this component is described by the black dashed curve. 
}\label{SIfig5}
\end{figure}

\endwidetext

\section{An analytically solvable topological model with Kerr-type nonlinear interactions}

We study an alternative model to provide additional verification of the nonlinear topological theory presented in this paper. The model is analytically solvable, in the sense that the nonlinear bulk modes as well as the dispersion relation can be exactly solved. The model is based on Eqs.(\ref{1v2}) with the Kerr-type nonlinearities~\cite{d2008ultraslow} on the field variables, 
\begin{eqnarray}\label{Analytic1}
f_i (x,y) = c_i y + d_i |y|^2 y,\quad i=1,2, 
\end{eqnarray}
where the parameters yield $0<c_1<c_2$ and $d_1>d_2>0$. This model is subjected to reflection symmetry. According to the main text, reflection symmetry demands the quantization of Berry phase of nonlinear bulk modes, regardless of the functional forms of interactions. This conclusion should remain valid for Kerr-type nonlinearities. To verify the quantization of Berry phase, we solve nonlinear traveling modes as below, 
\begin{eqnarray}\label{Analytic2}
\Psi_{n} = A(1, e^{-\mathrm{i}\phi_q})^\top e^{\mathrm{i}qn-\mathrm{i}\omega t},
\end{eqnarray}
where the dispersion relation is
\begin{eqnarray}\label{Analytic3}
\omega = \epsilon_0\pm\sqrt{c_1(A)^2+c_2(A)^2+2c_1(A)c_2(A)\cos q},\qquad
\end{eqnarray}
$c_i(A) = c_i + d_i A^2$, and the relative phase $\phi_q$ is
\begin{eqnarray}\label{Analytic4}
\phi_q = \arctan\left(\frac{-c_2(A)\sin q}{c_1(A)+c_2(A)\cos q}\right).
\end{eqnarray}
Following the convention of Fourier transformation in Eq.(\ref{B6}), the Fourier components of the sinusoidal nonlinear bulk mode are $\psi_{l,q}^{(1)} = \psi_{l,q}^{(2)} = A\delta_{l,1}$. According to Eq.(\ref{Analytic3}), the nonlinear bandgap never closes unless the wave amplitude hits the topological transition point $a_c$. Apart from $a_c$, Berry phase of nonlinear bulk modes is well-defined, and can be greatly simplified to the following result by employing the sinusoidal form of nonlinear waves,
\begin{eqnarray}\label{Analytic5}
\gamma(A)
=
\frac{1}{2}\mathrm{i}\oint_{\rm BZ} \mathrm{d}q \, \partial_q \ln [ c_1(A) +c_2(A)e^{\mathrm{i}q}].
\end{eqnarray}

According to our general theory, Berry phase is expected to be $\gamma(A<a_c) = \pi$ and $\gamma(A>a_c) = 0$, which holds true for arbitrary reflection-symmetric 1D systems and is independent of the functional forms of nonlinearities. This result is verified by evaluating Eq.(\ref{Analytic5}) for Kerr-type nonlinear interactions.

As stated by the nonlinear extension of bulk-boundary correspondence, nonlinear topological modes should emerge on the lattice open boundary when the bulk band is topologically non-trivial with Berry phase $\gamma=\pi$, whereas topological modes disappear when $\gamma=0$. Here we confirm this correspondence by studying the attributes of nonlinear topological edge modes. To this end, we Fourier transform the edge mode into frequency space, and truncating it to the fundamental harmonics, 
\begin{eqnarray}\label{Analytic6}
\Psi_n \approx  \psi_{-1,n} e^{\mathrm{i}\omega t}+\psi_{1,n} e^{-\mathrm{i}\omega t}.
\end{eqnarray}

Similarly, the nonlinear terms in the interactions are truncated as follows,
\begin{eqnarray}\label{Analytic7}
|\Psi_n^{(j)}|^2\Psi_n^{(j)} & \approx & (|\psi_{-1,n}^{(j)}|^2+2|\psi_{1,n}^{(j)}|^2) \psi^{(j)}_{-1,n} e^{\mathrm{i}\omega t} \nonumber \\
 & + & (2|\psi^{(j)}_{-1,n}|^2+|\psi^{(j)}_{1,n}|^2) \psi^{(j)}_{1,n} e^{-\mathrm{i}\omega t}.
\end{eqnarray}
Consequently, the equations of motion reduce to the following nonlinear recursion relations, 
\begin{eqnarray}\label{Analytic8}
 & {} & (\epsilon_0-s\omega) \psi_{s,n}^{(1)} +{C}_1(\psi_{s,n}^{(2)}) + {C}_2(\psi_{s,n-1}^{(2)})=0,\nonumber \\
 & {} & (\epsilon_0-s\omega)\psi_{s,n}^{(2)} +{C}_1(\psi_{s,n}^{(1)})+{C}_2(\psi_{s,n+1}^{(1)})=0,
\end{eqnarray}
where $s=\pm 1$, and 
\begin{eqnarray}\label{Analytic11}
{C}_i(\psi_{s,n}^{(j)})=c_i \psi_{s,n}^{(j)} + d_i (|\psi^{(j)}_{s,n}|^2+2|\psi^{(j)}_{-s,n}|^2) \psi^{(j)}_{s,n}.\qquad
\end{eqnarray}
We exploit the approximation $\psi^{(1)}_{s,n}\gg \psi^{(2)}_{s,n}$, which is numerically verified in Fig.\ref{SIfig5}(f). We solve Eqs.(\ref{Analytic8}) to find $\omega = \epsilon_0$, $\psi_{-1,n}^{(1)}=0$ for all $n$, ${\rm Arg\,}\psi_{1,n}^{(1)}={\rm Arg\,}\psi_{1,1}^{(1)}+(n-1)\pi$, and 
\begin{eqnarray}\label{Analytic10}
c_1(\psi_{1,n}^{(1)})|\psi_{1,n}^{(1)}| = c_2(\psi_{1,n+1}^{(1)})|\psi_{1,n+1}^{(1)}|, 
\end{eqnarray}
where $c_i(x) = c_i + d_i |x|^2$. Based on Eq.(\ref{Analytic10}), when $|\psi_{1,1}^{(1)}|<a_c$, an evanescent mode fades away from the lattice boundary, whereas for $|\psi_{1,1}^{(1)}|>a_c$, an unphysical mode quickly diverges to infinity and therefore cannot exist. The emergence and disappearance of edge modes are in accordance with topologically non-trivial and trivial Berry phases, which is the manifestation of bulk-boundary correspondence with Kerr-type nonlinearities.

%

\end{document}